\documentclass[11pt]{article}
\usepackage{mathptmx}
\usepackage{amssymb,amsmath}
\newtheorem{thm}{Theorem}[section]
\newtheorem{cor}[thm]{Corollary}
\newtheorem{lem}[thm]{Lemma}
\newtheorem{definition}[thm]{Definition}
\newtheorem{prop}[thm]{Proposition}
\def\proof{{\bf Proof. }}
\newtheorem{remark}[thm]{Remark}
\def\be{\begin{equation}}
\def\ee{\end{equation}}
\def\bea{\begin{eqnarray}}
\def\eea{\end{eqnarray}}
\def\bean{\begin{eqnarray*}}
\def\eean{\end{eqnarray*}}
\def\ea{\end{array}}
\def\ds{\displaystyle}
\def\nm{\noalign{\medskip}}
\def\Z{{\mathcal{Z}}}
\def\O{{\mathcal{O}}}
\def\W{{\mathcal{W}}}
\def\F{{\mathcal{F}}}
\def\P{{\mathcal{P}}}
\def\V{{\mathcal{V}}}
\def\L{{\mathcal{L}}}
\def\H{{\mathcal{H}}}
\def\S{{\mathcal{S}}}
\def\D{{\mathcal{D}}}

\def\p{{\mathbb{P}}}
\newcommand{\field}[1]{\mathbb{#1}}
\newcommand{\rz}{\field{R}}
\newcommand{\cz}{\field{C}}
\newcommand{\nz}{\field{N}}
\newcommand{\zz}{\field{Z}}

\def\ccup{\mathop{\cup}}
\def\ccap{\mathop{\cap}}

\def\d{{\rm{d}}}
\def\ra{{\rangle}}
\def\la{{\langle}}

\def\fin{{$\hfill\square$}}
\def\hbarr{{\varepsilon}}
\def\real{\textrm{Re}}
\def\Im{\textrm{Im}}
\def\Tr{\textrm{Tr}}
\begin{document}
\title{Mean field limit for bosons and infinite dimensional phase-space analysis}
\author{Z.~Ammari\thanks{Department of Mathematics
Universit{\'e} de Cergy-Pontoise UMR-CNRS 8088,
2, avenue Adolphe Chauvin
95302 Cergy-Pontoise Cedex France. Email: zied.ammari@u-cergy.fr}
\hspace{.3in} F.~Nier
\thanks{IRMAR, UMR-CNRS 6625, Universit\'e de Rennes I,
campus de Beaulieu, 35042 Rennes
Cedex, France. Email: francis.nier@univ-rennes1.fr} }

\renewcommand{\today}{november 2007}
\maketitle
\begin{abstract}
This article proposes the construction of Wigner measures in
the infinite dimensional bosonic quantum field theory, with
applications to the derivation of the mean field dynamics.
Once these asymptotic objects are well defined, it is shown how
they can be used to make connections between different kinds of results
or to prove new ones.
\end{abstract}
{\footnotesize{\it 2000 Mathematics subject  classification}: 81S30, 81S05, 81T10, 35Q55 }
{\footnotesize\tableofcontents}

\section{Introduction}
The bosonic quantum field theory relies on two different bases : On
one side the
quantization of a symplectic space, the approach followed for
example by Berezin in \cite{Ber}, Kree-Raczka in \cite{KrRa}; on
the other side the gaussian
stochastic processes presentation also known as the integral
functional point of view followed for example by Glimm-Jaffe
in \cite{GlJa} and Simon in \cite{Sim}. Both approaches have
to be handled in order to tackle on the most basic problems in
constructive
quantum field theory (see
\cite{BSZ}\cite{DeGe}).
The interaction of constructive quantum field theory with other fields of
mathematics like pseudodifferential calculus (see \cite{BeSh} or
\cite{Las}) or stochastic processes (see \cite{Mey}\cite{AtPa}) is
often instructive.\\
In the recent years the mean field limit of $N$-body quantum dynamics
has been reconsidered by various authors via a BBGKY-hierarchy
approach
(see \cite{ESY1}\cite{ESY2}\cite{FGS}\cite{FKP}\cite{BGGM}\cite{Spo} and
\cite{Ger} for a short presentation) mainly motivated by the
study of Bose-Einstein condensates (see \cite{Cas}).
Although this was present in earlier works around the so-called Hepp
method (see \cite{Hep} and \cite{GiVe}), the relationship with
the  microlocal or semiclassical analysis in  infinite
dimension has been neglected. Difficulties are known in this
direction : 1) The gap between the inductive and projective
construction of quantized observable in infinite dimension;
2) the difficulties to built algebras of pseudodifferential
operators which contain the usual hamiltonians and preserve some
properties of the finite dimensional calculus like a
Calderon-Vaillancourt theorem, a good notion of ellipticity or the
asymptotic positivity with a G{\aa}rding inequality;
3) even when step 2) is possible, no satisfactory Egorov theorem is
available.\\
Recall the example of an $N$-body Schr{\"o}dinger hamiltonian
$$
H_{N}=-\Delta+\frac{1}{N}\sum_{1\leq i<j\leq N}
V(x_{i}-x_{j})\quad,\qquad
\text{on}\ \rz^{dN}\,,
$$
and consider the time-evolved wave function
$$
\Psi_{N}(t)=e^{-itH_{N}}\, \psi^{\otimes N}\quad,\qquad \psi\in L^{2}(\rz^{d})\,.
$$
The $1$-particle marginal state, the quantum analogous of the one particle
empirical distribution in the classical $N$-body problem, is given by
$$
\Tr\left[A\varrho^{1}(t)\right]=\left\langle \Psi_{N}(t)\,,
  \frac{1}{N}\left[
\sum_{i=1}^{N}I\otimes\cdots I\otimes
I\otimes
\underbrace{A}_{i}\otimes I\otimes\cdots \otimes I\right]\Psi_{N}(t)\right\rangle
$$
The mean field limit says that in the limit $N\to\infty$, the marginal
state evolves according to a non-linear Hartree equation
\begin{eqnarray*}
  &&\varrho^{1}(t)=\left|z(t)\rangle\langle
    z(t)\right|+o(1)\quad,\qquad\text{as}\ N\to\infty\,,
\\
\text{with}
&&
\left\{
  \begin{array}[c]{l}
    i\partial_t z=-\Delta z+(V*\left|z\right|^{2})z\quad\text{on}\
    \rz_{t}\times\rz^{d}\\
    z(t=0)=\psi\,.
  \end{array}
\right.
\end{eqnarray*}
By setting $N=\frac{1}{\varepsilon}$ and in the Fock space framework with
$\varepsilon$-dependent CCR (i.e:
$\left[a(g),a^{*}(f)\right]=\varepsilon\left\langle
  g,\,f\right\rangle$), the problem becomes
\begin{eqnarray*}
  && H_{N}=\frac{1}{\varepsilon}\left[\int_{\rz^{d}}
\nabla a^{*}(x)\nabla
a(x)~dx+\int_{\rz^{2d}}V(x-y)a^{*}(x)a^{*}(y)a(x)a(y)~dxdy
\right]=\frac{1}{\varepsilon}H^{\varepsilon}
\\
&&
e^{-itH_{N}}=e^{-i\frac{t}{\varepsilon}H^{\varepsilon}}\,,\\
&&
\Tr\left[A\varrho^{1}(t)\right]\,
= \big\langle
\Psi_{N}(t)\,,\,d\Gamma(A)\Psi_{N}(t)\big\rangle
=
\big\langle
\Psi_{N}(t)\,,\,p_{A}(z)^{Wick}\Psi_{N}(t)\big\rangle\,,
\end{eqnarray*}
where $p_{A}$ is the polynomial $p_{A}(z)=\langle z\,, A z\rangle$\,.
Higher order marginals, taking into accounts correlations, can be
defined after using the polynomials $p_{A}(z)=\langle z^{\otimes k}\,,\,
A z^{\otimes k}\rangle$ with $A\in \mathcal{L}(L^{2}(\rz^{kd}))$\,.\\
On this example, the scaling of the hamiltonian, of the time scale and
of the observables as Wick  operators enters
formally in the $\varepsilon$-dependent semiclassical analysis.
The Hepp method concerns the evolution of squeezed coherent states
(\cite{Hep}\cite{GiVe}\cite{Cas}), which amounts in the finite
dimensional case to the phase-space evolution of a gaussian state  according
to the time dependent quadratic approximation of the non linear
hamiltonian, centered on the solution to the classical hamiltonian
equation.
We refer the reader to \cite{CRR} for accurate developments of such an
approach in the finite dimensional case.\\
In the nineties and as a byproduct of the development of microlocal analysis,
alternative and more flexible methods were introduced in order to
study the semiclassical limit with the help of Wigner (or
semiclassical) measures (see
\cite{Bur}\cite{Ger}\cite{HMR}\cite{LiPa}\cite{Tar}).
Such objects are defined by duality and rely
on the asymptotic positivity of the $\varepsilon$-dependent
quantizations.
It gives a weak but more flexible form of the principal term of the
semiclassical (here mean-field) approximation. Via the introduction of
probability measures on the symplectic phase-space, it provides an
interesting way to analyze the relationship between the two basic
approaches to quantum field theory.
Further in finite dimension, the Wick, anti-Wick and Weyl
quantizations are asymptotically equivalent in the limit
$\varepsilon\to 0$. This is not so obvious in infinite dimension.\\
Several attempts have been tried to develop an infinite dimensional
Weyl pseudodifferential calculus with an inductive approach.
Lascar in \cite{Las} introduced an algebra and a notion of ellipticity
in this direction, making more effective the general presentation of
\cite{KrRa}. The works of Helffer-Sj{\"o}strand in \cite{Hel2}\cite{HeSj} and
Amour-Kerdelhu{\'e}-Nourrigat in \cite{AKN} about the pseudodifferential calculus in large
dimension motivated by the analysis of the thermodynamical limit enter
in this category. With such an approach, it is not clear that the
infinite dimensional phase-space is well explored and that no
information is lost in the limit $\varepsilon\to 0$. Meanwhile this
inductive approach is limited by Hilbert-Schmidt type restriction like
in Shale's theorem about the quasi-equivalence of gaussian measures.
It is known after \cite{Gro} that the nonlinear
transformations which preserve the quasi-equivalence with a given
gaussian measure within the Schr{\"o}dinger representation are very
restricted and do not cover realistic models. Hence no Egorov
theorem can be expected with Weyl observables.\\
Simple remarks suggests alternative point of views. The Wick calculus
with  polynomial symbols present encouraging specificities: It
contains the standard hamiltonians, it makes an algebra under more general
assumptions (the Hilbert-Schmidt condition can be relaxed) and allows
some propagation results when tested on appropriate  states (see
\cite{FGS}\cite{FKP}).
Meanwhile the Wigner measures in the limit
$\varepsilon\to 0$ can be defined very easily via the separation
of variables as weak distribution, in a projective way which fits with
the stochastic processes point of view.\\
After reviewing and sometimes simplifying or improving known results
and techniques about the mean field limit, our aim is to show the
interests of the extension to the infinite dimensional case of Wigner
measures:
\begin{itemize}
\item After the introduction of the small parameter $\varepsilon\to 0$
  and the definition of Weyl operator $W(z)$, $z\in\Z$ the
  phase-space, choosing between the quantization of symplectic space
  and the stochastic processes point of view is no more a question of
  general principles nor of mathematical taste. It is a matter of
  scaling. The symplectic geometry arises when considering macroscopic
  phase-space translation $W(\frac{z}{\varepsilon})$, while the
  operator $W(z)$ is used with this scaling in the introduction of
  Wigner measures via their characteristic function.
  Corrections to the mean field limit considered for example in \cite{CCD} with a
  stochastic processes point of view can be interpreted within this
  picture: They attempt to give a better
  information on the shape of the state in a small phase-space scale.
\item Once the Wigner measures are well defined as Radon measures, it
  is possible to make explicit the relationship between different
  kinds of results and to extend them in a flexible way. It accounts
  for the propagation of chaos (result obtained via the BBGKY
  approach)  according to the classical hamiltonian
  dynamics in the phase-space. Actually we shall prove in a very
  general framework that the propagation of squeezed coherent states
  as derived via the Hepp method implies a weak version of the mean
  field limit for product states. Further propagation results can be
  obtained for some non standard mixed states without reconsidering a
  rather heavy analysis process.
\item The comparison between the Wick, Weyl and anti-Wick quantization
  can be analyzed accurately in the infinite dimensional case. With
  the Wick calculus, complete asymptotic expansions can be proved
  after testing with some specific states. The relationship of such
  results with the propagation of Wigner measures works in a rather
  general setting but has to be handled with care.
\item The gap between the projective and inductive approaches can be
  formulated accurately in the limit $\varepsilon\to 0$. We shall
  explain in the examples the possibility of a dimensional defect of compactness.
\end{itemize}
This work is presented and illustrated with examples simpler than more realistic models
considered in other works like \cite{GiVe}\cite{Hep}\cite{ESY1}\cite{ESY2}\cite{BGGM} with more
singular interaction potentials. That was our choice in order to make the correspondence between
various approaches more straightforward and to pave the way for further improvements.
We hope that this information will be valuable for other colleagues and useful
for further developments.

The outline of this articles is the following.
In Section~\ref{se.FspWick}, standard notions about the symmetric Fock
space are recalled and Wick calculus is specified.
In Section~\ref{se.WeylAWick} the Weyl and Anti-Wick calculus are
introduced in a projective way after recalling accurately (most of all the
scaling) of finite dimensional semiclassical calculus.
The Section~\ref{se.coherentprod} recalls the distinction between
coherent states and product or Hermite states, and their properties
when measured with different kinds of observables.
The two methods used to derive the mean field dynamics, the Hepp
method and the analysis through truncated Dyson expansions, are
reviewed within our formalism and with some variations
in Section~\ref{se.anex}. The Wigner measures are introduced in
Section~\ref{se.WigMeas} with the extension of some finite dimensional
properties and specific infinite dimensional phenomena. Finally
examples and applications are detailed in
Section~\ref{se.exampleappli}, in particular: 1) reconsidering a simple presentation
of the Bose-Einstein
condensation shows an interesting example of what we call the
dimensional defect of compactness; 2) a general result
says that the propagation of squeezed coherent states, which can be
attacked via the Hepp method, implies  a slightly weaker form of the propagation of chaos
(formulated with product states and Wick observables); 3) the mean field
dynamics can be easily derived for some  states which present some
asymptotically vanishing correlations.

\medskip
\noindent\textbf{Acknowledgements:} The authors would like to thank
V.~Bach, Y.~Coud{\`e}ne, J.~Fr{\"o}hlich, V.~Georgescu, C.~G{\'e}rard, P.~G{\'e}rard, S.~Graffi, T.~Jecko,
S.~Keraani and A.~Pizzo for profitable discussions related with this
work. This was partly completed while the first author had a sabbatical
semester in CNRS in spring 2007.

\section{Fock space and Wick quantization}
\label{se.FspWick}
After introducing the symmetric Fock space with
$\varepsilon$-dependent CCR's, an algebra of
observables  resulting from the
Wick quantization process is presented.

\subsection{Fock space}
Consider  a separable Hilbert space $\mathcal{Z}$ endowed with a
scalar product $\la .,.\ra$ which is anti-linear in the left
argument and linear in the right one and with the associated norm
$|z|=\sqrt{\la z,z\ra}$. Let $\sigma={\rm Im}\la ., .\ra$ and
$S={\rm Re} \la .,.\ra$ respectively denote the canonical symplectic
and
the real scalar product over $\mathcal{Z}$. The symmetric Fock space on $\mathcal{Z}$
is the Hilbert space
\begin{eqnarray*}
\mathcal{H}=\bigoplus_{n=0}^\infty\bigvee^n
\mathcal{Z}=\Gamma_{s}(\Z)\,,
\end{eqnarray*}
where $\bigvee^n \mathcal{Z}$ is
the $n$-fold symmetric tensor product. Almost all the direct sums
and tensor products are completed within the Hilbert framework. This
is omitted in the notation. On the contrary, a specific
$^{\textrm{alg}}$ superscript will be used for the algebraic direct
sums or tensor products.

For any $n\in \nz$, the orthogonal projection of $\bigotimes^n
\Z$ onto the closed subspace $\bigvee^n\Z$ will be denoted
by $\S_{n}$. For any $(\xi_{1},\xi_{2},\ldots,\xi_{n})\in \Z^{n}$,
 the vector $\xi_{1}\vee \xi_{2}\vee\cdots \vee\xi_{n}\in
 \bigvee^{n}\Z$
will be
$$
\xi_{1}\vee\xi_{2}\vee\cdots\vee\xi_{n}
=
\S_{n}(\xi_{1}\otimes\xi_{2}\cdots\otimes \xi_{n})
=\frac{1}{n!}\sum_{\sigma\in \Sigma_{n}}\xi_{\sigma(1)}\otimes\xi_{\sigma(2)}\cdots\otimes \xi_{\sigma(n)}
$$
The  family of vectors $(\xi_{1}\vee\cdots\vee \xi_{n})_{\xi_{i}\in
\Z}$ is a generating family of
$\bigvee^{n,alg}\Z$ and a total family of $\bigvee^{n}\Z$\,. Thanks to
the polarization identity
\begin{equation}
\label{eq.pola}
\xi_{1}\vee\xi_{2}\vee\cdots\vee\xi_{n}=\ds \frac{1}{2^n n!}
\sum_{\varepsilon_i=\pm 1} \varepsilon_1\cdots \varepsilon_n \;
\big( \sum_{j=1}^n \varepsilon_j
\xi_j)^{\otimes n}\,,
\end{equation}
the same property holds for the family $\left(z^{\otimes
    n}\right)_{n\in \nz, z\in \Z}$\,.

For two operators
$A_{k}:\bigvee^{i_{k}}\Z\to \bigvee^{j_{k}}\Z$, $k=1,2$, the
notation $A_{1}\bigvee A_{2}$ stands for
$$
A_{1}\bigvee A_{2}=\S_{j_{1}+j_{2}}\circ(A_{1}\otimes A_{2})\circ
\S_{i_{1}+i_{2}}\in
{\cal
  L}(\bigvee^{i_{1}+i_{2}}\Z, \bigvee^{j_{1}+j_{2}}\Z)\,.
$$
Any $z\in \Z$ is identified with the operator from
$\bigvee^{0}\Z=\cz\ni \lambda\mapsto \lambda z\in \Z=\bigvee^{1}\Z$
while $\la z|$ denotes the linear form $\Z\ni\xi\mapsto \la z\,,\,\xi\ra \in \cz$.
The creation and annihilation operators
 $a^*(z)$ and $a(z)$, parameterized by $\hbarr>0$,
are then defined by :
\begin{eqnarray*}
a(z)_{|\bigvee^n\Z}&=&\sqrt{\hbarr n} \; \;\la z|\otimes
I_{\bigvee^{n-1}\Z}\\
a^*(z)_{|\bigvee^n\Z}&=&\sqrt{\hbarr (n+1)}  \;\;\S_{n+1} \circ
(\;z\otimes I_{\bigvee^n\Z})=\sqrt{\hbarr(n+1)} \;z\bigvee
I_{\bigvee^{n}\Z}\,.
\end{eqnarray*}
Each of  $(a(z))_{z\in \Z}$ and $(a^*(z))_{z\in \Z}$
 are commuting families of operators and they
satisfy the canonical commutation relations (CCR):
\begin{eqnarray} \label{ccr}
[a(z_1),a^*(z_2)]=\hbarr\la z_1,z_2\ra I. \end{eqnarray}
We  also consider the
canonical quantization of the real
  variables
  $\Phi(z)=\frac{1}{\sqrt{2}} (a^*(z)+a(z))$ and
$\Pi(z)=\Phi(iz)=\frac{1}{i\sqrt{2}} (a(z)-a^*(z))$. They are
self-adjoint operators on $\mathcal{H}$ and satisfy the identities:
\begin{eqnarray*} [\Phi(z_1),\Phi(z_2)]=i \hbarr \sigma(z_1,z_2) I,
\hspace{.3in} [\Phi(z_1),\Pi(z_2)]=i\hbarr S(z_1,z_2) I. \end{eqnarray*} The
representation of the  Weyl commutation relations in the Fock space
\begin{eqnarray}
\label{eq.Weylcomm}
W(z_1) W(z_2)&=&e^{-\frac{i\hbarr}{2} \sigma(z_1,z_2)} W(z_1+z_2) \\
\nonumber
 &=& e^{-i \hbarr \sigma(z_1,z_2)} W(z_2) W(z_1),
  \end{eqnarray}
is obtained by setting $W(z)=e^{i\Phi(z)}$. The   generating functional
associated with this representation is given by
\begin{eqnarray*}
\la \Omega,
W(z)\Omega\ra= e^{-\frac{\hbarr}{4} |z|^2},
\end{eqnarray*}
where $\Omega$ is the vacuum vector $(1,0,\cdots)\in\H$.
The total family of
vectors
$E(z)=W\left(\frac{\sqrt{2}z}{i\hbarr}\right)\Omega=e^{\frac{1}{\hbarr}
\left[a^{*}(z)-a(z)\right]}\Omega$,
$z\in\Z$,  have the explicit form
\begin{eqnarray} \label{coherent-vect} E(z)&=&
e^{-\frac{|z|^{2}}{2\hbarr}}\sum_{n=0}^\infty \frac{1}{\hbarr^{n}}
\frac{a^*(z)^{n}}{n!}\Omega \nonumber\\
&=&e^{-\frac{|z|^{2}}{2\hbarr}} \sum_{n=0}^\infty  \hbarr^{-n/2}
\frac{z^{\otimes n}}{\sqrt{n!}}\,.
\end{eqnarray}
The   number operator is also parametrized by $\hbarr>0$,
\begin{eqnarray*}
N_{|\bigvee^n \Z}=\varepsilon{n} I_{|\bigvee^n \Z}.
\end{eqnarray*}
It is convenient to introduce the subspace
\begin{eqnarray*}
\H_{fin}=\bigoplus_{n\in\nz}^{\textrm{alg}}\bigvee^n\Z
\end{eqnarray*}
of $\H$, which is a set of analytic vectors for $N$.\\
For any contraction $S\in {\cal L}(\Z)$,
$\left|S\right|_{\mathcal{L}(\mathcal{H})}\leq 1$,
$\Gamma(S)$ is the
contraction in $\H$  defined by
$$
\Gamma(S)_{|\bigvee^{n}\Z}=S\otimes S\cdots\otimes S\,.
$$
More generally $\Gamma(B)$ can be defined by the same formula as an
operator on $\H_{fin}$ for any $B\in \mathcal{L}(\Z)$.
Meanwhile, for any self-adjoint operator $A:\Z\supset \D(A)\to \Z,$
the operator $d\Gamma(A)$ is the self-adjoint operator given by \begin{eqnarray*} &&
e^{\frac{it}{\varepsilon}d\Gamma(A)}=\Gamma(e^{itA})\\
&& d\Gamma(A)_{|\bigvee^{n,\textrm{alg}}\D(A)}=\hbarr\left[
\sum_{k=1}^{n}I\otimes\cdots\otimes\underbrace{A}_{k}\otimes
\cdots\otimes I\right]\,. \end{eqnarray*}
 For example $N=d\Gamma(I)$\,.

\subsection{Wick operators}
\label{sec.Wick} In this subsection we consider the Wick symbolic
calculus on (homogenous) polynomials. We will show some product and
commutation formulas  useful later for the application. For example
time evolved Wick observables can be expressed as
$\hbarr$-asymptotic expansion of quantized Wick symbols. For a
detailed exposition on  more general Wick polynomials we refer
the reader to \cite{DeGe}.

A $(p,q)$-homogeneous polynomial function of $z\in \Z$ is defined as
$P_{\ell}(z)=\ell(z^{\otimes q},z^{\otimes p})$, where $\ell$ is a sesquilinear form on
$(\bigotimes^{q,alg}\Z)\times(\bigotimes^{p,alg}\Z)$, with
$P_{\ell}(\lambda z)={\bar \lambda}^{q}\lambda^{p}P_{\ell}(z)$.
Owing to the polarization formula  (\ref{eq.pola}) and the identity
\begin{eqnarray*}
&&\ell(\eta^{\otimes q},\xi^{\otimes p})= \int_0^1\int_0^1 \ell(
[e^{2i\pi \theta} \eta+e^{2i\pi \varphi} \xi]^{\otimes q},[e^{2i\pi \theta} \eta+e^{2i\pi \varphi} \xi]^{\otimes p} ) \; e^{2i\pi(q\theta-p\varphi)}  \; d\theta \,d\varphi
\end{eqnarray*}
the correspondence $\ell\mapsto P_{\ell}$ is a bijection when the set of forms
is restricted to the sesquilinear forms on
$(\bigvee^{q,alg}\Z)\times
(\bigvee^{p,alg}\Z)$. Any of the continuity properties of $P_{\ell}$
are thus encoded by the continuity properties of the sesquilinear form
$\ell$ with the following hierarchy (from the weakest to the strongest)
\begin{eqnarray}
\nonumber
\left|\ell(\eta_{1}\vee\ldots\vee \eta_{q},\xi_{1}\vee\ldots \vee
  \xi_{p})\right|
\leq C_{\ell}\left|\eta_{1}\right|_{\Z}\ldots\left|\eta_{q}\right|_{\Z}
\left|\xi_{1}\right|_{\Z}\ldots\left|\xi_{p}\right|_{\Z},
&&
\eta_{i}\in \Z,\xi_{j}\in \Z
\\
\label{eq.contpol}
|\ell(\phi,\psi)|\leq
C_{\ell}\left|\phi\right|_{\bigvee^{q}\Z}\left|\psi\right|_{\bigvee^{p}\Z},
&&
\psi\in\bigvee^{p}\Z, \phi \in \bigvee^{q}\Z\\
\nonumber
|\sum_{1\leq i,j\leq K}c_{i,j}
\ell(\phi_{i},\psi_{j})|
\leq C_{\ell}\left|\sum_{1\leq i,j\leq K}c_{i,j}\la \phi_{i}|\otimes
  \psi_{j}\right|_{(\bigvee^{q}\Z)^{*}\otimes (\bigvee^{p} \Z)}\,,
&&
K\in\nz, c_{ij}\in\cz,\\
\nonumber
&& \phi_{i}\in \bigvee^{q}\Z,
\psi_{j}\in \bigvee^{p}\Z\,.
\end{eqnarray}
For example, when $p=q=1$ the two first ones define $\L(\Z)$, while the
third one defines the space of Hilbert-Schmidt operators. By Taylor expansion any $(p,q)$-homogenous
polynomial $P$ admits G\^{a}teaux differentials and we set
\begin{eqnarray*}
 \partial_{\overline{z}}^{k}\partial_{z}^{k'}P(z)[u_1,\cdots,u_k,v_1,\cdots,v_{k'}]=
 \bar\partial_{u_1}\cdots \bar\partial_{u_k}\partial_{v_1}\cdots \partial_{v_{k'}}P(z)
\end{eqnarray*}
where $\bar\partial_{u}, \partial_v$ are the complex directional derivatives relative to $u,v\in\Z$.
\begin{definition}
\label{de.hompol}
For $p,q\in \nz$, the set of $(p,q)$-homogeneous polynomial
functions on $\Z$ which satisfy the continuity condition
\eqref{eq.contpol} is denoted by $\P_{p,q}(\Z)$:
$$
\left(b(z)\in \P_{p,q}(\Z)\right) \Leftrightarrow \left\{
\begin{array}[c]{l}
  \tilde{b}=\frac{1}{p!}\frac{1}{q!}
\partial_{z}^{p}\partial_{\overline{z}}^{q}b(z)\in
   \L(\bigvee^p\Z, \bigvee^q\Z)\,,
\\
b(z)=\left\langle z^{\otimes q}\,,\,\tilde{b}z^{\otimes
p}\right\rangle\,.
  \end{array}
\right.
$$
The subspace of $\P_{p,q}(\Z)$ made of polynomials $b$ such that
$\tilde{b}$ is a compact operator $\tilde{b}\in
\mathcal{L}^{\infty}(\Z)$ (resp. $b\in \mathcal{L}^{r}(\Z)$) is
denoted by $\mathcal{P}^{\infty}_{p,q}(\Z)$ (resp. $\mathcal{P}^{r}_{p,q}(\Z)$).
\end{definition}
It will be sometimes convenient to consider $\tilde{b}$ as an operator
from $\bigotimes^{p}\Z$ into $\bigotimes^{q}\Z$ with the obvious
convention for symmetric operators $\tilde{b}=\mathcal{S}_{q}\tilde{b}\mathcal{S}_{p}$\,.
Owing to the condition $\tilde{b}\in  \L(\bigvee^p\Z,
\bigvee^q\Z)$ for $b\in \P_{p,q}(\Z)$, this definition implies that any
differential $\partial_{\overline{z}}^{j}\partial_{z}^{k}b(z)$ at the
point $z\in \Z$ equals
\begin{equation}
  \label{eq.differential}
\partial_{\overline{z}}^{j}\partial_{z}^{k}b(z)
=
\frac{p!}{(p-k)!}\frac{q!}{(q-j)!}(\la z^{\otimes q-j}|\bigvee
I_{\bigvee^{j}\Z})\tilde{b}(z^{\otimes p-k}\bigvee I_{\bigvee^k\Z})\quad
\in \L(\bigvee^{k}\Z,\bigvee^{j}\Z)\,.
\end{equation}
We will mainly work with fixed homogeneity degrees $p,q$ but the key
statement of this section (Proposition~\ref{symbcalc}) says
that  $\oplus_{p,q\in \nz}^{\textrm{alg}}\P_{p,q}(\Z)$ is an
algebra of symbols with the same explicit product formula as in the
finite dimensional case.

With any "symbol" $b\in \P_{p,q}(\Z)$, a Wick monomial $b^{Wick}$ can
be associated according to:
\begin{eqnarray} \label{def-wick}\nonumber
&&b^{Wick}:\H_{fin}\to\H_{fin},\\
&&b^{Wick}_{|\bigvee^n \Z}=1_{[p,+\infty)}(n)\frac{\sqrt{n!
(n+q-p)!}}{(n-p)!} \;\hbarr^{\frac{p+q}{2}} \;\left(\tilde{b}\bigvee
I_{\bigvee^{n-p} \Z}\right)\quad \in {\cal
L}(\bigvee^{n}\Z,\bigvee^{n+q-p}\Z)\,,
\end{eqnarray}
with
$\tilde{b}=(p!)^{-1}(q!)^{-1}\partial_{z}^{p}\partial_{\overline{z}}^{q}b(z)$\,.
\bigskip

Here are the basic symbol-operator correspondence:
\begin{eqnarray*}
&&\begin{array}{ccc}
 \la z,\xi\ra & \longleftrightarrow & a^*(\xi) \\
  \la \xi, z\ra & \longleftrightarrow & a(\xi) \\
\end{array}
\hspace{0.7cm}
\begin{array}{ccc}
\sqrt{2} S(\xi, z)  & \longleftrightarrow & \Phi(\xi)\\
\sqrt{2} \sigma(\xi, z)  & \longleftrightarrow &  \Pi(\xi)\,
\end{array}
\hspace{0.7cm}
\begin{array}{ccc}
\la z, A z\ra   & \longleftrightarrow &  d\Gamma(A) \\
\left|z\right|^{2}  & \longleftrightarrow &  N\,.
\end{array}
\end{eqnarray*}
Other examples can be derived from the next propositions.
The first one is a direct consequence of the definition
\eqref{def-wick}.
\begin{prop}
\label{wick_prop}
The following identities hold true on $\H_{fin}$ for every $b\in\P_{p,q}(\Z)$:\\
(i) $ \big(b^{Wick}\big)^*=\bar b^{Wick}$.\\
(ii) $\big(C(z) b(z) A(z)\big)^{Wick}=C^{Wick} b^{Wick} A^{Wick},$
if $A\in\P_{\alpha,0}(\Z)$, $
C\in\P_{0,\beta}(\Z).$\\
(iii) $e^{i \frac{t}{\hbarr} \d\Gamma(A)} b^{Wick} e^{-i
  \frac{t}{\hbarr} \d\Gamma(A)}=\left(b(e^{-it A}z)\right)^{Wick}$,
if $A$ is a self-adjoint operator on $\Z$.
\end{prop}

\begin{prop}
\label{pr.wickformules}
(i) The Wick operator associated with
$\ds b(z)=\prod_{i=1}^p \la z,\eta_i\ra \times \prod_{j=1}^q \la
\xi_j,z\ra$, $\eta_{i}, \xi_{j}\in \Z$, equals
\begin{eqnarray*}
b^{Wick}=a^*(\eta_1)\cdots a^*(\eta_p)a(\xi_1)\cdots a(\xi_q).
\end{eqnarray*}
(ii) For $b\in
  \P_{p,q}(\Z)$ and $z\in \Z$ the equality
\begin{eqnarray}
\label{eq.3} \la z^{\otimes j},b^{Wick} z^{\otimes k}\ra=
\delta^{+}_{k-p, j-q} \; \sqrt{\frac{k!j!}{(k-p)!
    (j-q)!}} \;
\hbarr^{\frac{p+q}{2}}\; \;  \left|z\right|^{k-p+j-q}b(z)
\end{eqnarray}
holds for any $k,j\in \nz$. The symbol
$\delta_{\alpha,\beta}^{+}$ denotes
$\delta_{\alpha,\beta}1_{[0,+\infty)}(\alpha)$ where
$\delta_{\alpha,\beta}$ is the standard Kronecker symbol.
\end{prop}
\proof \noindent (i) is a direct consequence of Proposition
\ref{wick_prop} with $(\la z\,,\,\xi\ra)^{Wick}=a^{*}(\xi)$ and
$(\la \xi, z\ra)^{Wick}=a(\xi)$\,.\\
(ii) This comes directly from the definition
\eqref{def-wick} of $b^{Wick}$\,.\fin
\
\bigskip

\noindent
The next result specifies the boundedness properties of $b^{Wick}$.
\begin{lem}
\label{wick-estimate}
For $b\in\P_{p,q}(\Z)$, the estimate
\begin{eqnarray}
\label{eq.2} \left|
  b^{Wick}\right|_{\mathcal{L}(\bigvee^{k}\Z,\bigvee^{j}\Z)}
 \leq
\delta_{k-p,j-q}^{+}\left( j\varepsilon
\right)^{\frac{q}{2}}(k\varepsilon)^{\frac{p}{2}} \left| \tilde{b}
\right|_{\L(\bigvee^p\Z,\bigvee^q\Z)}\,,\quad
\text{with}\;\tilde{b}=\frac{1}{p!q!}\partial_{z}^{p}\partial_{\bar
z}^{q}b\,,
\end{eqnarray}
holds for any $k,j\in \nz$.\\
This implies
\begin{equation}
  \label{eq.2bis}
\left|\left\langle N\right\rangle^{-\frac{q}{2}}b^{Wick}
\left\langle
  N\right\rangle^{-\frac{p}{2}}\right|_{\mathcal{L}(\mathcal{H})}\leq
\left|\tilde{b}\right|_{\L(\bigvee^p\Z,\bigvee^q\Z)}\,.
\end{equation}
\end{lem}
\proof A consequence of \eqref{eq.3} is
$b^{Wick}(\bigvee^{k}\Z)\subset \bigvee^{j}\Z$ with $j=k-p+q$. For
$\psi\in\bigvee^k\Z$ and $j=k-p+q$, write \begin{eqnarray*} \left|b^{Wick}
\psi\right|_{\bigvee^j\Z}&=& \frac{\sqrt{j!
    k!}}{(k-p)!} \hbarr^{\frac{p+q}{2}}
\left|\S_{j} (b\otimes I_{\bigotimes^{k-p}\Z}) \psi\right|_{\bigvee^j\Z} \\
&\leq & (j \varepsilon)^{\frac{q}{2}}(k \varepsilon)^{\frac{p}{2}}
\sqrt{\frac{j!}{(j-q)! j^q}} \;\sqrt{\frac{k!}{(k-p)! k^p}}
\;\;\left|b\otimes I_{\bigotimes^{k-p}\Z}\right|_{\L(\bigotimes^k\Z,
\bigotimes^j\Z)} \;\; \left|\psi\right|_{\bigvee^k\Z}\,. \end{eqnarray*} \fin

An important  property of our class of Wick polynomials is
that a composition of
$b_{1}^{Wick}\circ b_{2}^{Wick}$ with
$b_1,b_2\in\oplus_{p,q}^{\rm alg}\P_{p,q}(\Z)$
is a Wick polynomial with symbol in $\oplus_{p,q}^{\rm alg}\P_{p,q}(\Z)$.
 In the following we prove this result and specifies the Wick symbol of
 the product.\\
  For $b\in\P_{p,q}(\Z)$, specific cases with $j=0$ or $k=0$
of \eqref{eq.differential} imply
$$
\partial_{z}^{k}b(z)\in (\bigvee^{k}\Z)^{*}\quad\text{and}\quad
\partial_{\overline{z}}^{j}b(z)\in \bigvee^{j}\Z\,,
$$
for any fixed $z\in \Z$.
For two symbols $b_{i}\in \P_{p_{i},q_{i}}(\Z)$,
$i=1,2$, and any $k\in \nz$, the new symbol
$\partial_{z}^{k}b_{1}.\partial_{\bar
  z}^{k}b_{2}$ is now defined by
\begin{equation}
\partial_z^k b_1 \; .\;\partial_{\bar z}^k b_2 (z) =\la \partial_z^k b_1(z),
\partial_{\bar z}^k b_2(z)\ra_{(\bigvee^k \Z)^{*},\bigvee^{k}\Z}\quad.
\end{equation}
We also use the following notation for multiple Poisson brackets:
\begin{eqnarray*}
\{b_1,b_2\}^{(k)}&=&\partial^k_z b_1
.\partial^k_{\bar z} b_2 -\; \partial^k_z b_2 .\partial^k_{\bar z} b_1,\\
\{b_1,b_2\}&=&\{b_1,b_2\}^{(1)}.
\end{eqnarray*}
These operations with polynomials are easier to handle than there corresponding
versions for the operators $\tilde{b}_{i}\in
\L\left(\bigvee^{p_{i}}\Z, \bigvee^{q_{i}}\Z\right)$.
Nevertheless  their explicit operator expressions as
contracted products allow to check that
 $\oplus_{p,q}^{\rm alg}\P_{p,q}(\Z)$ is stable w.r.t these operations\,.
\begin{lem}
  \label{le.bdcontract}
Fix $p_{1},p_{2},q_{1}$ and $q_{2}$ in $\nz$.
For two polynomials $b_{i}\in \P_{p_{i},q_{i}}(\Z)$, $i=1,2$, set
$\tilde{b}_{i}=(p_{i}!q_{i}!)^{-1}$ $\partial_{z}^{p_{i}}\partial_{\bar
z}^{q_{i}}b_{i}$ and for any $k\in
\left\{0,\ldots,\min\{p_{1},q_{2}\}\right\}$
$$
\tilde{b}_{1}\mathop{\odot}^{k}\tilde{b}_{2}
=\frac{1}{(p_{1}+p_{2}-k)!(q_{1}+q_{2}-k)!}
\partial_{z}^{p_{1}+p_{2}-k}\partial_{\bar z}^{q_{1}+q_{2}-k}
\left[\partial_{z}^{k}b_{1}.\partial_{\bar z}^{k}b_{2}\right]\,.
$$
Then
\begin{eqnarray}
\label{form.oper}
\tilde{b}_{1}\displaystyle\mathop{\odot}^{k}\tilde{b}_{2}=\frac{p_{1}!}{(p_{1}-k)!}\frac{q_{2}!}{(q_{2}-k)!}
\;\S_{q_1+q_2-k}(\tilde b_1\otimes I_{\bigotimes^{q_2-k}\Z} )
(I_{\bigotimes^{p_1-k}}\otimes \tilde b_2)
\in \L(\bigvee^{p_{1}+p_{2}-k}\Z,\bigvee^{q_{1}+q_{2}-k}\Z),
\end{eqnarray}
with the estimate
\begin{equation}
  \label{eq.odotest}
\left| \tilde{b}_{1}\mathop{\odot}^{k}\tilde{b}_{2}
\right|_{\L(\bigvee^{p_{1}+p_{2}-k}\Z,\bigvee^{q_{1}+q_{2}-k}\Z)}
\leq \frac{p_{1}!}{(p_{1}-k)!}\frac{q_{2}!}{(q_{2}-k)!}
\left|\tilde{b}_{1}\right|_{\L(\bigvee^{p_{1}}\Z,\bigvee^{q_{1}}\Z)}
\left|\tilde{b}_{2}\right|_{\L(\bigvee^{p_{2}}\Z,\bigvee^{q_{2}}\Z)}
\quad .
\end{equation}
\end{lem}
\proof
For $\psi\in\bigvee^{p_1}\Z$ and $\phi\in\bigvee^{q_2}\Z$, introduce
the vector
$$
\langle z^{\otimes q_{2}-k}, \phi\rangle=
\left(\langle
z^{\otimes q_{2}-k}|\otimes I_{\bigotimes^{k}\Z}\right)
\phi=\frac{(q_{2}-k)!}{q_{2}!}\partial_{\overline{z}}^k b_{\phi}(z) \in \bigvee^{k}\Z
$$
with $b_{\phi}(z)=\la z^{q_{2}}\,,\, \phi\ra$ and the form
$$
\la\psi,z^{\otimes p_1-k}\ra :=\frac{(p_{1}-k)!}{p_{1}!}\partial_{z}^{k}b_{\psi}(z) \in
(\bigvee^{k}\Z)^{*}\,,\quad \text{with}\quad b_{\psi}(z)=\left\langle
  \psi\,,\,z^{\otimes p_{1}}\right\rangle.
$$
The identity
\begin{eqnarray}
\label{tensor-id}
\left<\la\psi,z^{\otimes p_1-k}\ra, \la z^{\otimes q_2-k},\phi\ra
\right>_{(\bigvee^k\Z)^*,\bigvee^k\Z}
=
\la \psi\otimes z^{\otimes q_2-k}, z^{\otimes p_1-k }\otimes
\phi\ra_{\bigotimes^{p_{1}+q_{2}-k}\Z}
\end{eqnarray}
is obviously true when $\psi=\xi^{\otimes p_{1}}$ and
$\phi=\eta^{\otimes q_{2}}$ with $\xi,\eta\in \Z$.
Since $(\xi^{\otimes n})_{\xi\in \Z}$
is a total space of $\bigvee^{n}\Z$ with the polarization identity
\eqref{eq.pola}, the identity
\eqref{tensor-id} holds for all $\phi\in \bigvee^{q_{2}}\Z$ and all
$\psi\in \bigvee^{p_{1}}\Z$\,.
After noticing the relations
$$
\partial_{z}^{k}b_{1}(z)=\frac{p_{1}!}{(p_{1}-k)!}\la
\psi,z^{\otimes p_{1}-k}\ra\quad,\quad
\partial_{\overline{z}}^{k}b_{2}(z)=\frac{q_{2}!}{(q_{2}-k)!}\la
z^{\otimes q_{2}-k}, \phi\ra\,,
$$
with $\psi=\tilde{b}_{1}^{*}z^{\otimes q_{1}}$ and
$\phi=\tilde{b}_{2}z^{\otimes p_{2}}$, the identity \eqref{tensor-id}
leads to
\begin{eqnarray*}
\partial_{z}^{k}b_{1}.\partial_{\bar z}^{k}b_{2}(z)=
\frac{p_{1}!}{(p_{1}-k)!}\frac{q_{2}!}{(q_{2}-k)!}
\left<  z^{\otimes q_1+q_2-k}, ( \tilde b_1\otimes I_{\bigotimes^{q_2-k}\Z}) \,
(I_{\bigotimes^{p_1-k}\Z}\otimes\tilde b_2) z^{\otimes p_2+p_1-k} \right>\,.
\end{eqnarray*}
Therefore $\partial_{z}^{k}b_{1}.\partial_{\bar z}^{k}b_{2}$ is a continuous homogenous polynomial in $\P_{p_1+p_2-k,q_1+q_2-k}(\Z)$
with the associated  operator given by (\ref{form.oper}). The estimate   \eqref{eq.odotest} follows immediately by
(\ref{form.oper}). \fin

\begin{prop}
\label{symbcalc}
The formulas
\begin{eqnarray}
\label{eq.wickproduct} (i)&&
b_1^{Wick} \; b_2^{Wick}=\left(\sum_{k=0}^{\min\{p_1,q_2\}}
\;\;\frac{\hbarr^k}{k!}  \;\;\;\partial^{k}_{z} b_1 .\partial^{k}_{\bar z}
b_2 \right)^{Wick}= \left(e^{\hbarr \la \partial_z,\partial_{\bar
\omega}\ra}
b_1(z) b_2(\omega)\left|_{z=\omega}\right. \right)^{Wick}\,,\\
\label{eq.wickpoisson} (ii)&&
[b_1^{Wick},b_2^{Wick}]=\left(\sum_{k=1}^{\max\{\min\{p_1,q_2\}\,,\,
\min\{p_{2},q_{1}\}\}} \;\;\frac{\hbarr^k}{k!}  \;\;\{b_1
,b_2\}^{(k)} \right)^{Wick}\,, \end{eqnarray} hold for any $b_{i}\in
\P_{p_{i},q_{i}}(\Z)$, $i=1,2$\, as identities on $\H_{fin}$.
\end{prop}
\begin{remark}
  This result has exactly the form of the finite dimensional formula.
Lemma~\ref{le.bdcontract} gives the relation with the writing which
can be found in \cite{FKP}.
\end{remark}
\proof The second statement (ii) is a straightforward consequence of
the first one (i). Let us focus on (i) which will be proved in
several
steps.\\
\noindent Step~0: Before proving the identity, first notice that
both sides are well defined. Actually, for any $b\in \P_{p,q}(\Z)$,
the operator $b^{Wick}$ sends $\H_{fin}$ into itself. Hence, the
product $b_{1}^{Wick}\circ b_{2}^{Wick}$ is well defined as an
operator $\H_{fin}\to \H_{fin}$. Finally we know from
Lemma~\ref{le.bdcontract} that
$e^{\varepsilon\la\partial_{z},\partial_{\overline{\omega}}\ra}b_{1}(z)b_{2}(\omega)\big|_{z=\omega}$
belongs to $\oplus_{p,q}^{\rm alg}\P_{p,q}(\Z)$.\\
\noindent Step~1: Consider $b_{1}(z)=\la \eta\,,\, z\ra$ and
$b_{2}(z)=\la z\,,\,\xi\ra^{q}$, $q\in \nz$. The formula
$$
a(\eta)a^{*}(\xi)^{q}=a^{*}(\xi)^{q}a(\eta)+ \varepsilon q \la
\eta\,,\,\xi\ra a^{*}(\xi)^{q-1}
$$
is exactly
$$
b_{1}^{Wick}b_{2}^{Wick}=(b_{1}b_{2})^{Wick}+
\varepsilon(\partial_{z}b_{1}.\partial_{\bar z}b_{2})^{Wick}\,.
$$
\noindent Step~2: Consider $b_{1}(z)=\beta_{p}(z)=\la
\eta\,,\,z\ra^{p}$ and $b_{2}(z)=\la z\,,\, \xi\ra^{q}$, $p,q\in
\nz$. The induction is already initialized for $p=1$
according to Step~1. Assume  that the formula is true for $p-1$ and
all $q\in \nz$ and compute
\begin{eqnarray*}
  \beta_{p}^{Wick}b_{2}^{Wick}&=&
\beta_{1}^{Wick}\left[\beta_{p-1}^{Wick}b_{2}^{Wick}\right]
=\beta_{1}^{Wick}\left[\sum_{k=0}^{\min\left\{p-1,q\right\}}
\frac{\varepsilon^{k}}{k!} \left\langle
\partial_{z}^{k}\beta_{p-1}\,,\,
\partial_{\bar
  z}^{k}b_{2}\right\rangle^{Wick}\right]\\
&=& a(\eta) \left[\sum_{k=0}^{\min\left\{p-1,q\right\}}
\frac{\varepsilon^{k}}{k!} \la \eta\,,\,\xi\ra^{k}
\frac{q!}{(q-k)!}\frac{(p-1)!}{(p-1-k)!}
a^{*}(\xi)^{q-k}a(\eta)^{p-1-k}\right]\\
&=& \sum_{k=0}^{\min\left\{p-1,q\right\}} \frac{\varepsilon^{k}}{k!}
\la \eta\,,\,\xi\ra^{k} \frac{q!(p-1)!}{(q-k)!(p-1-k)!}
\left[a^{*}(\xi)^{q-k}a(\eta)^{p-k}
\right.\\
&&\hspace{7cm} \left. + \varepsilon(q-k)\la \eta\,,\, \xi\ra
a^{*}(\xi)^{q-k}a(\eta)^{p-(k+1)}
\right]\\
&=& \sum_{k=0}^{\min\left\{p,q\right\}} \frac{\varepsilon^{k}\la
\eta,\xi\ra^{k} q!(p-1)!}{k!(q-k)!(p-1-k)!}
\left[1_{[0,p-1]}(k)+ \frac{k}{(p-k)}1_{[1,p]}(k) \right]
a^{*}(\xi)^{q-k}a(\eta)^{p-k}
\\
&=& \sum_{k=0}^{\min\left\{p,q\right\}} \frac{\varepsilon^{k}}{k!}
\left\langle \partial_{z}^{k}\beta_{p}\,,\,
\partial_{\bar
  z}^{k}b_{2}\right\rangle^{Wick}\,.
\end{eqnarray*}
We used several times the relation
$$
\partial_{z}^{j}\beta_{n}(z)= \frac{n!}{(n-j)!}
\la \eta\,,\, z\ra^{n-j} \la \eta\,|^{\otimes j}\,
$$
and its dual version for $\partial_{\bar z}^{j}b_{2}$\,.\\
\noindent Step~3: From Step~2, the statement (ii) of
Proposition~\ref{wick_prop} leads to
$$
a^*(\xi^1)^{q_1} a(\eta^1)^{p_1} \;\;
a^*(\xi^2)^{q_2}a(\eta^2)^{p_2} = \sum_{k=0}^{\min \{p_{1},q_{2}\}}
\frac{\hbarr^{k}}{k!}  \left(\partial_z^k \big( \la z,\xi^1\ra^{q_1}
\la \eta^1,z\ra^{p_1}\big) . \partial_{\bar z}^k  \big( \la
z,\xi^2\ra^{q_2}\la\eta^2,z\ra^{p_2}\big)\right)^{Wick}
$$
for any $\xi^{1},\xi^{2},\eta^{1},\eta^{2}\in \Z$ and any
$p_{1},q_{1},p_{2},q_{2}\in \nz$\,. Again the polarization
formula \eqref{eq.pola} in the form
\begin{eqnarray*} \prod_{i=1}^{n}a^\natural(\xi_i)=\ds \frac{1}{2^n n!}
\sum_{\varepsilon_i=\pm 1} \varepsilon_1\cdots \varepsilon_n \;
\left[a^\natural\big( \sum_{j=1}^n \varepsilon_j
\xi_j)\right]^{n}\,, \end{eqnarray*}
 yields the result for any
$$
b_{\ell}(z)=\prod_{i=1}^{p_{\ell}}\la z\,,\,\xi^{\ell}_{i}\ra
\prod_{j=1}^{q_{\ell}}\la \eta^{\ell}_{j}\,,\,z\ra \,,\quad
\ell=1,2\,,
$$
that is for any $\tilde{b}_{\ell}$ in the form
\begin{equation}
  \label{eq.rgfintyfin}
\tilde{b}_{\ell}=|\xi_{1}^{\ell}\vee\ldots \vee
\xi_{p_{\ell}}^{\ell}\ra
\la \eta_{1}^{\ell}\vee\ldots\vee \eta_{q_{\ell}}^{\ell}|\,,\quad \ell=1,2\,.
\end{equation}
\\
Step~4: We want to check the identity
$$
\left\langle \psi_{n'}\,,\, b_{1}^{Wick}\circ
  b_{2}^{Wick}\psi_{n}\right\rangle
=\sum_{p=0}^{\min\left\{p_{1},q_{2}\right\}}
\frac{\hbarr^{p}}{p!}
\left\langle \psi{_{n'}}\,,\,
(\partial_{z}^{p}b_{1}
\partial_{\overline{z}}^{p}b_{2})^{Wick}
\psi_{n}\right\rangle
$$
for any $\psi_{n}\in \bigvee^{n}\Z$ and any $\psi_{n'}\in
\bigvee^{n'}\Z$, $n,n'\in \nz$. \\
From the definition of $b^{Wick}$, the left-hand side equals
\begin{eqnarray*}
\left\langle \psi_{n'}\,,\, b_{1}^{Wick}\circ
  b_{2}^{Wick}\psi_{n}\right\rangle &=&
C_{n,n',p_{1,2},q_{1,2},\hbarr}
\left\langle \psi_{n'}\,,\, \left(\tilde{b}_{1}\bigvee
  I\big|_{\bigvee^{n+q_{2}-p_{2}-p_1}\Z}\right)
\left(\tilde{b}_{2}\bigvee I\big|_{\bigvee^{n-p_{1}}\Z}\right)\psi_{n}
\right\rangle
\\
&=&
C_{n,n',p_{1,2},q_{1,2},\hbarr}
\left\langle
\left(\tilde{b}_{1}^{*}\bigvee
  I\big|_{\bigvee^{n'-q_{1}}\Z}\right)\psi_{n'}\,,\,
\left(\tilde{b}_{2}\bigvee I\big|_{\bigvee^{n-p_{1}}\Z}\right)\psi_{n}
\right\rangle\,.
\end{eqnarray*}
Similarly and owing to Lemma~\ref{le.bdcontract},
 every term of the right-hand side satisfies
\begin{eqnarray*}
&&
\left\langle \psi{_{n'}}\,,\,
(\partial_{z}^{p}b_{1}
\partial_{\overline{z}}^{p}b_{2})^{Wick}
\psi_{n}\right\rangle
\\
&&
=C'_{n,n',p,p_{1,2},q_{1,2},\hbarr}
\left\langle \psi{_{n'}}\,,\,
\left[\left(\tilde{b}_{1}\otimes I_{\bigotimes^{q_{2}-p}\Z}\right)
\left( I_{\bigotimes^{p_{1}-p}\Z}\otimes \tilde{b}_{2}\right)\bigvee
I_{\bigvee^{n-p_{1}-p_{2}+p}\Z}
\right]
\psi_{n}\right\rangle
\\
&&
=C'_{n,n',p,p_{1,2},q_{1,2},\hbarr}
\left\langle \left(\tilde{b}_{1}^{*}
\otimes I_{\bigotimes^{n'-p_{1}}\Z}\right) \psi{_{n'}}\,,\,
\left( I_{\bigotimes^{p_{1}-p}\Z}\otimes \tilde{b}_{2}\otimes I_{\bigvee^{n-p_{1}-p_{2}+p}\Z}\right)
\psi_{n}\right\rangle.
\end{eqnarray*}
Hence for fixed $\psi_{n}, \psi_{n'}\in \H_{fin}$, both side are
sesquilinear continuous expression of
$(\tilde{b}_{1},\tilde{b_{2}})$ when the first factor is considered
with the $*$-strong topology of operators and the second one with the
strong topology. The operators \eqref{eq.rgfintyfin} for which the
equality is true, form a total family for these topologies: In two
steps, approximate first
any finite rank operators and then bounded operators by
finite rank operators. Thus
the equality holds for any $b_{\ell}\in \P_{p_{\ell},q_{\ell}}(\Z)$,
$\ell=1,2$\,.
\fin
\begin{remark}
The formulas \eqref{eq.wickproduct} and \eqref{eq.wickpoisson} make
sense with $\hbarr$-dependent symbols. One can work with polynomials
in $\hbarr$
$$
b(z,\hbarr)=\sum_{\alpha=0}^n \hbarr^\alpha b_{\alpha}(z),\quad
b_{\alpha}\in \P_{p,q}(\Z)
$$
or with asymptotic sums
$$
b(z,\hbarr)\sim
\sum_{\alpha=0}^{\infty}\varepsilon^{\alpha}b_{\alpha}(z)\, \quad
b_{\alpha}\in \P_{p,q}(\Z)\,.
$$
The expression \eqref{eq.wickproduct} and \eqref{eq.wickpoisson}
take
 then the form
\begin{eqnarray*} b_1^{Wick}b_2^{Wick} &\sim & \sum_{j=0}^{\infty} \hbarr^j \;
\left(\sum_{\alpha+\beta+k=j} \frac{1}{k!} \big(\partial_z^k
b_{1,\alpha} .\partial_{\bar z}^k
b_{2,\beta}\big) \right)^{Wick}\\
\left[b_1^{Wick},\,b_2^{Wick}\right] &\sim & \sum_{j=1}^{\infty}
\hbarr^j \; \left(\sum_{\alpha+\beta+k=j} \frac{1}{k!}
\big(\partial_z^k b_{1,\alpha} .\partial_{\bar z}^k b_{2,\beta}-
\partial_z^k b_{2,\beta} .\partial_{\bar z}^k b_{1,\alpha} \big) \right)^{Wick}\,,
\end{eqnarray*} for $b_{1}\sim \sum_{\alpha}\hbarr^{\alpha}b_{1,\alpha}\in
\P_{p_{1},q_{1}}(\Z)$ and $b_{2}\sim
\sum_{\beta}\hbarr^{\beta}b_{2, \beta}\in
\P_{p_{2},q_{2}}(\Z)$\,.
Here $(p_{1},q_{1})$ (resp. $(p_{2},q_{2})$) does not depend on
$\alpha$ (resp. $\beta$).
\end{remark}
We have the following useful result.
\begin{prop}
  \label{pr.transla}
  For any $b\in \oplus_{p,q\in \nz}^{\rm alg}\P_{p,q}(\Z)$ we have:\\
(i) $b^{Wick}$ is closable with
$$
\H_0={\rm vect}\{W(z)\phi, \phi\in\H_{fin},z\in\Z\}
$$
a core  of the closure.\\
(ii) By setting $E(z)=W(\frac{\sqrt{2}z}{i\varepsilon})\Omega$
  according to \eqref{coherent-vect}, the identity
  \begin{equation}
    \label{eq.wicksymbeq}
b(z)=\left\langle E(z)\,,b^{Wick}E(z)\right\rangle
\end{equation}
holds for every $z\in\Z$\,.\\
(iii) For any $z_{0}\in \Z$ the identity
$$
W(\frac{\sqrt{2}}{i\varepsilon}z_{0})^{*}b^{Wick}W(\frac{\sqrt{2}}{
i\varepsilon}z_{0})=(b(z+z_{0}))^{Wick}
$$
holds on $\H_0$ where $b(\cdot+z_{0})\in \oplus_{p,q\in \nz}^{\rm alg}\P_{p,q}(\Z)$\,.
\end{prop}
\proof
(i)  $b^{Wick}$ is closable by Proposition \ref{wick_prop} (i). It is enough to consider
$b\in\P_{p,q}(\Z)$ when we prove that $\H_0$ is a core for  the closure of $b^{Wick}$.
The last statement is deduced from the estimate
\begin{multline}
\label{estdzero}
\sum_{n=0}^\infty \frac{1}{n!} \; \left|b^{Wick} \, \Phi(z)^n \varphi^{(k)}\right|_{\mathcal{H}}
\leq  |\tilde{b}|_{\mathcal{L}(\bigvee^{p}\Z,\bigvee^{q}\Z)} \,
|\varphi^{(k)}|_{\bigvee^{k}\Z}\times
\\
 \sum_{n=0}^\infty \frac{(\sqrt{2\varepsilon})^n}{n!} \sqrt{
\frac{(n+k)!}{k!}} [\varepsilon (n+k+q)]^{ \frac{p+q}{2}} \, |z|^n<\infty
\end{multline}
for any $\varphi^{(k)}\in\bigvee^k\Z$ and $z\in\Z$. In order to prove
(\ref{estdzero}), use Lemma
\ref{wick-estimate} and estimate the action of $b^{Wick}$ on $\Phi(z)^n \varphi^{(k)}$ by
$\ds\max_{p\leq r\leq k+n} |b^{Wick}|_{\L(\bigvee^{r}\Z,\bigvee^{r-p+q})}$ and bound the norm
of $\Phi(z)^n \varphi^{(k)}$ by
 $|\varphi^{(k)}|\, |z|^n
 \,\sqrt{\frac{(2\varepsilon)^n(n+k)!}{k!}}$.\\
(ii) One writes for $b\in \P_{p,q}(\Z)$ and $z\in
\Z$
\begin{eqnarray*}
\la E(z)\,,\,b^{Wick}E(z)\ra &=&
e^{-\frac{|z|^{2}}{\varepsilon}}\sum_{n_{1},n_{2}\in \nz}
\frac{\la z^{\otimes n_{1}}\,,\, b^{Wick}z^{\otimes
    n_{2}}\ra}{\sqrt{n_{1}!}
\sqrt{n_{2}!}}
\\
&=& e^{-\frac{|z|^{2}}{\varepsilon}}\sum_{n_{1},n_{2}\in \nz}
\delta^{+}_{n_{1}-q,n_{2}-p}\frac{\varepsilon^{\frac{p+q}{2}}|z|^{n_{1}-p+n_{2}-q}}{
\sqrt{(n_{1}-q)!}\sqrt{(n_{2}-p)!}\varepsilon^{\frac{n_{1}+n_{2}}{2}}}b(z)=b(z)\,.
\end{eqnarray*}
(iii) The fact that $b(.+z_{0})$ remains in the class
$\oplus_{p,q\in \nz}^{\rm alg}\P_{p,q}(\Z)$ come from the Taylor
expansion and \eqref{eq.differential}. In order to prove the equality,
differentiate $A(t)=\left[
W(\frac{\sqrt{2}}{i\varepsilon}tz_{0})b(z+tz_{0})^{Wick}W(\frac{\sqrt{2}}{i\varepsilon}tz_{0})^{*}\right]$
in a weak sense on $\H_0$. Proposition~\ref{symbcalc} implies
\begin{eqnarray*}
i\partial_{t}A(t)&=&
W(\frac{\sqrt{2}}{i\varepsilon}tz_{0})
\left[-[\Phi(\frac{\sqrt{2}}{i\varepsilon}z_{0}), b(z+tz_{0})^{Wick}]
+i\partial_{t}b(z+tz_{0})^{Wick}
\right]
W(\frac{\sqrt{2}}{i\varepsilon}tz_{0})^{*}\\
&=&
W(\frac{\sqrt{2}}{i\varepsilon}tz_{0})
\left[
\langle i z_{0},\partial_{\overline{z}}b(z+tz_{0})\rangle
-\langle \partial_{z}b(z+z_{0})\,,\,iz_{0}\rangle
+i\partial_{t}b(z+tz_{0})
\right]^{Wick}
W(\frac{\sqrt{2}}{i\varepsilon}tz_{0})^{*}
=0\,.
\end{eqnarray*}
\fin
\begin{remark}
  \label{re.wickgen}
The relation \eqref{eq.wicksymbeq} allows to define easily the Wick
symbol of an operator which is defined as a series, when it makes
sense, instead of a Wick polynomial. For example the Wick symbol of
the Weyl operator $W(\xi)$ equals
\begin{equation}
  \label{eq.wicktrans}
\la E(z)\,,\,W(\xi)E(z)\ra= \la
\Omega\,,\,e^{-i\hbarr\sigma(\xi,\frac{\sqrt{2}z}{i\hbarr})}W(\xi)\Omega\ra
= e^{i\sqrt{2}S(\xi,z)}e^{-\frac{\hbarr |\xi|^{2}}{4}}\,.
\end{equation}
\end{remark}
A variation of Proposition~\ref{pr.transla}
  ensures that $b(Az+z_{0})$ can be Wick quantized for any bounded
  \underline{complex} affine transformation in $\Z$ when $b\in
  \P_{p,q}(\Z)$.
Actually real symplectic affine transformations of symbols in $ \P_{p,q}(\Z)$ may
also be Wick quantized  but only under a Hilbert-Schmidt
  condition on $A$ which agrees  with Shale's theorem or the presentation
of general Bogoliubov transformations (see \cite{Ber}). The following
result will be useful in Subsection \ref{se.hepp}.

\begin{prop}
  \label{pr.realtrans}
Let $B\in \L(\Z)$ and let $B_{2}\in \L^{2}(\Z)$ be an Hilbert-Schmidt
operator on
$\Z$ and let $J:\Z\ni z\mapsto Jz=:\overline{z}\in \Z$ be any anti-unitary operator
on $\Z$. Then for any $b\in \P_{p,q}(\Z)$ the polynomial
$b(Bz+B_{2}\overline{z})$ belongs to $\oplus_{p'+q'=p+q}\P_{p',q'}(\Z)$
with the estimate
$$
\left|\partial_{\overline{z}}^{q'}\partial_{z}^{p'}b(Bz+B_{2}\overline{z})
\right|_{\L(\bigvee^{p'}\Z,\bigvee^{q'}\Z)}
\leq C_{p,q}\left(\left|B\right|_{\L(\Z)}+\left|B_{2}\right|_{\L^{2}(\Z)}\right)^{p+q}
\left|\tilde{b}\right|_{\L(\bigvee^{p}\Z,\bigvee^{q}\Z)}\,.
$$
\end{prop}
\proof
 For $b\in \P_{p,q}(\Z)$ write, after recalling
 $\tilde{b}=\mathcal{S}_{q}\tilde{b}\mathcal{S}_{p}$ in
 $\L(\bigotimes^{p}\Z,\bigotimes^{q}\Z)$,
\begin{eqnarray*}
b(Bz+B_{2}\overline{z})
&=&
\left\langle (Bz+B_{2}\overline{z})^{\otimes
    q}\,,\,\tilde{b}(Bz+B_{2}\overline{z})^{\otimes p}\right\rangle\\
&=&\sum_{j=0}^{q}\sum_{k=0}^{p}C_{q}^{j}C_{p}^{k}
\left\langle (Bz)^{\otimes q-j}\otimes (B_{2}\overline{z})^{\otimes j}
  \,,\, \tilde{b}(B_{2}\overline{z})^{\otimes k}\otimes
(Bz)^{\otimes p-k}\right\rangle
\\
&=&
\sum_{j=0}^{q}\sum_{k=0}^{p}C_{q}^{j}C_{p}^{k}\ell_{j,k}(z^{\otimes q+k-j}
\,,\, z^{\otimes p+j-k})\,.
\end{eqnarray*}
The sesquilinear form $\ell_{j,k}$ is defined on
$(\bigotimes^{q-j}\Z \otimes^{alg} \bigotimes^{k}\Z)\times
(\bigotimes^{j}\Z\otimes^{alg} \bigotimes^{p-k}\Z)$ by
$$
\ell_{j,k}\left(\phi_{1}\otimes \phi_{2} , \psi_{1}\otimes
  \psi_{2}\right)
=
\left\langle (B^{\otimes q-j}\phi_{1})\otimes (B_{2}^{\otimes j}
\overline{\psi_{2}})\,,\,
\tilde{b}
(B_{2}^{\otimes k}\overline{\phi_{2}})\otimes (B^{\otimes p-k})\psi_{1}
\right\rangle
$$
It satisfies for $\Phi=\sum_{\alpha=1}^{N}\phi_{1,\alpha}\otimes
\phi_{2,\alpha}$ and $\Psi=\sum_{\beta=1}^{N}\psi_{1,\beta}\otimes
\psi_{2,\beta}$
\begin{eqnarray*}
\ell_{j,k}\left(\Phi, \Psi\right)
&=&
\sum_{\beta=1}^{N}\left\langle
(B_{2}^{\otimes j}
\overline{\psi_{2,\beta}})\,,\, C_{\Phi} (B^{\otimes
  p-k})\psi_{1,\beta}\right\rangle
\\
&=&
\sum_{\beta=1}^{N}
\left\langle
\overline{\psi_{2,\beta}}\,,\, (B_{2}^{*})^{\otimes j}C_{\Phi} (B^{\otimes
  p-k})\psi_{1,\beta}\right\rangle
\end{eqnarray*}
with
$$
C_{\Phi}=\sum_{\alpha=1}^{N}
(\langle B^{\otimes q-j}\phi_{1,\alpha}|\otimes I_{\bigotimes^{j}\Z})
\tilde{b}\left(|B_{2}^{\otimes
    k}\phi_{2,\alpha}\rangle
\otimes I_{\bigotimes^{p-k}\Z}\right)\in
\L(\bigotimes^{p-k}\Z ,\bigotimes^{j}\Z)\,.
$$
Since $B_{2}^{\otimes j}$ is a Hilbert-Schmidt operator the estimate
$$
\left|\ell_{j,k}\left(\Phi , \Psi\right)\right|\leq
\left|B_{2}\right|_{\L^{2}(\Z)}^{j}
\left|B\right|_{\L(\Z)}^{p-k}\left|C_{\Phi}\right|_{\L(\bigotimes^{p-k}\Z
  ,\bigotimes^{j}\Z)}\left|\Psi\right|_{\bigotimes^{p-k+j}(\Z)}
$$
holds for any $\Psi\in \bigotimes^{j}\Z\otimes^{alg}
\bigotimes^{p-k}\Z$\,. In order to estimate $\left|C_{\Phi}\right|_{\L(\bigotimes^{p-k}\Z
,\bigotimes^{j}\Z)}$ take any $U\in \bigotimes^{j}\Z$ and any $V\in
\bigotimes^{p-k}\Z$ and compute
\begin{eqnarray*}
  \left|\left\langle U\,,\,C_{\Phi}V\right\rangle\right|
&=&
\left|\sum_{\alpha=1}^{N}\left\langle B^{\otimes
      q-j}\phi_{1,\alpha}\otimes U\,,\, \tilde{b} (B_{2}^{\otimes
      k}\phi_{2,\alpha}\otimes V)
\right\rangle
\right|
\\
&=&
\left|\sum_{\alpha=1}^{N}\left\langle
\phi_{1,\alpha}\,,\,  (B^{*})^{\otimes
      q-j}C_{UV}
B_{2}^{\otimes k}\phi_{2,\alpha}
\right\rangle
\right|\\
\text{with}&&
C_{UV}=(I_{\bigotimes^{q-j}\Z}\otimes\langle
U|)\tilde{b}(I_{\bigotimes^{k}\Z}\otimes |V\rangle)\in
\L(\bigotimes^{k}\Z,\bigotimes^{q-j}\Z)\,.
\end{eqnarray*}
Again the Hilbert-Schmidt condition implies
$$
  \left|\left\langle U\,,\,C_{\Phi}V\right\rangle\right|
\leq
\left|B_{2}\right|_{\L^{2}(\Z)}^{k}\left|B\right|_{\L(\Z)}^{q-j}
\left|U\right|_{\bigotimes^{j}\Z}
\left|\tilde{b}\right|_{\L(\bigvee^{p}\Z,\bigvee^{q}\Z)}
\left|V\right|_{\bigotimes^{p-k}\Z}\left|\Phi\right|_{\bigotimes^{q-j+k}\Z}\,.
$$
We have proved an estimate for $\left|C_{\Phi}\right|$ which implies
that the estimate
$$
\left|\ell_{j,k}(\Phi, \Psi)\right|\leq
\left|B_{2}\right|_{\L^{2}(\Z)}^{j+k}\left|B\right|_{\L(\Z)}^{p+q-k-j}
\left|\tilde{b}\right|_{\L(\bigvee^{p}\Z,\bigvee^{q}\Z)}
\left|\Phi\right|_{\bigotimes^{q-j+k}\Z}\left|\Psi\right|_{\bigotimes^{p-k+j}}\,,
$$
extends continuously to any $\Phi\in\bigotimes^{q-j+k}\Z$ and any $\Psi\in
\bigotimes^{p-k+j}\Z$. It holds in particular when
$\Phi\in\bigvee^{q-j+k}\Z$ and $\Psi\in
\bigvee^{p-k+j}\Z$. Hence $\ell_{j,k}(z)\in \P_{p-k+j, q-j+k}(\Z)$ holds
for any $(j,k)$, $j\leq q$ and $k\leq p$, with
a norm estimate which yields the final result.
\fin
\section{Weyl  and  Anti-Wick quantization}
\label{se.WeylAWick}
Our extension of the Weyl and Anti-Wick pseudodifferential calculus
to the infinite dimensional case is based on a separation of variables
approach within a projective setting. This is slightly different than
the one developed by B.~Lascar in \cite{Las} where the inductive
approach leads to a natural Hilbert-Schmidt condition and restricts the
exploration of the infinite dimensional phase-space $\Z$.

\subsection{Cylindrical functions and Weyl quantization}
\label{se.cyl}
Let  $\p$ denote the set of all finite rank orthogonal projections
on $\Z$ and  for a given $p\in\p$ let $L_{p}(dz)$ denote the
Lebesgue measure on the finite dimensional subspace $p\Z$. A
function $f:\Z\to\cz$ is said  cylindrical if there
exists $p\in\p$ and a function $g$ on $p\Z$ such that $ f(z)=g(pz),$
for all $z\in\Z$. In this case we say that $f$ is based on the
subspace $p\Z$. We set $\S_{cyl}(\Z)$ to be the   cylindrical
Schwartz space:
$$
(f\in \S_{cyl}(\Z))\Leftrightarrow
\left(\exists p\in \p,\exists g\in \S(p\Z), \quad f(z)= g(pz)\right)\,.
$$
\smallskip
It is well known that the Fourier-Wigner transform defined by the
expression
\begin{eqnarray*}
z\mapsto\V[\phi,\psi](z)=\la \psi,W(\sqrt{2}\pi z)
\phi\ra,
\end{eqnarray*}
for any $\phi,\psi\in\H$, belongs to
$L^2(p\Z,L_{p}(dz))\cap C_0(p\Z)$ for every $p\in\p$. Introduce the
Fourier transform of a function $f\in\S_{cyl}(\Z)$ based on the
subspace $p\Z$ as
\begin{eqnarray*}
\F[f](z)=\int_{p\Z} f(\xi) \;\;e^{-2\pi i
\,S(z,\xi)}~L_{p}(d\xi)
\end{eqnarray*}
and its inverse Fourier transform is
\begin{eqnarray*}
f(z)=\int_{p\Z} \F[f](z) \;\;e^{2\pi i
\,S(z,\xi)}~L_{p}(dz)\,.
\end{eqnarray*}
Therefore the so-called Wigner
transform can be written  as $\W[\phi,\psi]=\F^{-1}[\V[\phi,\psi]]$.
With any
 symbol $b\in\S_{cyl}(\Z)$ based on $p\Z$,
a {\it Weyl observable} can be associated  according to
\begin{eqnarray}
\label{weyl-obs} b^{Weyl}=\int_{p\Z} \F[b](z) \;\;\; W(\sqrt{2}\pi
z)~L_{p}(dz)\,.
\end{eqnarray}
It can be expressed as a quadratic form in the
following way
\begin{eqnarray*}
\la \psi, b^{Weyl} \phi\ra_{\H}&= &\int_{p\Z}
\F[b](z)
\;\;\V[\phi,\psi](z)~L_{p}(dz)\\
&=&\int_{p\Z} b(z) \;\;\W[\phi,\psi](z)~L_{p}(dz)\,.
\end{eqnarray*}
Note that
$b^{Weyl}$ is a well defined bounded operator on $\H$ for all
$b\in\S_{cyl}(\Z)$ since $\V[\phi,\psi](z)$ is a bounded function
and $\F[b](z)$ is in $L^1(p\Z,L_{p}(dz))$. Remember also that this
quantization of cylindrical symbols depends on the parameter
$\hbarr$ like the Weyl operators $W(\sqrt{2}\pi z)$\,.

The next estimate  will be useful. A similar inequality can be found
in \cite{DeGe}.

\begin{lem}
\label{est.weyl} For any $\delta\in
[0,1]$ there exists a constant $C_\delta>0$ such that the
estimate
\begin{eqnarray*}
\left|[W(z_1)-W(z_2)] (N+1)^{-\delta/2} \right| \leq
C_\delta \; \,\left|z_1-z_2\right|^{\delta} \;
[\min(\hbarr |z_1|, \hbarr |z_2|)^{\delta}+\max(1,\hbarr)^\delta],
\end{eqnarray*}
holds for  all $\hbarr>0$,  and all $z_1,z_2\in\Z$\,.
\end{lem}
\proof
 We have by Weyl's relation
\begin{eqnarray}
\label{r.eq1}
\left|[W(z_1)-W(z_2)] (N+1)^{-\delta/2}\right| \leq
\left|[W(z_1-z_2)-I] (N+1)^{-\delta/2}\right|+
\left|e^{i\hbarr\sigma(z_1,z_2)}-1\right|.
\end{eqnarray}
The estimate
$|e^{is}-1|\leq C_\delta \,|s|^\delta$,  leads to
\begin{eqnarray*}
\left|e^{i\hbarr\sigma(z_1,z_2)}-1\right|=
\left|e^{i\hbarr\sigma(z_1-z_{2},z_2)}-1\right|=
\left|e^{i\hbarr\sigma(z_1,z_2-z_{1})}-1\right|\leq C_\delta
\,\hbarr^\delta \; \left|z_1-z_2\right|^\delta \;
\min(|z_1|,|z_2|)^{\delta}.
\end{eqnarray*}
The first part of the r.h.s. in (\ref{r.eq1}) is estimated via a complex
interpolation argument.  Indeed, for $\delta=0$ notice that
$\left|W(z_{1}-z_{2})-I\right|\leq 2$ and for $\delta=1$
 the estimate $\left|e^{is}-1\right|\leq C_{1}|s|$ combined with
the spectral theorem yields
\begin{eqnarray*}
\left|[W(z_1-z_2)-I]
(N+1)^{-1/2}\psi\right|&\leq& C_{1} \left|
|\Phi(z_1-z_2)| (N+1)^{-1/2}\psi\right|\\
&\leq & C_{1} \left|\Phi(z_1-z_2)(N+1)^{-1/2}\psi\right|.
\end{eqnarray*}
Now by the number estimate (\ref{eq.2bis}) we obtain
\begin{eqnarray*}
\left|[W(z_1-z_2)-I]
(N+1)^{-1/2}\right|
&\leq & C \;\, \max(1,\hbarr) \; \left|z_1-z_2\right| \,.
\end{eqnarray*}
\fin
\subsection{Finite dimensional Weyl quantization}
\label{se.weylfin}
The finite dimensional Weyl calculus provides us a collection
of results on the Weyl quantization. We specify here the relation
between the Weyl quantization defined on $\Z$ via \eqref{weyl-obs}
and the usual semiclassical
 Weyl quantization within the Schr{\"o}dinger representation
on $\rz^{d}$. \\
For $p\in\mathbb{P}$ the orthogonal projector
$I-p$ is denoted by $p^{\bot}$. Let $\Gamma_s(p\Z)$ denotes the symmetric Fock space over
$p\Z$. The separation of variables in finite dimensions
extends to  general symmetric Fock spaces owing to the
 canonical isomorphism of Fock spaces
\begin{eqnarray}
\label{eq.Tunit}
T_p:\H=\Gamma_s(\Z)\to\Gamma_s(p\Z)\otimes\Gamma_s(p^\bot\Z),
\end{eqnarray}
for any finite dimensional projector $p\in \mathbb{P}$, with
 $T_p\Omega=\Omega^{p\Z}\otimes\Omega^{p^\bot \Z}$ when
$\Omega^{p\Z}$ and $\Omega^{p^\bot \Z}$ are the vacuum vectors of the
corresponding Fock spaces. We will  omit
the notation $T_p$ and identify directly the tensor products.\\
Fix $p\in \p$. The tensor decomposition of the Weyl quantization comes
from the Weyl relation which implies
$$
W(\xi+\xi')=W(\xi)W(\xi')=W_{p}(\xi)\otimes W_{p^{\bot}}(\xi')
$$
for any $(\xi,\xi')\in p\Z\times p^{\bot}\Z$. The symbols $W_{p}$
stands for the Weyl operator defined on the Fock space
$\Gamma_{s}(p\Z)$ and the Weyl quantization of $b\in \S(F)$, for
any finite dimensional complex subspace $F$ of $\Z$, is denoted by $b^{Weyl}_{F}$.
Hence the Weyl quantization of $b\in \S_{cyl}(\Z)$ based on $p\Z$ equals
$$
b^{Weyl}=\int_{p\Z}\mathcal{F}[b](z)W(\sqrt{2}\pi z)~L_{p}(dz)=
b_{p\Z}^{Weyl}\otimes I_{\Gamma_{s}(p^{\bot}\Z)}\,.
$$

In order to apply directly the finite dimensional results on Weyl
quantization, we need to specify the correspondence of
representations.\\
On $\rz^{d}$ the Weyl quantization is often introduced as
$$
b^{Weyl}(x,hD_{x})u (x)=
\int_{\rz^{d}}e^{i\frac{(x-y).\xi}{h}}
b(\frac{x+y}{2},\xi)u(y)~\frac{d\xi dy}{(2\pi h)^{d}}\,.
$$
By a simple conjugation with a dilatation, it becomes
$a^{Weyl}(\sqrt{h}x,\sqrt{h}D_{x})$ where the position
($x$) and frequency ($\xi$) variables play the same role.
 An equivalent definition can be given with
the help of the phase translations~:
$$
\tau_{(x_{0},\xi_{0})}=e^{i(\xi_{0}x-x_{0}D_{x})}=
\left(e^{i(\xi_{0}x-x_{0}\xi)}\right)^{Weyl}
\,,
\quad [\tau_{x_{0},\xi_{0}}u](x)=e^{i\xi(2x-x_{0})/2}u(x-x_{0})\,.
$$
It reads
\begin{eqnarray*}
b^{Weyl}(\sqrt{h}x,\sqrt{h}D_{x})
&=& \int_{T^{*}\rz^{d}}\mathcal{F}[b](y,\eta)e^{2i\pi
  (y.(\sqrt{h}x) + \eta.\sqrt{h}D_{x} )}~dyd\eta
\\
&=&
\int_{T^{*}\rz^{d}}\mathcal{F}[b](y,\eta)
\tau_{(-2\pi\sqrt{h}\eta,2\pi\sqrt{h}y)}~dyd\eta\,.
\end{eqnarray*}
The symplectic form $[\![~,~]\!]$ and the scalar product  $(~,~)$ on
$T^{*}\rz^{d}$ are defined according to
\begin{eqnarray*}
  &&
[\![(x,\xi),(y,\eta)]\!]=\xi.y-x.\eta=-\Im\left\langle x+i\xi\,,\, y+i\eta\right\rangle=-\sigma(x+i\xi,y+i\eta)
\\
&&
((x,\xi),(y,\eta))= x.y+\xi.\eta=\real\left\langle x+i\xi\,,\, y+i\eta\right\rangle=S(x+i\xi,y+i\eta)\,.
\end{eqnarray*}
After noting
$$
\left[\sqrt{h}x+\sqrt{h}\partial_{x},\sqrt{h}x-\sqrt{h}\partial_{x}\right]
=2h\,,
$$
the correspondence with the definition \eqref{weyl-obs} is
summarized in the next table
\begin{eqnarray*}
    p\Z\sim \cz^{d}&&T^{*}\rz^{d}\\
\Gamma_s(p\Z)\sim \Gamma_s(\cz^{d})
\,, &&L^{2}(\rz^{d})\\
\langle z_{1},z_{2}\rangle
=S(z_{1},z_{2})+i\sigma(z_{1},z_{2})
&
z=e^{i\theta}(x+i\xi)
&
((x_{1},\xi_{1})\,,\,(x_{2},\xi_{2}))=\xi_{1}.\xi_{2}+x_{1}.x_{2}=S(z_{1},z_{2})\quad
\\
&&
[\![(x_{1},\xi_{1}),(x_{2},\xi_{2})]\!]=\xi_{1}.x_{2}-x_{1}.\xi_{2}=-\sigma(z_{1},z_{2})
\\
   a(z)=a(\sum_{j=1}^{d}\alpha_{j}e_{j})
  &&
  a(z)=\sum_{j=1}^{d}\overline{\alpha_{j}}(\sqrt{h}\partial_{x_{j}}+\sqrt{h}x_{j})
\\
a^{*}(z)=a^{*}(\sum_{j=1}^{d}\alpha_{j}e_{j})
&&
 a^{*}(z)=\sum_{j=1}^{d}\alpha_{j}(-\sqrt{h}\partial_{x_{j}}+\sqrt{h}x_{j})
\\
 \left[a(z_{1}),a^{*}(z_{2})\right]=\varepsilon\left\langle
     z_{1}\,,\, z_{2}\right\rangle
&\quad{ \varepsilon=2h}&
 \left[a(z_{1}),a^{*}(z_{2})\right]=2h\left\langle
     z_{1}\,,\, z_{2}\right\rangle
\\
\Phi(z_{0})=\frac{1}{\sqrt{2}}(a(z_{0})+a^{*}(z_{0}))
&
z_{0}=x_{0}+i\xi_{0}& \sqrt{2h}(x_{0}.x + \xi_{0}.D_{x})\\
W(z_{0})=e^{i\Phi(z_{0})} & \theta=0&
\tau_{(
  -\sqrt{2h}
  \xi_{0},\sqrt{2h}x_{0})}
\\
 E(z_{0})=W(\frac{\sqrt{2}}{i\varepsilon}z_{0})\Omega
&\frac{z_{0}}{i}=\xi_{0}-ix_{0}&
\tau_{(\frac{x_{0}}{\sqrt{h}}, \frac{\xi_{0}}{\sqrt{h}})}(\pi^{-d/4}e^{-\frac{x^{2}}{2}})
\\
z_{0}^{\otimes n}, |z_{0}| =1 && \textrm{Hermite function}\\
&&
\quad
(n!)^{-1/2}[z_{0}.(-\partial_{x}+x)]^{n}(\pi^{-d/4}e^{-\frac{x^{2}}{2}})
\\
 \ccap_{k\in\nz}D(\langle N_{p\Z}\rangle^{k})\,,\quad
\ccup_{k\in \nz}D(\langle N_{p\Z}\rangle^{k})^{*}
&&
\mathcal{S}(\rz^{d})\,,\quad \mathcal{S}'(\rz^{d})
  \end{eqnarray*}

Once this is fixed, the general results on the semiclassical
Weyl-H{\"o}rmander pseudodifferential calculus
(\cite{BoLe}\cite{BoCh}\cite{HeNi}\cite{Hor}\cite{Mar}\cite{NaNi}\cite{Rob})
can be applied for any fixed $p\in \mathbb{P}$. The notion of slow
and temperate metric and weight depend only on the symplectic
structure which is given by $\sigma(z_{1},z_{2})=\Im\langle
z_{1}\,,\,z_{2}\rangle$.
With such a metric the gain function $\lambda$ is given on $p\Z$ by
$$
\lambda^2(z)=\inf_{T\in
  p\Z\setminus\left\{0\right\}}\frac{g^{\sigma}_{z}(T)}{g_{z}(T)}\,\quad
\text{with}\quad
g^{\sigma}_{z}(T)=\sup_{S\in p\Z\setminus\left\{0\right\}}
\frac{\left|[\![T,S]\!]\right|^{2}}{g(S)}=\sup_{S\in p\Z\setminus\left\{0\right\}}
\frac{\left|\sigma(T,S)\right|^{2}}{g(S)}
\,.
$$
With a slow and temperate metric $g$ and a slow and temperate weight
$m$, is associated a symbol class usually denoted $S(m,g)$.\\
After writing $X=(x,\xi)\in T^{*}\rz^{d}$ for the complete phase-space
variable, the differential operator $D_{X}$ is
$(D_{x},D_{\xi})=(i^{-1}\partial_{x},i^{-1}\partial_{\xi})$.
In the composition formula of symbols, the differential operator
$\frac{ih}{2}[\![D_{X_{1}},D_{X_{2}}]\!]$ appears.
After recalling
$$
\partial_{\overline{z}}=\frac{1}{2}(\nabla_{x}+i\nabla_{\xi})
\quad\text{and}\quad
\partial_{z}=\frac{1}{2}(\nabla_{x}-i\nabla_{\xi})
$$
it equals
\begin{eqnarray*}
\frac{ih}{2}[\![D_{X_{1}},D_{X_{2}}]\!]
=
\frac{\varepsilon}{2}\left(\partial_{z_{1}}.\partial_{\overline{z}_{2}}-\partial_{\overline{z}_{1}}.\partial_{z_{2}}\right)\,.
\end{eqnarray*}
We refer to \cite{NaNi} for an explicit
semiclassical writing of the Weyl-H{\"o}rmander calculus within the
Bony-Lerner  presentation (\cite{BoLe}) and with a general version
of the Beals criterion following Bony-Chemin (\cite{BoCh})\,.
\begin{prop}
\label{pr.weylcalc}
Let $g$ be a slow and temperate metric on $p\Z$,
$\textrm{dim}_{\cz}(p\Z)=d$
and let
$m_{1}$ and $m_{2}$ be two slow and temperate weights for $g$.
For $b_{\ell}\in S_{p\Z}(m_{\ell},g)$,$\ell=1,2$,
 the operator $b^{Weyl}_{\ell,p\Z}$ acts
continuously on
$\ccap_{k\in\nz}D(\langle N_{p\Z}\rangle^{k})$
and on $\ccup_{k\in \nz}D(\langle N_{p\Z}\rangle^{k})^*$.\\
The symbol $b_{1}\#^{\varepsilon/2}b_{2}$ of $b_{1,p\Z}^{Weyl}\circ
b_{2,p\Z}^{Weyl}$ satisfies
\begin{eqnarray*}
b_{1}\#^{\varepsilon/2}b_{2}(z)
&=&
e^{\frac{\varepsilon}{2}\left(\partial_{z_{1}}.\partial_{\overline{z}_{2}}-
\partial_{\overline{z}_{1}}.\partial_{z_{2}}\right)}
 b_{1}(z_{1})b_{2}(z_{2})
\Big|_{z_{1}=z_{2}=z}
\\
&=&
\sum_{0 \leq j<\nu}
\frac{1}{j!}
\left(
\frac{\varepsilon}{2}\left(\partial_{z_{1}}.\partial_{\overline{z}_{2}}-
 \partial_{\overline{z}_{1}}.\partial_{z_{2}}\right)
\right)^{j}
b_{1}(z_{1})b_{2}(z_{2})
\Big|_{z_{1}=z_{2}=z}
+
\varepsilon^{\nu}R_{\nu}(b_{1},b_{2};\varepsilon)
\end{eqnarray*}
where $R_{\nu}(b_{1},b_{2};\varepsilon)$ is uniformly bounded w.r.t
$\varepsilon$ in the Fr{\'e}chet space
$S_{p\Z}(\frac{m_{1}m_{2}}{\lambda^{\nu}},g)$\,.
The Calderon-Vaillancourt theorem
$$
\left|b^{Weyl}_{p\Z}\right|_{\L(\Gamma_{s}(p\Z))}\leq Cp_{k_{d}}(b)
$$
and the G{\aa}rding inequality
$$
\left(b\geq 0\right) \Rightarrow \left(b^{Weyl}_{p\Z}\geq
  -C'p_{k_{d}}'(b)\varepsilon\right)
$$
respectively for $b\in S_{p\Z}(1,g)$ and $b\in S_{p\Z}(\lambda,g)$\,.
The index $k_{d}$
for the seminorms $p_{k_{d}}$ and $p_{k_{d}}'$)  recalls the
dimension dependent number of derivatives required in the estimates.
\end{prop}
The typical example H{\"o}rmander metrics, which will be used here,
are $|dz|^{2}=dx^{2}+d\xi^{2}$ ($\lambda(z)=1$) and
$\frac{|dz|^{2}}{\left\langle z\right\rangle^{2}}=
\frac{dx^{2}}{\left\langle (x,\xi)\right\rangle^{2}}+
\frac{d\xi^{2}}{\left\langle (x,\xi)\right\rangle^{2}}$
($\lambda(z)=1+\left|z\right|^{2}$)\,.
Both of them  split up in the $(x,\xi)$ coordinates and the Beals
criterion of Bony-Chemin \cite{BoCh} translated in the semiclassical case
in \cite{NaNi}-Appendix-A can be applied. Following the method recalled in
\cite{HeNi}-Chapter-4, this allows to check that functions of fully elliptic
self-adjoint pseudodifferential operators are pseudodifferential
operators, with an explicit knowledge of their principal symbol.
In particular, this can be applied with $1+\frac{\varepsilon\textrm{dim}p}{2}+N_{p\Z}=
(1+\left|z\right|^{2})^{Weyl}_{p\Z}$ while noticing that
$1+\frac{\varepsilon\textrm{dim}p}{2}+N_{p\Z}$ is a fully elliptic operator in
$S(\left\langle z\right\rangle^{2},\frac{|dz|^{2}}{\left\langle
    z\right\rangle^{2}})$
(such a result with $\varepsilon=1$ can be found also in \cite{Hel1}).
\begin{prop}
\label{pr.calcfunc}
Fix $p\in\p$, fix the exponent $s\in \rz$ and let
$N_{p\Z}=d\Gamma(I_{p\Z})$ be the number operator on
$\Gamma_{s}(p\Z)$. For any $s\in \rz$, $(1+\frac{\varepsilon\textrm{dim}p}{2}+
N_{p\Z})^{s/2}$ can be written $(b(s,\varepsilon))^{Weyl}_{p\Z}$ with
$\varepsilon^{-1}(b(z;s,\varepsilon)-\left\langle z\right\rangle^{s})$
uniformly bounded in $S(\left\langle z\right\rangle^{s-2}, \frac{|dz|^{2}}{\left\langle z\right\rangle^{2}})$\,.
\end{prop}

\subsection{Weyl quantization and Laguerre connection}
In this paragraph, the relationship between the Wick and Weyl calculus
is checked in the infinite dimensional setting. It specifies the
relation between the representation of the Weyl algebra, generated
by the $W(\xi)$, and the number representation which puts the stress
on Wick symbols or Hermite states $z^{\otimes k}$. This relies
on the introduction of Hermite and Laguerre polynomials, recalled below.

Let $h_{n}(x)$ denote, for any $n\in \nz$, the $n$-th Hermite
polynomial in $\cz$:
\begin{equation}
  \label{eq.hermite}
  h_{n}(x)=(-1)^{n}e^{x^{2}}\frac{d^{n}}{dx^{n}}(e^{-x^{2}})=\sum_{r=0}^{[n/2]}
(-1)^{r}\frac{n!}{r!(n-2r)!}(2x)^{n-2r}\,.
\end{equation}
Those classical polynomials are also given by the generating
function
\begin{eqnarray}
  \label{eq.genhermite}
\sum_{n=0}^{\infty}\frac{t^{n}}{n!}h_{n}(x)=e^{x^{2}}\left[\sum_{n=0}^{\infty}
\frac{(-t\partial_{x})^{n}}{n!}e^{-x^{2}}\right]=e^{x^{2}}
e^{-t\partial_{x}}[e^{-x^{2}}]=e^{2tx-t^{2}}\,.
\end{eqnarray}
\begin{lem}
\label{weyl-estimate}\ \\
(i) For any $\xi\in\Z$, the following identity holds in $\H_{fin}$:
\begin{eqnarray*}
\ds W(\xi) &=&\sum_{n=0}^\infty \frac{|\sqrt{\hbarr}\xi|^n}{2^n
n!} \;\; h_n\left(\frac{i\sqrt{2}
S(\xi,z)}{|\sqrt{\hbarr}\xi|}\right)^{Wick}\, .
\end{eqnarray*}
(ii) For any
 $n,j,k\in \nz$ the estimate
\begin{eqnarray*}
\ds \left| 1_{\left\{j\hbarr\right\}}(N)\circ h_n\left(i\sqrt{2}
    S(\xi,z)\right)^{Wick}
\circ 1_{\left\{k\hbarr\right\}}(N)\
\right|_{\L(\bigvee^k\Z,\bigvee^j\Z)}\leq (
1+2\sqrt{2(k+j)\varepsilon}\; \left|\xi\right|)^{n}\;
\frac{n!}{[n/2]!} \,,
 \end{eqnarray*}
holds for any $\xi\in \Z$\,.
\end{lem}
\proof
Using the generating function \eqref{eq.genhermite}
with $t=\frac{\sqrt{\hbarr}|\xi|}{2}$ and
$x=\frac{i\sqrt{2}S(\xi,z)}{\sqrt{|\hbarr}\xi|}$ implies
the equality of the Wick symbols
$$
e^{i\sqrt{2}S(\xi,z)}e^{-\frac{\varepsilon|\xi|^{2}}{4}}=
e^{i\frac{2\sqrt{2}S(\xi,z)}{\sqrt{\varepsilon}|\xi|}\frac{\sqrt{\varepsilon}|\xi|}{2}}
e^{-\frac{\varepsilon|\xi|^{2}}{4}} =
\sum_{n=0}^{\infty}\frac{(\sqrt{\varepsilon}|\xi|)^{n}}{2^{n}n!}h_{n}
\left(\frac{i\sqrt{2}S(\xi,z)}{\left|\sqrt{\varepsilon}\xi\right|}\right)\,.
$$
Nevertheless the equality of the the series of Wick quantized
operators has to be checked.\\
Recall that elements of $\H_{fin}$ are analytic vectors with infinite radius of convergence
 for the field operators. Hence the sum
\begin{eqnarray*}
W(\xi)\psi=\sum_{n=0}^\infty \frac{i^n}{n!} \; \Phi(\xi)^n\psi,  \;\;\;\; \psi\in\H_{fin},
\end{eqnarray*}
is absolutely convergent for all $\xi\in\Z$. Therefore to prove $(i)$ it is enough to compute the Wick symbol of
$\Phi(\xi)^n$ for all $n$. Indeed using the Wick ordering rules, we have
\begin{eqnarray*}
\Phi(\xi)^n&=&\sum_{r=0}^{[n/2]} \frac{n!}{\sqrt{2^n} r! (n-2r)!} \frac{|\xi|^{2r}}{2^r}\; \hbarr^r \;
\sum_{s=0}^{n-2r} C_{n-2r}^s \; a^*(\xi)^s \; a(\xi)^{n-2r-s}\\
&=&\frac{|\xi|^n}{2^n}
\;\sum_{r=0}^{[n/2]} \frac{n!}{r! (n-2r)!} \; \hbarr^r \; \left(\frac{\sqrt{2^{n-2r}}}{|\xi|^{n-2r}} \;\; \sum_{s=0}^{n-2r} C_{n-2r}^s
\la z,\xi\ra^s \; \la \xi,z\ra^{n-2r-s} \right)^{Wick}\\ \nm
&=& \frac{|\xi|^n}{2^n}
\;\left(\sum_{r=0}^{[n/2]} \frac{n!}{r! (n-2r)!} \; \hbarr^r \;
\big(\frac{2\sqrt{2} S(\xi,z)}{|\xi|}\big)^{n-2r} \right)^{Wick}.
\end{eqnarray*}
To prove the second statement $(ii)$, take $\psi_{k}\in \bigvee^{k}\Z$
and $\psi_{j}\in \bigvee^{j}\Z$ and write
\begin{eqnarray*}
\la \psi_{j}\,,\, h_n\left(i\sqrt{2} S(\xi,z)\right)^{Wick}\psi_{k}\ra
=\sum_{r=0}^{[n/2]}
\frac{n!}{(n-2r)! r!} \; \la \psi_{j}\,,\,\left(\big(2i\sqrt{2}
S(\xi,z)\big)^{n-2r}\right)^{Wick}
\psi_{k}\ra\,.
 \end{eqnarray*}
Using Lemma~\ref{wick-estimate} one obtains
\begin{eqnarray*}
\left|\la \psi_{j}\,,\,
h_n\left(i\sqrt{2} S(\xi,z)
\right)^{Wick}
\psi_{k}\ra
\right|
&\leq &\left|\psi_{j}\right|_{\bigvee^{j}\Z}\left|\psi_{k}\right|_{\bigvee^{k}\Z}\sum_{r=0}^{[n/2]}
\frac{n!}{(n-2r)! r!} \;
(2\sqrt{2(k+j)\varepsilon}\left|\xi\right|)^{n-2r}
\\
&\leq &
\left|\psi_{j}\right|_{\bigvee^{j}\Z}\left|\psi_{k}\right|_{\bigvee^{k}\Z}
\sum_{s=0}^{n} \frac{n!}{(n-s)! s!} \;
(2\sqrt{2(k+j)\varepsilon}\left|\xi\right|)^{n-s} \frac{s!}{[s/2]!}
\\
&\leq&
\left|\psi_{j}\right|_{\bigvee^{j}\Z}\left|\psi_{k}\right|_{\bigvee^{k}\Z}
(1+2\sqrt{2(k+j)\varepsilon}\left|\xi\right|)^{n} \;
\frac{n!}{[n/2]!}\,.
\end{eqnarray*}
\hfill
\fin

The Laguerre polynomials are defined by the formula
\begin{eqnarray*}
L_k^{(j)}(t)=\sum_{m=0}^k (-1)^m \frac{(k+j)!}{(k-m)! (j+m)!  m!} \;
t^m, \quad t \in\cz.
\end{eqnarray*}
The following proposition gives the Laguerre connection (see \cite{Fol},\cite{Rip}).
\begin{prop}
\label{laguerre}
For $z,\xi\in\Z$ with $\left|z\right|=1$, the next equalities hold
according to the ordering of $j$ and $k\in \nz$,
\begin{equation}
  \label{eq.lag1}
\V[z^{\otimes k},z^{\otimes j}](\frac{\xi}{\pi\sqrt{2\hbarr}})
=\left\{
  \begin{array}[c]{ll}
(i)^{k-j} \sqrt{\frac{j!}{k!}} L_j^{(k-j)}(|\la \xi,z\ra|^2) \la
\xi,z\ra^{k-j} e^{-|\xi|^2/2} &\mbox{ \rm if }  k\geq j\,,\\
(i)^{j-k} \sqrt{\frac{k!}{j!}} L_k^{(j-k)}(|\la \xi,z\ra|^2) \la
z,\xi\ra^{j-k} e^{-|\xi|^2/2} &\mbox{ \rm if } j\geq k\,.
\end{array}
\right.
\end{equation}
\end{prop}
\proof Let us establish the expression of $\V[z^{\otimes
k},z^{\otimes j}]$ in the case $k\geq j$. The case $j\leq k$ is
similar. Using Lemma~\ref{weyl-estimate}  one obtains
\begin{eqnarray*}\ds
\V[z^{\otimes k},z^{\otimes j}](\frac{\xi}{\pi\sqrt{2\hbarr}})&=&
\la z^{\otimes j}, W(\sqrt{\frac{2}{\hbarr}}\xi) z^{\otimes k}\ra\\
\nm &=&\sum_{n=0}^\infty \frac{|\xi|^n}{\sqrt{2^n}n!} \;\;\la
z^{\otimes j},\; h_n\left(\frac{ i
S(\sqrt{\frac{2}{\hbarr}}\xi,.)}{|\xi|} \right)^{Wick}z^{\otimes
k}\ra\, .
\end{eqnarray*}
Now let use the explicit form of $h_n$ and Proposition
\ref{pr.wickformules}. We obtain  for $\left|z\right|=1$,
\begin{eqnarray*} \V[z^{\otimes k},z^{\otimes
j}](\frac{\xi}{\pi\sqrt{2\hbarr}})&=& \sum_{n=0}^\infty
\sum_{r=0}^{[n/2]} \sum_{s=0}^{n-2r} \frac{i^n|\xi|^{2r}}{2^r r!
(n-2r)!} C_{n-2r}^s
 \;  \; \hbarr^{r-\frac{n}{2}} \;\;\la z^{\otimes j},
\left(\la \xi,.\ra^s \la .,\xi\ra^{n-2r-s} \right)^{Wick}z^{\otimes
k}\ra\\ \nm \ds &=& \sum_{n=0}^\infty \sum_{r=0}^{[n/2]}
\sum_{s=0}^{n-2r}  \frac{i^n|\xi|^{2r}}{2^r r! (k-j+s)!s!}
 \;  \;
|\la \xi,z\ra|^{2s} \; \la\xi,z\ra^{k-j} \frac{\sqrt{k!
j!}}{(j-s)!}\;
\delta^{+}_{k-n+2r+s,j-s} \\
&=& (i)^{k-j} \sqrt{\frac{j!}{k!}}  \sum_{s=0}^{j} \sum_{r=0}^\infty
\frac{(-1)^{r} |\xi|^{2r}}{2^r r!} \frac{(-1)^{s} k!}{s!
(k-j+s)!(j-s)!}
 \;  \; |\la \xi,z\ra|^{2s} \; \la\xi,z\ra^{k-j} \;.
\end{eqnarray*}
The last term gives the claimed identity. \hfill\fin

\subsection{Anti-Wick Operators}

The Anti-Wick quantization is introduced by a separation of variables
process like the Weyl quantization. For a given $p\in \p$, set $p
^{\bot}=1-p$, and use the tensor decomposition \eqref{eq.Tunit}. The
Weyl operators on $p\Z$ and $p^{\bot}\Z$ are denoted by
$W_{p}(\xi_{1})$ and $W_{p^{\bot}}(\xi_{2})$ with
$W(\xi_{1}\oplus^{\bot} \xi_{2})=W_{p}(\xi_{1})\otimes
W_{p^{\bot}}(\xi_{2})$\,.
For any $\xi\in p\Z$, the coherent state
$E_{p}(\xi)$ is defined by $E_{p}(\xi)=W_{p}(\frac{\sqrt{2}\xi}{i\hbarr}) \Omega^{p\Z}$\,.
Introduce the projector $P_{\xi}$  on $\H$ after tensorization with $I_{\Gamma_{s}(p^{\bot}\Z)}$:
\begin{eqnarray*}
p\Z\ni
\xi\mapsto P^\hbarr_\xi=
\left(|E_{p}(\xi)\ra \la E_{p}(\xi)|\right)\otimes
I_{\Gamma_s(p^\bot\Z)}\,.
\end{eqnarray*}

The  Anti-Wick operator associated with  a symbol
$b\in\mathcal{S}_{cyl}(\Z)$ based on $p\Z$ is then defined  by
\begin{eqnarray*}
\label{eq.AWickquant}
b^{A-Wick}=
\int_{p\Z} b(\xi) \;\; P^\hbarr_\xi \;\;\;
\frac{L_{p}(d\xi)}{(\pi\hbarr)^{\textrm{dim}p\Z}}
=b^{A-Wick}_{p\Z}\otimes I_{\Gamma_{s}(p^{\bot}\Z)}\,.
\end{eqnarray*}
The above formula can be first considered in a weak sense or as a Bochner
integral when  $b\in\S(p\Z)$ and
the bounded projector $P^\hbarr_\xi$ is continuous w.r.t. $\xi$.
The finite dimensional identification of the Weyl symbol of
$|W_{p}(\frac{\sqrt{2}\xi}{i\hbarr}) \Omega^{p\Z}\ra \la
W_{p}(\frac{\sqrt{2}\xi}{i\hbarr}) \Omega^{p\Z}|$, can be deduced
after completing the table of correspondences in
Subsection~\ref{se.weylfin}:
\begin{eqnarray*}
    p\Z\sim \cz^{d}&
z=x+i\xi
&T^{*}\rz^{d}\\
\Gamma_s(p\Z)\sim \Gamma_s(\cz^{d})
\,, &
\varepsilon=2h
&L^{2}(\rz^{d})\\
 E_{p}(z_{0})=W_{p}(\frac{\sqrt{2}}{i\varepsilon}z_{0})\Omega^{p\Z}
&\frac{z_{0}}{i}=\xi_{0}-ix_{0}&
\tau_{(\frac{x_{0}}{\sqrt{h}}, \frac{\xi_{0}}{\sqrt{h}})}(\pi^{-d/4}e^{-\frac{x^{2}}{2}})
\\
|\Omega^{p\Z}\ra \la\Omega^{p\Z}|=\gamma^{Weyl}
&&
(\pi)^{-d/2}
e^{-\frac{x^{2}}{2}-\frac{y^{2}}{2}}=g^{Weyl}(\sqrt{h}x,\sqrt{h}D_{x})
\\
\gamma(z)=2^{d}e^{-\frac{\left|z\right|_{p\Z}^{2}}{\varepsilon/2}}
&
\Leftarrow
&\text{with}~g(x,\xi)=2^{d}e^{-\frac{x^{2}+\xi^{2}}{h}}
\end{eqnarray*}
From the conjugation
$$
\tau_{(\frac{x_{0}}{\sqrt{h}},
  \frac{\xi_{0}}{\sqrt{h}})}a^{Weyl}(\sqrt{h}x,\sqrt{h}D_{x})
\tau_{(\frac{x_{0}}{\sqrt{h}}, \frac{\xi_{0}}{\sqrt{h}})}^{*}=
a(.-x_{0},.-\xi_{0})^{Weyl}(\sqrt{h}x,\sqrt{h}D_{x})
$$
the above correspondence gives
$$
|E_{p}(\xi)\ra \la
E_{p}(\xi)|=\gamma_{\xi}^{Weyl}\quad\text{with}\quad
\gamma_{\xi}(z)=2^{d}e^{-\frac{\left|z-\xi\right|_{p\Z}^{2}}{\varepsilon/2}}\,.
$$
Hence the usual finite dimensional relation between the Weyl and
Anti-Wick quantization now reads (after tensorization with $I_{\Gamma_{s}(p^{\bot}\Z)}$)
\begin{eqnarray}
\label{eq.AWW1}
b^{A-Wick}&= &
\left(b\mathop{*}_{p\Z}\frac{e^{-\frac{|z|_{p\Z}^{2}}{\hbarr/2}}}{(\pi\varepsilon/2)^{\textrm{dim}p\Z}}\right)^{Weyl}
\\
\label{eq.AWW2}
&=&
\int_{p\Z}
\F[b](\xi)\;\;W(\sqrt{2}\pi\xi)\;\;e^{-\frac{\hbarr \pi^2}{2} |\xi|^2_{p\Z}}
\;L_{p}(d\xi)\,,
\end{eqnarray}
for any $b\in \mathcal{S}(p\Z)$ by setting
$$
b\mathop{*}_{p\Z}\gamma(z)=\int_{p\Z}b(z)\gamma(z-z')~L_{p}(dz')\,.
$$
From \eqref{eq.AWW1}, the Anti-Wick quantization can be extended to symbols in
$S(1,|dz|^{2})$ with the next properties (see \cite{HMR}).
\begin{prop}
\label{pr.Awick}
Fix $p\in \p$.
Let $b\in S_{p\Z}(1,|dz|^{2})$, the following statements hold true:\\
(i) If $b\geq 0$ then $b^{A-Wick}\geq 0$.\\
(ii) $\left| b^{A-Wick}\right|_{\L(\H)} \leq \left|b\right|_{L^\infty(p\Z)}$.\\
(iii) The comparison with the Weyl quantization is given by
\eqref{eq.AWW1} with the estimate
$$
\left|b^{A-Wick}-b^{Weyl}\right|_{\mathcal{L}(\H)}\leq C_{d}p_{k_{d}}(b)\varepsilon
$$
where the constant $C_{d}>0$ and the seminorm $p_{k_{d}}$ depend
essentially on the dimension $d=\textrm{dim}p\Z$.
\end{prop}
A variation of it holds when $b\in
\mathcal{F}^{-1}\left(\mathcal{M}_{b}(p\Z)\right)$, when
$\mathcal{M}_{b}(p\Z)$ denotes the set of bounded (Radon) measures on
$p\Z$ and comes directly from \eqref{eq.AWW2}.
\begin{prop}
  \label{pr.AWW2}
For any $p\in \p$ and any $b\in
\mathcal{F}^{-1}\left(\mathcal{M}_{b}(p\Z)\right)$, the Anti-Wick and Weyl
observables are asymptotically the same:
$$
\lim_{\varepsilon\to 0}\left|b^{A-Wick}-b^{Weyl}\right|_{\L(\H)}=0\,.
$$
\end{prop}
\proof
Recall that $b^{Weyl}$ can be defined for any $b\in \mathcal{S}'(p\Z)$
as a continuous operator from $\ccap_{k\in \nz}D(N_{p\Z}^{k})$  $\sim
\mathcal{S}(\rz^{d})$
to $\ccup_{k\in \nz}D(N_{p\Z}^{k})^*\sim \mathcal{S}'(\rz^{d})$, with
$d=\textrm{dim}p\Z$
 and \eqref{eq.AWW2} is still valid for such a symbol. Assume
 $\mathcal{F}b=\nu\in \mathcal{M}_{b}(p\Z)$. The identity
$$
\left\langle \psi\,,(b^{Weyl}-b^{A-Wick})\varphi\right\rangle
=\int_{p\Z}\left\langle
  \psi\,,\,W(\sqrt{2}\pi\xi)\varphi\right\rangle\left(1-e^{-\frac{\varepsilon\pi^{2}}{2}\left|\xi\right|^{2}}\right)
d\nu(\xi)
$$
holds for any $\varphi,\psi\in
 \ccap_{k\in \nz}D(N_{p\Z}^{k})$. This implies
$$
\left|b^{Weyl}-b^{A-Wick}\right|_{\L(\H)}\leq
\int_{p\Z}\left(1-e^{-\frac{\varepsilon\pi^{2}}{2}\left|\xi\right|^{2}}\right)
d\left|\nu\right|(\xi)
\stackrel{\varepsilon\to 0}{\to}0\,.
$$
\fin
\subsection{Weyl quantization and specific Wick observables}

In finite dimension, that is for any fixed $p\in \p$,
 polynomially bounded symbols can be introduced
after considering the class of symbols
$\ccup_{s\in\rz}S_{p\Z}(\langle z\rangle^{s}, g_{p})$
 where $g_{p}$ is either the metric  $|dz|^{2}$   or
 $\frac{|dz|^{2}}{\langle z\rangle^{2}}$ on $p\Z$.
According to
 Proposition~\ref{pr.weylcalc} it is an algebra with the
 Moyal product, $\#^{\varepsilon/2}$,  associated with the composition of
 Weyl quantized observable with a complete asymptotic expansion
of $b_{1}\#^{\varepsilon/2}b_{2}$. For any $m,q\in \nz$, the finite
dimensional space
$\P_{m,q}(p\Z)$ of $(m,q)$-homogeneous polynomials on $\Z$ is
contained in $\mathcal{S}_{p\Z}(\langle z\rangle^{m+q},g_{p})$. The
comparison between the Weyl and Wick quantizations is symmetric to
\eqref{eq.AWW1} (see\cite{BeSh}):
$$
\forall b\in \oplus_{m,q}^{\rm alg}\P_{m,q}(p\Z),\quad b_{p\Z}^{Weyl}=
\left(b\mathop{*}_{p\Z}\frac{e^{-\frac{|z|_{p\Z}^{2}}{\hbarr/2}}}{(\pi\varepsilon/2)^{\textrm{dim}p\Z}}\right)^{Wick}\,.
$$
For polynomials the deconvolution is possible and we get for any
$m,q\in \nz$ and any $b\in\P_{m,q}(p\Z)$
$$
\varepsilon^{-1}(b_{p\Z}^{Wick}-b_{p\Z}^{Weyl})=c_{p\Z}(\varepsilon)^{Weyl}
$$
where the symbol $c(\varepsilon)$ equals
$$
c(\varepsilon)=\varepsilon^{-1}
\left[
\left(
b\mathop{*}_{p\Z}\frac{e^{\frac{|z|_{p\Z}^{2}}{\hbarr/2}}}{(\pi\varepsilon/2)^{\textrm{dim}p\Z}}\right)-b
\right]
$$
and is uniformly bounded in $S_{p\Z}(\langle z\rangle^{m+q-2},g_{p})$
w.r.t $\varepsilon\in (0,\overline{\varepsilon})$.

The space
$\P_{m,q}(p\Z)$ is identified with a subspace of $\P_{m,q}(\Z)$
and even of any $\P_{m,q}^{r}(\Z)$ for any $r\in [1,+\infty]$ with
\begin{eqnarray*}
\forall b\in \P_{m,q}(p\Z),
&&
\quad\forall z\in \Z,\;
b(z)=b(pz+p^{\bot}z)=b(pz)\\
&&
\quad
\tilde{b}=p^{\otimes q}\circ\tilde{b}\circ p^{\otimes m}=\Gamma_{s}(p)\tilde{b}\Gamma_{s}(p)\,.
\end{eqnarray*}
After tensoring the finite dimensional comparison with $I_{\Gamma_{s}(p^{\bot}\Z)}$, we have proved
\begin{prop}
\label{pr.weylwick}
For any $p\in \p$, any $m,q\in\nz$, the class of symbols
$\P_{m,q}(p\Z)$ is contained in $\P_{m,q}^{1}(\Z)$ $ \cap S_{p\Z}(\langle
z\rangle^{m+q},g_{p})$. Moreover the
operator
$\varepsilon^{-1}(b^{Wick}-b^{Weyl})$ can be written
$c_{\varepsilon}^{Weyl}$ with $c_{\varepsilon}$ uniformly bounded in
$S_{p\Z}(\langle
z\rangle^{m+q-2},g_{p})$ w.r.t $\varepsilon\in
(0,\overline{\varepsilon})$. ( The metric $g_{p}$ can be either  $|dz|^{2}$   or
 $\frac{|dz|^{2}}{\langle z\rangle^{2}}$ on $p\Z$.)
\end{prop}

\section{Coherent and product states}
\label{se.coherentprod}
We distinguish the coherent states
$E(z)=W(\frac{\sqrt{2}}{i\hbarr}z)\Omega$ (resp. the projector
$|E(z)\ra\la E(z)|$) from the product or Hermite state $z^{\otimes
  k}$ (resp. the projector $|z^{\otimes k}\ra\la z^{\otimes k}|$).
Although they are intimately related, the asymptotics  of coherent
state $E(z)$ tested on Wick, Weyl or Anti-Wick observables
encoded exactly the geometry of the phase-space $\Z$, while the
asymptotics of the product states $z^{\otimes k}$, $k\hbarr\to 0$
keeps track of the gauge invariance
$$
\forall \theta\in [0,2\pi]\,,\quad |(e^{i\theta}z)^{\otimes k}\ra
\la (e^{i\theta}z)^{\otimes k}|=|z^{\otimes k}\ra\la z^{\otimes k}|
$$
with variations according to the
quantization.

\begin{prop}
\label{pr.limV}
Fix  $z,\xi\in\Z$ with $\left|z\right|=1$.\\
$(i)$ The convergence
\begin{eqnarray*}
\lim_{\tiny \begin{array}{c} \hbarr\to 0
\\
k\hbarr\to 1 \end{array}}
\V[z^{\otimes k},z^{\otimes k-m}](\xi) =\frac{1}{2\pi}
\int_0^{2\pi} e^{2\pi i S(z^\theta,\xi)} e^{-im\theta}
\,d\theta\,,
\end{eqnarray*}
holds for any fixed $m\in \nz$  by setting
$z^\theta=e^{i\theta} z$\,.\\
$(ii)$ The coherent state
$E(z)=W(\frac{\sqrt{2}}{i\varepsilon}z)\Omega$  satisfies
$$
\V\left[E(z),E(z)\right](\xi)=e^{2\pi i
  S(\xi,z)}e^{-\frac{\hbarr|\xi|^{2}}{2}}
\stackrel{\hbarr \to 0}{\to}e^{2\pi i
  S(\xi,z)}\,.
$$
\end{prop}
\proof
$i)$ Set $j=k-m$ and compute $\V[z^{\otimes k},z^{\otimes
j}](\xi)$ with $\xi=\frac{\xi'}{\sqrt{2}\pi}$
according to Proposition \ref{laguerre} :
\begin{eqnarray*}
&&\V[z^{\otimes k},z^{\otimes j}](\frac{\xi'}{\sqrt{2}\pi}) = (i)^{m}
\sqrt{\frac{j!}{k!}} L_j^{(m)}(\frac{\hbarr}{2}|\la \xi',z\ra|^2)
(\frac{\hbarr}{2})^{m/2}
\la \xi',z\ra^{m} e^{-\hbarr|\xi'|^2/4}\\
&&\quad= (i)^{m} \sum_{s=0}^\infty
 \frac{(-1)^s}{s! (s+m)!} 1_{[0,j]}(s)\sqrt{\frac{j!}{(j-s)! k^s}}
\sqrt{\frac{k!}{(j-s)! k^{m+s}}} (\frac{\hbarr
k}{2})^{\frac{2s+m}{2}}|\la \xi',z\ra|^{2s} \la \xi',z\ra^{m}
e^{-\hbarr|\xi'|^2/4}\,.
\end{eqnarray*}
The bounds $(\hbarr k)\leq C$ and
$\sum_{s=0}^{\infty}\frac{C^{s}}{s!(s+m)!}<\infty$ imply
\begin{eqnarray*}
\lim_{\tiny \begin{array}{c} \hbarr\to 0
\\
k\hbarr\to 1 \end{array}}
\V[z^{\otimes k},z^{\otimes j}](\frac{\xi'}{\sqrt{2}\pi}) = (i)^{m}
\sum_{s=0}^\infty \frac{(-1)^s}{2^{\frac{2s+m}{2}} s! (s+m)!}  |\la
\xi',z\ra|^{2s} \la \xi',z\ra^{m}\,,
\end{eqnarray*}
owing to Lebesgue's theorem. A simple series expansion
$e^{t}=\sum_{k=0}^{\infty}\frac{t^{k}}{k!}$  for $t=i\sqrt{2}
S(z^\theta,\xi')$ gives
\begin{eqnarray*}
\frac{1}{2\pi} \int_0^{2\pi}
e^{i\sqrt{2} S(z^\theta,\xi')} e^{-im\theta} \,d\theta=(i)^{m}
\sum_{s=0}^\infty \frac{(-1)^s}{2^{\frac{2s+m}{2}}s! (s+m)!}
 |\la \xi',z\ra|^{2s} \la \xi',z\ra^{m}.
\end{eqnarray*}
$ii)$ is a straightforward consequence of \eqref{eq.wicktrans}.
\hfill\fin

\noindent The next result specifies the similarity and the
differences between the product states and the coherent states in
the mean-field or semiclassical limit.
\begin{thm}
\label{th.prodcoh}
Let $z\in \Z$ and  $m\in \nz$ be fixed with $|z|=1$ and set
$z^{\theta}=e^{i\theta}z$ for $\theta\in [0,2\pi]$. The next limits
exist as $\hbarr\to 0$, $k\hbarr \to 1$.
\label{main-1} \\
(i) For $b\in\S_{cyl}(\Z)$,
$$\ds
\lim_{\tiny \begin{array}{c} \hbarr\to 0
\\
k\hbarr\to 1 \end{array}}
 \la z^{\otimes k-m}, b^{Weyl}\,z^{\otimes k}\ra
=
\lim_{\tiny \begin{array}{c} \hbarr\to 0
\\
k\hbarr\to 1 \end{array}}
 \la z^{\otimes k-m}, b^{A-Wick}\,z^{\otimes k}\ra
 =\frac{1}{2\pi} \int_0^{2\pi} b(z^\theta) e^{-im\theta}
 d\theta\,.
$$
Meanwhile the coherent state $E(z)$ satisfies
$$
\lim_{\hbarr \to 0}\langle E(z)\,,\, b^{Weyl}E(z)\rangle
=\lim_{\hbarr \to 0}\langle E(z)\,,\, b^{A-Wick}E(z)\rangle=b(z)\,.
$$
(ii)  For $b\in\P_{p,q}(\Z)$, with $p,q\in \nz$ fixed,
$$
\lim_{\tiny \begin{array}{c} \hbarr\to 0
\\
k\hbarr\to 1 \end{array}}
\la z^{\otimes k-m}, b^{Wick}\, z^{\otimes k} \ra=\delta_{p-q,m}
\;\,b(z) = \frac{1}{2\pi} \int_0^{2\pi} b(z^\theta) e^{-im\theta}
 d\theta
\,.
$$
Meanwhile the coherent state $E(z)$ satisfies
$$
\forall \varepsilon >0\,,\quad \langle E(z)\,,\, b^{Wick}E(z)\rangle  = b(z)\,.
$$
\end{thm}
\proof
Set $j=k-m$, with $m\in\nz$ fixed.\\
For $(i)$, fix $b\in \S_{cyl}(\Z)$. The definition of
$b^{Weyl}$ in (\ref{weyl-obs}), says
\begin{eqnarray*}
\la z^{\otimes j},
b^{Weyl}\,z^{\otimes k}\ra&=&\int_{p\Z} \F[b](\xi) \;
\la z^{\otimes j},W(\sqrt{2}\pi \xi)\,z^{\otimes k}\ra~L_{p}(d\xi)\\
&=&\int_{p\Z} \F[b](\xi) \; \V[z^{\otimes k},z^{\otimes
j}](\xi)~L_{p}(d\xi)\,.
\end{eqnarray*}
Since $\F[b]\in\S(p\Z)$ and
$\V[z^{\otimes k},z^{\otimes j}](\xi)$ converges pointwise according
to Proposition~\ref{pr.limV},
 Lebesgue's theorem yields
 \begin{eqnarray*}
 \lim_{\tiny \begin{array}{c} \hbarr\to 0
\\
k\hbarr\to 1 \end{array}}
\la z^{\otimes j}, b^{Weyl}\,z^{\otimes k}\ra &=&
 \int_{p\Z} \F[b](\xi)\left(
 \frac{1}{2\pi} \int_0^{2\pi} e^{i2\pi
S(z^\theta,\xi)} e^{-im\theta} \,d\theta\right)~L_{p}(d\xi)
\\
&=&
 \frac{1}{2\pi}  \int_0^{2\pi} b(z^\theta) e^{-im\theta} \,d\theta.
\end{eqnarray*}
The limit with  Anti-Wick observables is a consequence of
the formula \eqref{eq.AWW2}:
\begin{eqnarray*}
\la z^{\otimes j}, b^{A-Wick}\,z^{\otimes k}\ra=\int_{p\Z}
\F[b](\xi)\;\; \la z^{\otimes j}, W(\sqrt{2}\pi\xi)z^{\otimes k}\ra
\;\;e^{-\frac{\hbarr \pi^2}{2} |\xi|^2_{p\Z}}~L_{p}(d\xi)\,.
\end{eqnarray*}
The statement about the coherent state $E(z)$ is even simpler by
referring to Proposition~\ref{pr.limV}~$(ii)$.\\
Let us
prove $(ii)$. The statement $(ii)$ of
Proposition~\ref{pr.wickformules}
gives
\begin{eqnarray*} \la z^{\otimes j}, b^{Wick}\, z^{\otimes k} \ra&=&
\delta^{+}_{k-p, j-q} \; \sqrt{\frac{k!j!}{(k-p)! (j-q)!}} \;
\hbarr^{\frac{p+q}{2}}\; \; \la z^{\otimes q}, b z^{\otimes p}\ra\\
&=& \delta_{m, p-q} \; \sqrt{\frac{k!}{(k-p)! k^p}} \;
\sqrt{\frac{j!}{ (j-q)! k^q}}(\hbarr k)^{p+q}\; \; \la z^{\otimes
q}, b z^{\otimes p}\ra. \end{eqnarray*} We conclude again with
$\sqrt{\frac{k!}{(k-p)! k^p}} \; \sqrt{\frac{j!}{ (j-q)! k^q}}\to 1$
as $k\to\infty$. \hfill \fin

\section{An example of a dynamical mean-field limit}
\label{se.anex}
In order to illustrate the general presentation, the
standard example of the mean field derivation of the Hartree
equation from the non relativistic Hamiltonian of bosons with a
quartic interaction is considered. Two standard methods are considered:
The coherent state method  (see
\cite{Hep}\cite{GiVe} or \cite{Cas} for a rapid
presentation) also known as Hepp method and the propagation of chaos
approach with a truncated
Dyson expansion according \cite{FGS}\cite{FKP}\cite{ESY1}\cite{ESY2}\cite{Spo}.

Consider $\Z=L_{\cz}^2(\rz^d,dx)$ and take $V\in
L^{\infty}_{\rz}(\rz^d,dx)$ such that $V(-x)=V(x)$.
The polynomial $Q(z)=\la z^{\otimes 2}\,,\,\tilde{Q}z^{\otimes
2}\ra$ is associated with the operator
$\tilde{Q}\in\L(\bigotimes^{2}\Z)$
 defined by
\begin{eqnarray*}
\tilde{Q}:\otimes^{2}\Z&\to& \otimes^{2}\Z\,,\\
u(x_{1})w(x_{2})&\mapsto& \frac{1}{2}
V(x_{1}-x_{2}) \, u(x_{1})w(x_{2}).
\end{eqnarray*}
The Hamiltonian is
defined as
\begin{eqnarray*} H_\hbarr=\d\Gamma(-\Delta)+
Q^{Wick},
\end{eqnarray*} where $-\Delta$ is  the Laplacian of $\rz^d$,
while
 $H_\hbarr^0$ denotes the free Hamiltonian
$\d\Gamma(-\Delta)$. It is well known that $H_\hbarr$ is a
self-adjoint operator on $\H$ (see \cite{GiVe}) and  the quantum
time-evolution group  is denoted by
$U_\hbarr(t)=e^{-i\frac{t}{\hbarr}
  H_\hbarr}$ while
$U_{\hbarr}^{0}(t)=e^{-i\frac{t}{\hbarr}H_{0}}=\Gamma(e^{it
\Delta})$ stands for the free dynamics.
Although the Wick quantization of non continuous
  polynomials has not been introduced here, this Hamiltonian appears
  as the Wick quantization of the energy functional
\begin{equation}
    \label{eq.enfunct}
    h(z)=\int_{\rz^{d}}\left|\nabla z\right|^{2}~dx+ Q(z)\,.
\end{equation}
It is also well known that the
mean field limit  is the  nonlinear dynamics issued from the
{\it Hartree equation}
\begin{eqnarray}
\label{hartree}
i \partial_t z_t=- \Delta z_t+ V*|z_t|^2 z_t=\partial_{\overline{z}}h(z_{t})
\end{eqnarray}
with initial condition $z_0=z\in \Z$.\\
An important property of the dynamical
groups $U_{\hbarr}(t)$ and $U_{\hbarr}^{0}(t)$ is that they
preserve the number
$$
U_{\hbarr}(t)^{*}NU_{\hbarr}(t)=N\,,\quad
[H_{\hbarr},N]=[H^{0}_{\hbarr},N]=[Q^{Wick},N]=0\,.
$$
\begin{remark}
All the results of this section can be easily adapted with a self-adjoint
operator $A$ on $\Z$ and a polynomial $Q(z)\in \oplus_{n\in\nz}^{\rm alg}
\P_{n,n}(\Z)$. Nevertheless it is better to stick to this concrete
presentation which fits better with a widely studied problem.
\end{remark}

\subsection{Propagation of squeezed coherent states (Hepp method)}
\label{se.hepp}

In finite dimension it is nothing but checking the propagation of
gaussian wave packets. Although it works only for some specific states
it is a direct and very flexible method.
Moreover it agrees very well with the phase-space
geometric intuition. Extensions with more singular potentials or
about the long time behaviour have been carried out in
\cite{Hep}\cite{GiVe}.
\begin{prop}
\label{pr.hepp}
  For any $z_{0}\in \Z$, the estimate
$$
\left|e^{-i\frac{t}{\varepsilon}H_{\varepsilon}}E(z_{0})-e^{i\frac{\omega(t)}{\hbarr}}
W(\frac{\sqrt{2}}{i\varepsilon}z_{t})U_{2}(t,0)\Omega\right|_{\mathcal{H}}\leq
C \; e^{C\left|V\right|_{L^{\infty}}\langle z_{0}\rangle^{2}(\left|t\right|+1)}\;\varepsilon^{1/2}
$$
holds with
\begin{eqnarray}
\label{eq.eqclass}
&&i\partial_{t}z_{t}=-\Delta z_{t}+
(V*\left|z_{t}\right|^{2})z_{t}\quad,\quad z_{t=0}=z_{0}\\
\label{eq.omega}
&&\omega(t)=\int_{0}^{t} Q(z_{s})~ds\\
\label{eq.U2}
&&i\varepsilon\partial_{t}U_{2}(t,0)=[d\Gamma(-\Delta)+
Q_{2}(t)^{Wick}]U_{2}(t,0)\quad,\quad
U_{2}(0,0)=I\,,\\
\label{eq.Q2}
&&
Q_{2}(t,z)=\frac{1}{2}\left[\langle \partial_{z}^{2}Q(z_{t})\,,z^{\otimes
  2}\rangle  +
\langle
z^{\otimes 2}\,,\,\partial_{\overline{z}}^{2}Q(z_{t})\rangle
+2 \left\langle
  z\,,\, \partial_{\overline{z}}\partial_{z}Q(z_{t})z\right\rangle
\right]
\,,\\
\nonumber
&&
\langle \partial_{z}^{2}Q(z_{t})\,,z^{\otimes
  2}\rangle = 2\left\langle \tilde Q \, z_{t}^{\otimes 2}\,,\,
  z^{\otimes 2}\right\rangle\in \P_{2,0}(\Z)\,,\\
&&
\nonumber
\langle z\,,\,\partial_{\overline{z}}\partial_{z}Q(z_{t})z\rangle =4
\left\langle z\vee z_{t}\,,\,\tilde Q \, z\vee
z_{t}\right\rangle\in \P_{1,1}(\Z)\,.
\end{eqnarray}
\end{prop}
\proof This proposition says that the evolution of a coherent state is
well described after applying a time dependent (real) affine
Bogoliubov transformation like the ones considered in Proposition
\ref{pr.realtrans}.\\
It is sufficient that
$$
e^{i\frac{t}{\varepsilon}
H_{\varepsilon}}e^{i\frac{\omega(t)}{\varepsilon}}
W(\frac{\sqrt{2}}{i\varepsilon}z_{t}) U_{2}(t,0)\Omega
=
e^{i\frac{t}{\varepsilon}H_{\varepsilon}}\Gamma(e^{it\Delta})
e^{i\frac{\omega(t)}{\varepsilon}}W(\frac{\sqrt{2}}{i\varepsilon}e^{-it\Delta}z_{t
}) \Gamma(e^{-it\Delta})U_{2}(t,0)\Omega
$$
remains close enough to $\Omega$\,. The quantities
$\hat{U}_{\varepsilon}(0,t)=e^{i\frac{t}{\varepsilon}H_{\varepsilon}}\Gamma(e^{it\Delta})$,
$\hat{U}_{2}(t,0)=\Gamma(e^{-it\Delta})U_{2}(t,0)$ and
$\hat{z}_{t}=e^{-it\Delta}z_{t}$ solve the differential equations
\begin{eqnarray}
 \label{eq.hatU} &&
i\varepsilon\partial_{t}\hat{U}_{\varepsilon}(0,t)= -
\hat{U}_{\varepsilon}(0,t)
\Gamma(e^{-it\Delta})Q^{Wick}\Gamma(e^{it\Delta})=
-\hat U_{\varepsilon}(t,0) \hat{Q}(t)^{Wick}\,,\\
\label{eq.hatU2}
&&
i\varepsilon \partial_{t}\hat{U}_{2}(t,0)=
\Gamma(e^{-it\Delta})Q_{2}(t)^{Wick}
\Gamma(e^{it\Delta})\hat{U}_{2}(t,0)=
\hat{Q}_{2}(t)^{Wick}\hat{U}_{2}(t,0)\,, \\
\label{eq.ut}
&&i\partial_{t}\hat{z}_{t}=
e^{-it\Delta}(V*\left|e^{it\Delta}\hat{z}_{t}\right|^{2})e^{it\Delta}
\hat{z}_{t}=
\partial_{\overline{z}}\hat{Q}(t,\hat{z}_{t})\quad,\quad
\hat{z}_{0}=z_0\,,
\end{eqnarray}
after setting
\begin{equation}
  \label{eq.transfoQ}
\hat{Q}(t,z)=Q(e^{it\Delta}z) \quad \text{and}\quad \hat{Q}_{2}(t,z)=Q_{2}(t,e^{it\Delta}z)\,.
\end{equation}
The main properties of $\hat{U}_{2}(t,0)$ are derived in \cite[Proposition 4.1]{GiVe} and in particular
we know that $\hat{U}_{2}(t,0) \Omega$ belongs to the domain of the closure of any $b^{Wick}$
with $b\in\oplus_{p,q\in\mathbb{N}}^{\rm alg}\P_{p,q}(\Z)$.

The differentiation of the Weyl relation \eqref{eq.Weylcomm} on $\H_{fin}$ says
\begin{eqnarray*}
i\varepsilon\partial_{t}W(\frac{\sqrt{2}}{i\varepsilon}\hat{z}_{t})
&=&
\left[-\real\langle\hat{z}_{t}\,,\, i\partial_{t}\hat{z}_{t}\rangle
+\sqrt{2}\Phi(i\partial_{t}\hat{z}_{t})
\right]W(\frac{\sqrt{2}}{i\varepsilon}\hat{z}_{t})\\
&=&\left[-\real\langle \hat{z}_{t}\,,\,
  \partial_{\overline{z}}\hat{Q}(t,\hat{z}_{t})\rangle +
 a^{*}(\partial_{\overline{z}}\hat{Q}_{t}(\hat{z}_{t}))+
 a(\partial_{\overline{z}}\hat{Q}_{t}(\hat{z}_{t}))\right]
W(\frac{\sqrt{2}}{i\varepsilon}\hat{z}_{t})\\
&=&
\left[
-\real\langle \hat{z}_{t}\,,\,
  \partial_{\overline{z}}\hat{Q}(t,\hat{z}_{t})\rangle
+ \real\langle
z\,,\, \partial_{\overline{z}}\hat{Q}_{t}(\hat{z}_{t})\rangle^{Wick}
\right]W(\frac{\sqrt{2}}{i\varepsilon}\hat{z}_{t})\,.
\end{eqnarray*}
The translation property (iii) of
Proposition~\ref{pr.transla} then leads to
$$
e^{i\frac{t}{\varepsilon}
H_{\varepsilon}}e^{i\frac{\omega(t)}{\varepsilon}}
W(\frac{\sqrt{2}}{i\varepsilon}z_{t}) U_{2}(t,0)\Omega -\Omega
=\frac{1}{i\varepsilon}\int_{0}^{t}
\hat{U}_{\varepsilon}(0,s)e^{i\frac{\omega(s)}{\varepsilon}}
W(\frac{\sqrt{2}}{i\varepsilon}\hat{z}_{s}) \mathcal{A}(s)^{Wick}\hat{U}_{2}(s,0)\Omega~ds
$$
after testing both sides on $\mathcal{H}_{fin}$ and  setting
\begin{eqnarray*}
  \mathcal{A}(s,z)&=& -\hat{Q}(s,z+\hat{z}_{s})
-\omega'(s)+
\real\langle \hat{z}_{s}\,,\,
 \partial_{\overline{z}}\hat{Q}(s,\hat{z}_{s})\rangle
+ \real\langle
z\,,\, \partial_{\overline{z}}\hat{Q}_{s}(\hat{z}_{s})\rangle
+\hat{Q}_{2}(s,z)\\
&=&  -\hat{Q}(s,z+\hat{z}_{s})
+\hat{Q}(\hat{z}_{s})
+\langle
z\,,\,
\partial_{\overline{z}}\hat{Q}_{s}(\hat{z}_{s})\rangle
+
\langle
\partial_{z}\hat{Q}_{s}(\hat{z}_{s})\,,\, z\rangle
+\hat{Q}_{2}(s,z)\,.
\end{eqnarray*}
The last equality is given by our choice of $\omega(t)$ in
\eqref{eq.omega}. It suffices to find a uniform estimate w.r.t $s\in
[0,t]$
of  the squared norm
\begin{equation}
  \label{eq.sqnorm}
\left|\varepsilon^{-1}\mathcal{A}(s)^{Wick}\hat{U}_{2}(s,0)\Omega\right|_{\H}^{2}
=
\varepsilon^{-2}\left\langle \Omega\,,\,
  \hat{U}_{2}(0,s)\mathcal{A}(s)^{Wick,*}\mathcal{A}(s)^{Wick}\hat{U}_{2}(s,0)
\Omega\right\rangle\,.
\end{equation}
The important point is that the symbol
$\mathcal{A}(s)$ vanishes at the second order at $z=0$. More precisely
it can be written
\begin{eqnarray*}
  &&
  \mathcal{A}(s)=\mathcal{A}_{1,2}(s)+\mathcal{A}_{2,1}(s)+\mathcal{A}_{2,2}(s)\\
\text{with}&&
\mathcal{A}_{p,q}(s)\in \P_{p,q}(\Z)\\
\text{and}&&
\left|\tilde{\mathcal{A}}_{p,q}(s)\right|_{\L(\bigvee^{p}\Z,\bigvee^{q}\Z)}\leq
C_{p,q}\left|V\right|_{L^{\infty}}\left|z_{0}\right|^{4-p-q}\,.
\end{eqnarray*}
Owing to Proposition \ref{symbcalc} and Lemma~\ref{le.bdcontract} the operator
$\mathcal{A}(s)^{Wick,*}\mathcal{A}^{Wick}(s)$ takes the form
\begin{eqnarray*}
&&
\mathcal{A}(s)^{Wick,*}\mathcal{A}(s)^{Wick}
=\sum_{k=0}^{2}\varepsilon^{k}\sum_{6-2k\leq p+q\leq
  8}\mathcal{B}_{k,p,q}(s)^{Wick}
\\
\text{with}
&&
\left|\tilde{\mathcal{B}}_{k,p,q}(s)\right|_{\mathcal{L}(\bigvee^{p}\Z,\bigvee^{q}\Z)}\leq
C_{k,p,q}\left|V\right|_{L^{\infty}}^{2}\left\langle z_{0}\right\rangle^{2}\,.
\end{eqnarray*}
The estimate of every term
\begin{eqnarray*}
\varepsilon^{k-2}\left\langle
  \Omega\,,\hat{U}_{2}(0,s)\mathcal{B}_{k,p,q}(s)^{Wick}\hat{U}_{2}(s,0)\Omega
\right\rangle\quad,\quad p+q\geq 6-2k
\end{eqnarray*}
is given by the Lemma~\ref{le.bogo} below and yields the result.
\fin
\begin{lem}
  \label{le.bogo}
Consider the time dependent Wick operator $\hat{Q}_{2}$ defined by
\eqref{eq.Q2} \eqref{eq.transfoQ} and parametrized by $z_{0}\in \Z$.
Consider  the associated unitary operator
$\hat{U}_{2}(s,0)$ defined by \eqref{eq.hatU2}. For any $p,q\in \nz$,
there exists a constant $C_{p,q}$ such that the estimate
$$
\left|\left\langle \Omega\,,\,
    \hat{U}_{2}(0,s)b^{Wick}\hat{U}_{2}(s,0)\Omega\right\rangle\right|
\leq C_{p,q}\;
e^{C_{p,q}\left|V\right|_{L^{\infty}}\langle z_{0}\rangle^{2}(|s|+1)}
\left|\tilde{b}\right|_{\L(\bigvee^{p}\Z,\bigvee^{q}\Z)}
\varepsilon^{\frac{p+q}{2}}
$$
holds for any $b\in \P_{p,q}(\Z)$ and any $s\in \rz$\,.
\end{lem}
\proof
By introducing an anti-unitary operator $Jz=\overline{z}$. The
$\rz$-linear operator $\partial_{\overline{z}}\hat{Q}_{2}(t)$ can be
written
$$
\partial_{\overline{z}}\hat{Q}_{2}(t)z=R(t)z+R_{2}(t)\overline{z}\,.
$$
The definitions \eqref{eq.Q2}\eqref{eq.transfoQ} ensure that $R(t)$ is a
bounded operator strongly continuous with respect to $t\in\rz$ and
that $R_{2}(t)$ is a Hilbert-Schmidt operator which depends
continuously on $t\in\rz$ in the Hilbert-Schmidt norm. Moreover the
following uniform estimates hold
$$
\left|R(t)\right|_{\L(\Z)}\leq
2\left|V\right|_{L^{\infty}}\left|z_{0}\right|^{2}
\quad ,\quad
\left|R_{2}(t)\right|_{\L^{2}(\Z)}\leq 2\left|V\right|_{L^{\infty}}\left|z_{0}\right|^{2}\,.
$$
Hence the equation
$$
i\partial_{t}\Phi_{2}=\partial_{\overline{z}}\hat{Q}_{2}(t)\Phi_{2}
= R(t)\Phi_{2}+R_{2}(t)J\Phi_{2}
$$
defines a dynamical system of bounded $\rz$-linear operators with the
estimate
$$
\left|\Phi_{2}(t_{2},t_{1})\right|_{\L_{\rz}(\Z)}\leq
e^{4\left|t_{2}-t_{1}\right|
\left|V\right|_{L^{\infty}}\left|z_{0}\right|^{2}}\,.
$$
More precisely the Duhamel formula
$$
\Phi_{2}(t_{2},t_{1})=Te^{-i\int_{t_{1}}^{t_{2}}R(s)~ds}
-i\int_{t_{1}}^{t_{2}}Te^{-i\int_{t}^{t_{2}}R(s)~ds}R_{2}(t)J\Phi_{2}(t,t_{1})~dt
$$
implies that the $\rz$-linear operator $\Phi_{2}(t_{2},t_{1})$ can be written
\begin{eqnarray*}
&&\Phi_{2}(t_{2},t_{1})=B(t_{2},t_{1})+B_{2}(t_{2},t_{1})J\\
\text{with}&&
\left|B(t_{2},t_{1})\right|_{\L(\Z)}+\left|B_{2}(t_{2},t_{1})\right|_{\L^{2}(\Z)}\leq
C\left|V\right|_{L^{\infty}}\left|z_{0}\right|^{2}(\left|t_{2}-t_{1}\right|+1)
e^{C\left|t_{2}-t_{1}\right|\left|V\right|_{L^{\infty}}\left|z_{0}\right|^{2}}\,.
\end{eqnarray*}
According to Proposition~\ref{pr.realtrans}, for any $c\in
\oplus_{p+q=m}\P_{p,q}(\Z)$ and any $t\in \rz$, the polynomial
$c(t,z)=c(\Phi_{2}(0,t)z)$
belongs to $\oplus_{p+q=m}\P_{p,q}(\Z)$ with
\begin{eqnarray*}
 \sum_{p+q=m}
 \left|\partial_{\overline{z}}^{q}\partial_{z}^{p}c(t,z)
\right|_{\L(\bigvee^{p}\Z,\bigvee^{q}\Z)}
\leq
C^{1}_{m}e^{C_{m}^{1}\left|V\right|_{L^{\infty}}\langle
    z_{0}\rangle^{2}(|t|+1)}
\sum_{p+q=m}
\left|\partial_{\overline{z}}^{q}\partial_{z}^{p}c(z)\right|_{\L(\bigvee^{p}\Z,\bigvee^{q}\Z)}\,.
\end{eqnarray*}
Applying the characteristic method, that is differentiating
$c(z)=c(t,\Phi_{2}(t,0)z)$, shows that $c(z,t)$ solves the equation
$$
i\partial_{t}c(t,z)
+\partial_{z}c(t,z).\partial_{\overline{z}}\hat{Q}_{2}(t,z)
-
\partial_{z}\hat{Q}_{2}(t,z)\partial_{\overline{z}}c(t,z)=0\,.
$$
Thanks to the Wick calculus in Proposition~\ref{symbcalc} and the fact that
$\hat{U}_{2}(t,0)\Omega\in\cap_{k\in\mathbb{N}}\D(N^k)$ (see \cite[Proposition 4.1]{GiVe}), this leads to
\begin{eqnarray*}
i\partial_{t}\hat{U}_{2}(0,t)c(t)^{Wick}\hat{U}_{2}(t,0)\Omega
&=&
\hat{U}_{2}(0,t)\left(\varepsilon^{-1}[c^{Wick}(t),\hat{Q}_{2}(t)^{Wick}]
+i\partial_{t}c(t)^{Wick}\right)\hat{U}_{2}(t,0)\Omega
\\
&=&
\hat{U}_{2}(0,t)\frac{\varepsilon}{2}\left(\left\{c(t),\hat{Q}_{2}(t)\right\}^{(2)}\right)^{Wick}
\hat{U}_{2}(t,0)\Omega\,.
\end{eqnarray*}
Take $b\in \oplus_{p+q=m_{0}}\P_{p,q}(\Z)$ and apply
this result with $c$ defined by $c(s,z)=b(z)$, which means
\begin{eqnarray*}
&&
  c(\Phi_{2}(0,s)z)=c(s,z)=b(z)\\
\text{or}&&
  c(z)=b(\Phi_{2}(s,0)z)\in
  \oplus_{p+q=m_{0}}\P_{p,q}(\Z)
\\
\text{with}&&
\sum_{p+q=m_{0}}
\left|\partial_{\overline{z}}^{q}\partial_{z}^{p}c(z)\right|_{\L(\bigvee^{p}\Z,\bigvee^{q}\Z)}
\leq C^{1}_{m_{0}}e^{C_{m_{0}}^{1}\left|V\right|_{L^{\infty}}\langle
    z_{0}\rangle^{2}(|s|+1)}
\sum_{p+q=m_{0}}
\left|\partial_{\overline{z}}^{q}\partial_{z}^{p}b(z)\right|_{\L(\bigvee^{p}\Z,\bigvee^{q}\Z)}\,.
\end{eqnarray*}
This leads to
\begin{eqnarray*}
  \left\langle
    \Omega\,,\,\hat{U}_{2}(0,s)b^{Wick}\hat{U}_{2}(s,0)\Omega\right\rangle
&=&
\left\langle \Omega\,, c^{Wick} \Omega\right\rangle
+
\int_{0}^{s}
\left\langle \Omega\,,\,
\partial_{t}\left(\hat{U}_{2}(0,t)c(t)^{Wick}\hat{U}_{2}(t,0)\right)\Omega\right\rangle~dt\\
&=&
-\frac{i\varepsilon}{2}\int_{0}^{s}
\left\langle \Omega\,,\,
\hat{U}_{2}(0,t)\left(\left\{c(t),\hat{Q}_{2}(t)\right\}^{(2)}\right)^{Wick}
\hat{U}_{2}(t,0)\Omega\right\rangle~dt\,.
\end{eqnarray*}
By noticing that the symbol
$\left\{c(t),\hat{Q}_{2}(t)\right\}$
vanishes when $m_{0}< 2$ or  belongs to
$\oplus_{p+q=m_{0}-2}\P_{p,q}(\Z)$
with
\begin{eqnarray*}
&&
\sum_{p+q=m_{0}-2}
\left|\partial_{\overline{z}}^{q}\partial_{z}^{p}\left\{c(t),\hat{Q}_{2}(t)\right\}^{(2)}
\right|_{\L(\bigvee^{p}\Z,\bigvee^{q}\Z)}
\leq
C\left|V\right|_{L^{\infty}}\left|z_{0}\right|^{2}
\sum_{p+q=m_{0}}
\left|\partial_{\overline{z}}^{q}\partial_{z}^{p}c(t)
\right|_{\L(\bigvee^{p}\Z,\bigvee^{q}\Z)}
\\
&&
\hspace{2cm}
\leq
C\left|V\right|_{L^{\infty}}\left|z_{0}\right|^{2}
C^{1}_{m_{0}}e^{C_{m_{0}}^{1}\left|V\right|_{L^{\infty}}\langle
    z_{0}\rangle^{2}(2|s|+1)}
\sum_{p+q=m_{0}}
\left|\partial_{\overline{z}}^{q}\partial_{z}^{p}b
\right|_{\L(\bigvee^{p}\Z,\bigvee^{q}\Z)}
\end{eqnarray*}
the result is proved by induction on $m_{0}$ and by
using
$x^{n}\leq n!e^{x}$ for $x>0$.
\fin
\subsection{Truncated Dyson expansion}
\label{se.truncDy}
 We focus  now on the propagation of
chaos point of view which has been considered
 by several authors in \cite{ESY1}\cite{ESY2}\cite{BGGM}\cite{FGS}.
In the bosonic setting Hermite states tested on some Wick
observable is exactly the BBGKY hierarchy. For example the reduced one
particle density matrix can be defined as
$\Tr[\varrho_{1}A]=\Tr[\varrho d\Gamma(A)]=\Tr[\varrho
\mathcal{A}^{Wick}]$
with $\mathcal{A}(z)=\langle z\,,\,Az\rangle$\,.
 While reproducing the Dyson expansion analysis
of \cite{FGS}, we check here that a full asymptotic expansion can be
written, when  Wick observables are tested after the suitable
number truncation.

The strategy of the proof in \cite{FGS}
relies on an analysis of the
 Schwinger-Dyson expansion of a time evolved observable $U_\hbarr(t)^*
\O \; U_\hbarr(t)$ is given  by
\begin{eqnarray} \label{dyson} U_\hbarr(t)^* \O
\; U_\hbarr(t)=\O_t+\sum_{n=1}^\infty (\frac{i}{\hbarr})^n \int_0^t
dt_1\cdots\int_0^{t_{n-1}}dt_n [Q^{Wick}_{t_n}, \cdots
[Q^{Wick}_{t_1}, \O_t]\cdots ]
\end{eqnarray}
 where $\O_t=U_\hbarr^0(t)^* \O\;
U_\hbarr^0(t)$, $Q^{Wick}_{s}=U_\hbarr^0(s)^* Q^{Wick}
\;U_\hbarr^0(s)$. The commutation relation in
Proposition~\ref{wick_prop} (iii) yields \begin{eqnarray*}\ds Q^{Wick}_s=\left(\la (e^{is
\Delta} z)^{\otimes 2}, Q (e^{is \Delta}z)^{\otimes
2}\ra\right)^{Wick}\,,
\end{eqnarray*}
or shortly $Q_{s}(z)=Q(e^{is\Delta}z)$
and we shall set more generally for $b\in \P_{p,q}(\Z)$ and $s\in \rz$
$$
b_{s}\in \P_{p,q}(\Z)\,:\qquad
\forall z\in \Z,\;
b_{s}(z)=b(e^{is\Delta}z)\,.
$$
Although the convergence of the series
can be proved as an operator acting on $\bigvee^{k}\Z$, with $k\in
\nz$ fixed, the $\hbarr$-asymptotic analysis is done with its
truncated version
\begin{eqnarray} \label{dyson-2} U_\hbarr(t)^* \hspace{-.1in}&\O
\hspace{-.1in} &U_\hbarr(t)=\O_t+\sum_{n=1}^{\ell-1}
(\frac{i}{\hbarr})^n \int_0^t dt_1\cdots\int_0^{t_{n-1}}dt_n
[Q^{Wick}_{t_n}, \cdots [Q^{Wick}_{t_1}, \O_t]\cdots ]\nonumber \\
&+&(\frac{i}{\hbarr})^{\ell}\int_0^t
dt_1\cdots\int_0^{t_{\ell-1}}dt_{\ell} \;\;
U_\hbarr(t_{\ell})^*U^0_\hbarr(t_\ell) [Q^{Wick}_{t_\ell}, \cdots
[Q^{Wick}_{t_1}, \O_t]\cdots ] U^{0}_\hbarr(t_\ell)^*
U_\hbarr(t_{\ell}).
\end{eqnarray}
The Poisson brackets analogue of the multicommutators will be necessary.
\begin{definition}
 \label{de.Cnr} For $n,r\in \nz$, $r\leq n$  and any fixed $b\in \P_{p,q}(\Z)$,
 the polynomial $C^{(n)}_{r}(t_{1},\ldots, t_{n})$  is defined by
\begin{eqnarray}
\label{cnr} C^{(n)}_r(t_n,\cdots,t_1,t)=\frac{1}{2^r}\; \sum_{\sharp
\{i:\; \varepsilon_i=2\}=r}\;  \{Q_{t_n}, \cdots, \{Q_{t_1},b_t
\underbrace{\}^{(\varepsilon_1)}\cdots\}^{(\varepsilon_n)}}_{
\varepsilon_i\in\{1,2\}}
\in\P_{p-r+n,q-r+n}(\Z)\,,
\end{eqnarray}
and $C^{(n)}_{r}(t_{1},\ldots, t_{n},t,z)$ denotes its values at $z\in
\Z$
while $\widetilde{C^{(n)}_{r}}(t_{1},\ldots, t_{n},t)$ or simply
$\widetilde{C^{(n)}_{r}}$ denotes the associated operator according to
Definition~\ref{de.hompol}\,.
\end{definition}
We shall prove.
\begin{thm}
\label{main-2}
Fix $p,q\in \nz$ and assume $b\in\P_{p,q}(\Z)$. Then the asymptotic expansion
$$
U_\hbarr (t)^* b^{Wick}U_\hbarr (t)
=\sum_{r=0}^{\ell-1}\hbarr^{r}\sum_{n=0}^{\infty}  i^n
 \int_0^t dt_1\cdots\int_0^{t_{n-1}} dt_n \;
 \left[C_r^{(n)}(t_n,\cdots,t_1,t)\right]^{Wick}
+\varepsilon^{\ell}R_{\ell}(\varepsilon,t)
$$
holds for any $\ell\in\nz$ and any $\delta>0$
in $\L(\bigvee^{k}\Z,\bigvee^{k-p+q}\Z)$
 with the uniform estimate
$$
\left|R_{\ell}(\varepsilon,t)\right|_{\L(\bigvee^{k}\Z,\bigvee^{k-p+q}\Z)}\leq~C_{\ell,\delta}
\quad\text{when}  \quad k\hbarr\leq 1+\delta/2
\quad\text{and}\quad
4(1+2\delta)|t|\left|V\right|_{L^{\infty}}\leq 1\,.
$$
\end{thm}
A particular case takes a more explicit form.
\begin{thm}
  \label{th.main2.2}
Take $b\in \P_{p,q}(\Z)$.
Let $z\in \Z$ be such that $|z|=1$ and call $z_t$ the
solution to (\ref{hartree}) with $z_{0}=z$. \\
(i) Then the expansion
\begin{eqnarray}
\label{eq.expanbeta}
\la z^{\otimes k-m}, U_\hbarr (t)^*
b^{Wick}U_\hbarr (t)\, z^{\otimes k} \ra=
\delta_{p-q,m}\left[\sum_{r=0}^{\ell-1}
 \hbarr^r \,  \; \beta^{(r)}(t,z,k,\varepsilon) +O_{t}(\hbarr^{\ell})\right]\,,
\end{eqnarray}
holds as $\hbarr\to 0$, $k\hbarr\to 1$ by setting
\begin{eqnarray}
\label{btr}
\hspace{-.2in}\beta^{(0)}(t,z,k,\varepsilon)&=& b(z_t), \nonumber\\
\hspace{-.2in}\beta^{(r)}(t,z,k,\varepsilon)&=&\sum_{n=r}^{k-p+r}
i^n \, \frac{\sqrt{k!(k-m)! \,
\hbarr^{p+q+2(n-r)}}} {(k-(p+n-r))!}
\int_0^t
dt_1\cdots\int_0^{t_{n-1}}dt_n \;C^{(n)}_r(t_n,\cdots,t_1,t;z) \,,
\end{eqnarray}
and as soon as $ 4|t|\, |V|_{L^\infty}<1$\,.\\
(ii) More generally the limit
\begin{eqnarray*}
\lim_{\tiny \begin{array}{c} \hbarr\to 0,
 \\ k\hbarr\to 1 \end{array}} \la z^{\otimes k-m}, U_\hbarr
(t)^* b^{Wick}U_\hbarr (t)\, z^{\otimes k} \ra=\delta_{p-q,m} \;
b(z_t)\,
\end{eqnarray*}
holds for all times $t\in\rz$.
\end{thm}
\begin{cor}
\label{cor-2}
In the specific case $m=0$, $q=p$, the expansion \eqref{eq.expanbeta}
takes the form
\begin{eqnarray*} \la z^{\otimes k}, && \hspace{-.2in}U_\hbarr (t)^*
b^{Wick}U_\hbarr (t)\, z^{\otimes k} \ra=\sum_{s=0}^{\ell-1}
 \hbarr^s \, \sum_{n=0}^\infty i^n\;
 \int_0^t dt_1\cdots\int_0^{t_{n-1}} dt_n \; \\ &&[\sum_{j=0}^{s}
\alpha_{j}^{s-j,n}(k\varepsilon)
C^{(n)}_{s-j}(t_n,\cdots,t_1,t;z)]   +O(\hbarr^{\ell}),
\end{eqnarray*}
where the coefficients $\alpha_{j}^{r,n}(\kappa)$ are polynomials
in $\kappa$  given by
\begin{eqnarray*}
\sum_{j=0}^{p+n-r-1} \alpha^{r,n}_j(\kappa)
\hbarr^j=\kappa(\kappa-\hbarr) (\kappa-2\hbarr)
\cdots (\kappa-(p+n-r-1)\hbarr), \end{eqnarray*}
and the convention that $\alpha_j^{r,n}=0$ when
$\;j\geq(p+n-r)\;$ or $r>n$.
\end{cor}
\proof We are considering the particular case $p=q$, $m=0$.
Setting $\kappa=k\varepsilon=(k-m)\varepsilon$ gives:
\begin{eqnarray*}
\frac{k!\hbarr^{p+(n-r)}}{(k-(p+n-r))!}
    =\kappa(\kappa-\hbarr) (\kappa-2\hbarr) \cdots
(\kappa-(p+n-r-1)\hbarr).
\end{eqnarray*}
Putting  together the terms of order $\hbarr^s$, $s$ less than
$\ell-1$ in Thm.~\ref{main-2}(ii), yields the result.
\hfill
\fin

Before proving Theorem~\ref{main-2} and Theorem~\ref{th.main2.2},
let us collect  some technical preliminaries.
\begin{lem}
\label{tech.lem.1} For $b\in\P_{p,q}(\Z)$
the identity
\begin{eqnarray*}
\frac{1}{\hbarr^n}\;[Q_{t_n}^{Wick} , \cdots , [Q_{t_1}^{Wick} ,
b^{Wick}_t ]]= \sum_{r=0}^n \hbarr^r \,
\left(C^{(n)}_r(t_n,\cdots,t_1,t)\right)^{Wick}\,,
\end{eqnarray*}
holds with the symbols $C^{(n)}_{r}(t_{1},\cdots,t_{n},t)$ defined according to \eqref{cnr}
in Definition~\ref{de.Cnr}.
\end{lem}
\proof Proposition~\ref{symbcalc}  provides  the  induction formula
\begin{eqnarray} \label{eq.8} C^{(n)}_r=\{Q_{t_n},C^{(n-1)}_r\}+\frac{1}{2}
\{Q_{t_n},C^{(n-1)}_{r-1}\}^{(2)}, \end{eqnarray} with $C^{(l)}_r=0$ if $l<r$
or $r<0$. In particular, we get \begin{eqnarray*}
C^{(n)}_0=\{Q_{t_n},\cdots,\{Q_{t_1},b_t\}\}. \end{eqnarray*}
 A simple iteration of  (\ref{eq.8}) yields the result.
\hfill\fin

\begin{lem}
\label{tech.lem.2}
Let $b$ belong to $\P_{p,q}(\Z)$.\\
(i) The estimate
\begin{eqnarray*}
\left|\widetilde{\Xi_{1}}\right|_{\mathcal{L}(\bigvee^{p+1}\Z,\bigvee^{q+1}\Z)}
\leq  \;(p+q) \;|V|_{L^\infty} \;|b|_{\mathcal{L}(\bigvee^p\Z,
\bigvee^q\Z)}\,,
\end{eqnarray*}
holds by setting
$\widetilde{\Xi_{1}}=\frac{1}{(p+1)!}\frac{1}{(q+1)!}\partial_z^{p+1}\partial_{\bar
z}^{q+1}\{Q_s,b_t\}^{(1)}\in
\mathcal{L}(\bigvee^{p+1}\Z,\bigvee^{q+1}\Z)$.\\
(ii) Similarly, the inequality
\begin{eqnarray*}
\left|\widetilde{\Xi_{2}}\right|_{\mathcal{L}(\bigvee^{p}\Z,\bigvee^{q}\Z)}
\leq  \; [p(p-1)+q(q-1)] \;|V|_{L^\infty} \;|b|_{\mathcal{L}(
\bigvee^p\Z, \bigvee^q\Z)}\,.
\end{eqnarray*}
holds with $\widetilde{\Xi_{2}}=\frac{1}{p!}\frac{1}{q!}\partial_z^{p}\partial_{\bar z}^{q}\{Q_s,b_t\}^{(2)}$\,.\\
(iii) For any $n\in \nz$ and $r\in
\left\{0,1,\ldots,n\right\}$, the operator $\widetilde{C^{(n)}_{r}}$
associated with the symbol $C_{r}^{(n)}(t_{n},\ldots, t_{1},t)\in
\P_{p+n-r,q+n-r}(\Z)$ according to Definition~\ref{de.Cnr} satisfies
\begin{eqnarray*}
\left|\widetilde{C^{(n)}_r}\right|_{\mathcal{L}(\bigvee^{p+n-r}\Z,
  \bigvee^{q+n-r}\Z)}\leq
2^{n-r} C_n^r \; (p+n-r)^{2r}
\;\frac{(p+n-r-1)!}{(p-1)!} \;|V|_{L^\infty}^n
\;|b|_{\mathcal{L}(\bigvee^p\Z, \bigvee^q\Z)}\,, \end{eqnarray*} when $p\geq
q$ with a similar expression when $q\geq p$ (replace $(p+n-r, p-1)$
with  $(q+n-r, q-1)$)\,.
\end{lem}
\proof The statements (i) and (ii) are particular cases of Lemma
\ref{le.bdcontract}. The estimate in (iii) is a consequence of
(i)(ii) and the definition (\ref{cnr}). \hfill\fin

\noindent
\textbf{Proof of Theorem~\ref{main-2}.} Set $j=k-p+q$. Since
$U_{\varepsilon}(t)$
and $U_{\varepsilon}^{0}(t)$ preserve the number
like $Q_{t}^{Wick}$ the equality
\begin{eqnarray*}
U_\hbarr (t)^* b^{Wick}U_\hbarr
(t)&=& \sum_{n=0}^{\ell-1} \left(\frac{i}{\hbarr}\right)^n \int_0^t
dt_1\cdots\int_0^{t_{n-1}} dt_n
 [Q^{Wick}_{t_n}, \cdots [Q^{Wick}_{t_1}, b_t^{Wick}]\cdots ]\\
&& \hspace{-.4in} + \left(\frac{i}{\hbarr}\right)^{\ell} \int_0^t
dt_1\cdots\int_0^{t_{\ell-1}} dt_\ell \; U_\varepsilon(t_\ell)^*
U^0_\varepsilon(t_\ell) [Q^{Wick}_{t_\ell}, \cdots [Q^{Wick}_{t_1},
b_t^{Wick}]\cdots ] U_\varepsilon^0(t_\ell)^* U_\varepsilon(t_\ell)\,,
\end{eqnarray*}
derived from (\ref{dyson-2}) holds in
$\L(\bigvee^k\Z,\bigvee^j\Z)$.
Then Lemma \ref{tech.lem.1} implies
\begin{eqnarray}
\label{eq.10.1} U_\hbarr
(t)^* b^{Wick}U_\hbarr (t) &=&\sum_{n=0}^{\ell-1}
 i^n  \;\;
\int_0^t dt_1\cdots\int_0^{t_{n-1}} dt_n \; \sum_{r=0}^{n} \hbarr^r
\left[C_r^{(n)}(t_n,\cdots,t_1,t)\right]^{Wick} \\ \label{eq.10.2}
&&\hspace{-1in} + i^{\ell} \int_0^t
dt_1\cdots\int_0^{t_{\ell-1}} dt_\ell \; U_\varepsilon(t_\ell)^*
U^0_\varepsilon(t_\ell) \hbarr^\ell
\left[C^{(\ell)}_\ell(t_\ell,\cdots,t_1,t)\right]^{Wick}
U_\varepsilon^0(t_\ell)^* U_\varepsilon(t_\ell) \\ \label{eq.10} &&
\hspace{-1in} + i^{\ell} \int_0^t
dt_1\cdots\int_0^{t_{\ell-1}} dt_\ell \; U_\varepsilon(t_\ell)^*
U^0_\varepsilon(t_\ell) \sum_{r=0}^{\ell-1} \hbarr^{r}
\left[C^{(\ell)}_r(t_\ell,\cdots,t_1,t)\right]^{Wick}
U_\varepsilon^0(t_\ell)^* U_\varepsilon(t_\ell). \end{eqnarray}
Keep untouched the part (\ref{eq.10.1})-(\ref{eq.10.2}) and iterate  the Dyson
series on the third term (\ref{eq.10}). While doing so, use the
formula
\begin{eqnarray} \label{eq.21} [\frac{Q_{t_{n+1}}^{Wick}}{\hbarr},
\sum_{r=0}^{\ell-1}\hbarr^r
\left[C^{(n)}_r(t_{n},\cdots,t_1,t)\right]^{Wick}]&=&
\sum_{r=0}^{\ell-1} \hbarr^{r} \left[
C_r^{(n+1)}(t_{n+1},\cdots,t_1,t)\right]^{Wick}\\
&&+
 \frac{\hbarr^\ell}{2} \left[ \{Q_{t_{n+1}},C_\ell^{(n)}(t_{n+1},\cdots,t_1,t)\}^{(2)}\right]^{Wick},\nonumber
\end{eqnarray}
inductively for  $n=\ell,\ell+1,\ldots, M-1$.
After $M-\ell$ steps, collecting the factors of $\varepsilon^{\ell}$
yields
\begin{eqnarray} U_\hbarr (t)^* b^{Wick}U_\hbarr (t)
&=&\sum_{n=0}^{M-1}  i^n  \;\;
 \int_0^t dt_1\cdots\int_0^{t_{n-1}} dt_n \; \sum_{r=0}^{\min(\ell-1,n)}
 \hbarr^r \left[C_r^{(n)}(t_n,\cdots,t_1,t)\right]^{Wick} \label{eq.10.3}\\
&&\hspace{-1.7in} + \sum_{n=\ell}^{M} i^{n}
\int_0^t dt_1\cdots\int_0^{t_{n-1}} dt_n\; U_\varepsilon(t_n)^*
U^0_\varepsilon(t_n) \frac{\hbarr^\ell}{2}
\left[\{Q_{t_{n}},C^{(n-1)}_{\ell-1}(t_{n-1},\cdots,t_1,t)
\}^{(2)}\right]^{Wick}
U_\varepsilon^0(t_n)^* U_\varepsilon(t_n)\label{eq.10.4} \\
&& \hspace{-1.7in} + i^{M} \int_0^t
dt_1\cdots\int_0^{t_{M-1}} dt_M \; U_\varepsilon(t_M)^*
U^0_\varepsilon(t_M) \sum_{r=0}^{\ell-1} \hbarr^{r}
\left[C^{(M)}_r(t_M,\cdots,t_1,t)\right]^{Wick} U_\varepsilon^0(t_M)^*
U_\varepsilon(t_M). \label{eq.10.5}
\end{eqnarray}
Assume that for  $\delta>0$ there exists
a constant $C_{\delta}$ such that
\begin{eqnarray}
\label{mefta}
\sum_{n=\ell}^{\infty}
(1+\delta)^{n}  \sum_{r=0}^{\ell}
 \int_0^t dt_1\cdots\int_0^{t_{n-1}} dt_n \;
  \left|\widetilde{C_r^{(n)}}(t_n,\cdots,t_1,t)\right|_{\L(\bigvee^{p+n-r} \Z,\bigvee^{q+n-r}\Z)}<C_{\delta}\,.
\end{eqnarray}
According to Lemma~\ref{wick-estimate}, the first term
\eqref{eq.10.3} of
\eqref{eq.10.3}\eqref{eq.10.4}\eqref{eq.10.5}
provides in
$U_\hbarr (t)^* b^{Wick}U_\hbarr (t)\big|_{\bigvee^{k}\Z}$
the partial sum of a convergent
 series in $\L(\bigvee^{k}\Z,\bigvee^{k-p+q}\Z)$ when $k\varepsilon\leq 1+\frac\delta 2$.
 With the
same argument the remainder term \eqref{eq.10.5} vanishes
as $M\to \infty$ and $k\varepsilon\leq 1+\frac \delta 2$.
By referring to Lemma~\ref{tech.lem.2}~(ii) and again to
Lemma~\ref{wick-estimate} the factor of $\varepsilon^{\ell}$
in  \eqref{eq.10.4} is
associated with a series which converges in
$\L(\bigvee^{k}\Z,\bigvee^{k-p+q}\Z)$
 as $M\to\infty$
uniformly w.r.t. $(k,\varepsilon)$ when $k\varepsilon\leq 1+\frac
\delta 2$. The sum of the series is simply denoted by $R_{\ell}(t,\varepsilon)$.
Let us prove (\ref{mefta}) to finish the proof of (ii).  Lemma
\ref{wick-estimate} and Lemma \ref{tech.lem.2} say
 \begin{eqnarray*}
&&\sum_{n=\ell}^{\infty}
(1+\delta)^{n}  \sum_{r=0}^{\ell}
 \int_0^t dt_1\cdots\int_0^{t_{n-1}} dt_n \;
   \left|\widetilde{C_r^{(n)}}
(t_n,\cdots,t_1,t)\right|_{\L(\bigvee^{p+n-r} \Z,\bigvee^{q+n-r}\Z)}
 \\ &&\leq \sum_{n=\ell}^{\infty}
 (1+\delta)^{n}  \sum_{r=0}^{\ell}
 \; \frac{|t^n|}{n!}
 \max_{t_{n}\leq \ldots\leq t_{1}\leq t}
\left|\widetilde{C_r^{(n)}}(t_n,\cdots,t_1,t)\right|_{\L(\bigvee^{p+n-r}
\Z,\bigvee^{q+n-r}\Z)}
\\
&&\leq  \sum_{n=\ell}^{\infty}
 (1+\delta)^{n}
\sum_{r=0}^{\ell} \,\frac{2^{n-r}|t^n|}{ n!} C_n^r \;
 [(p+n-r)  (p+n-r-1)]^r \;\frac{(p+n-r-1)!}{(p-1)!}
\;|V|_{L^\infty}^n \;|\tilde b|_{\L(\bigvee^p\Z, \bigvee^q\Z)}\\
&&\leq   \sum_{n=\ell}^{\infty}
(1+\delta)^{n}|t|^{n}
\sum_{r=0}^{\ell} \,\frac{2^{n-r}}{r!} (p+n)^{2r} C_{n-r+p-1}^{p-1} \;
\;|V|_{L^\infty}^n \;|\tilde b|_{\L(\bigvee^p\Z, \bigvee^q\Z)}\\
&&
 \leq 2^p \sum_{n=\ell}^{\infty}
 (1+\delta)^{n}4^{n}|t|^{n} (n+p)^{2\ell}
\;|V|_{L^\infty}^n \;|\tilde b|_{\L(\bigvee^p\Z, \bigvee^q\Z)}\,. \end{eqnarray*}
The last
r.h.s. is finite whenever $4|t| |V|_{L^\infty}<(1+\delta)^{-1}$. The
condition $(1+2\delta)4|t| |V|_{L^\infty}\leq 1$ is sufficient and
provides the uniform bound $C_{\delta}$ in \eqref{mefta}\,.
\fin

\noindent\textbf{Proof of Theorem~\ref{th.main2.2}:}
Set $j=k-m$. By Theorem~\ref{main-2}, the right-hand side of
\eqref{eq.expanbeta} vanishes when $m\neq p-q$ and the convergence of
the series in $\L(\bigvee^{k}\Z,\bigvee^{k-p+q}\Z)$ combined with
Proposition~\ref{pr.wickformules}-ii)
implies
\begin{eqnarray*}
&&\la z^{\otimes j}, U_\hbarr (t)^* b^{Wick}U_\hbarr
(t)\,z^{\otimes k} \ra
\\&&=\sum_{r=0}^{\ell-1} \hbarr^r \,\sum_{n=0}^{\infty}
i^n
\sqrt{\frac{k!j! \, \hbarr^{p+q+2(n-r)}} {(k-(p+n-r))!
(j-(q+n-r))!}} \delta_{k-(p+n-r),\,j-(q+n-r)}^{+}
\\
&&
\hspace{3cm}
\times\int_0^t dt_1\cdots\int_0^{t_{n-1}} dt_n \;
C_r^{(n)}(t_n,\cdots,t_1,t;z)
+O_{\delta}(\hbarr^\ell)\,,
\end{eqnarray*}
when $k\varepsilon\leq 1+\frac \delta 2$, for any $\delta>0$. By
considering the limit $\varepsilon\to0$, $k\varepsilon\to 1$ every
factor
$$
\sqrt{\frac{k!j! \, \hbarr^{p+q+2(n-r)}}{(k-(p+n-r))!
(j-(q+n-r))!}}
$$
converges to 1.
Therefore this proves (ii) for small times $t$ such that $4 |t| |v|_{L^\infty}<1$
up to the identification of the first term as
  $b(z_{t})$. From our definitions we know
$$
b(z_{t})=\left\langle z_{t}^{\otimes q}\,,\, \tilde{b}z_{t}^{\otimes
    p}\right\rangle = b_{t}(e^{-is\Delta}z_{s})\big|_{s=t}\,.
$$
By setting  $w_{s}=e^{-is\Delta}z_{s}$, the quantity $b(z_{t})$
equals
$$
b(z_{t})=b_{t}(w_{0})+\int_{0}^{t}\partial_{s}[b_{t}(w_{s})]~ds
=
b_{t}(w_{0})+\int_{0}^{t}\overline{\partial_{s}w_{s}}.\partial_{\overline
z}b_{t}(w_{s})+ \partial_{z}b_{t}(w_{s}).\partial_{s}w_{s}~ds
$$
Moreover the equation  \eqref{hartree} has the equivalent form with
the vector $w_{s}=e^{-is\Delta}z_{s}$ and $\overline{w_{s}}$
$$
i\partial_{s}w_{s}=e^{-is\Delta}\partial_{\overline{z}}
Q(z_{s})=\partial_{\overline{z}}Q_{s}(w_{s})
\,\qquad
-i\partial_{s}\overline{w_{s}}=\partial_{z}Q_{s}(w_{s})\,.
$$
Hence we get
$$
b(z_{t})=b(w_{0})+i\int_{0}^{t}\left\{Q_{t_{1}},b_{t}\right\}(w_{t_{1}})~dt_{1}\,.
$$
An induction with $w_{0}=z$ and the convergence of the series already
checked yields
$$
b(z_{t})=\sum_{n=0}^{\infty}\int_{0}^{t}d_{t_{1}}\cdots\int_{0}^{t_{n-1}}dt_{n}\;\;
C^{(n)}_{0}(t_{n},\ldots,t_{1},t;z)\,.
$$

Now let us prove the limit (i) for all times by following the argument
in \cite{FGS}, \cite{Spo}. Assume that the result is true for $|t|\leq
\frac{K}{4|V|_{L^{\infty}}}$. Let $s$ be such that
$|s|<1/4|V|_{L^\infty}$. The convergence of the series given in
Theorem~\ref{main-2} and the fact that $U_{\varepsilon}(t)$ preserves
the number gives
\begin{eqnarray}
\label{eq.11}
&&
\hspace{-1cm}\la z^{\otimes j}\,,\,U_\hbarr (t+s)^*
b^{Wick}U_\hbarr (t+s)\,z^{\otimes k} \ra\nonumber
\\&&
=\sum_{n=0}^{\infty}
 i^n  \;\; \sum_{r=0}^{n} \hbarr^r
\int_0^s ds_1\cdots\int_0^{s_{n-1}} ds_n \;  \la z^{\otimes
j},U_\hbarr (t)^*
[C_r^{(n)}(s_n,\cdots,s_1,s)]^{Wick}  U_\hbarr (t)\,z^{\otimes k} \ra\nonumber\\
&&
=\sum_{n=0}^{\infty}
 i^n  \;
\int_0^s ds_1\cdots\int_0^{s_{n-1}} ds_n \;  \la z^{\otimes
j},U_\hbarr (t)^* [C_0^{(n)}(s_n,\cdots,s_1,s)]^{Wick}  U_\hbarr
(t)\,z^{\otimes k} \ra+O_{s}(\hbarr)
\end{eqnarray}
with an absolutely and uniformly convergent
series in the (\ref{eq.11}) when $k\varepsilon$ is close to $1$.
 Hence  the limit
$\hbarr\to 0$, $\hbarr k\to 1$ and the sum $\sum_{n=0}^\infty$ in
(\ref{eq.11}) can be interchanged when $4|s|
|V|_{L^\infty}<1$. An induction on $K=0,1,2\ldots$ finishes the proof.\hfill\fin

\subsection{Coherent states and Wick observables}
We show here that information on the propagation of
coherent states can be directly deduced from the results about Hermite
states.
\begin{prop}
\label{pr.cohwick}
For any $z_{0}\in\Z$ and any $b\in \P_{p,q}(\Z)$, the limit
$$
\lim_{\hbarr\to 0}\left\langle
  U_{\varepsilon}(t)E(z_{0})\,,\,b^{Wick}U_{\varepsilon}(t)E(z_{0})\right\rangle
=b(z_{t})
$$
holds for any $t\in \rz$ when $z_{t}$ denotes the solution to the Hartree
equation \eqref{hartree}.
\end{prop}
\proof
By symmetry, one can assume $m=p-q\geq 0$. Recall that $
\ds E(z_{0})=e^{-\frac{|z_{0}|^{2}}{2\varepsilon}}\sum_{n=0}^{\infty}\frac{\varepsilon^{-n/2}}{\sqrt{n!}}z_{0}^{\otimes
n}$ and start first with $\left|z_{0}\right|=1$.
Since $U_{\varepsilon}(t)$ preserves the number, one gets
\begin{eqnarray*}
&&
\left\langle
    U_{\varepsilon}(t)E(z_{0})\,,\,b^{Wick}U_{\varepsilon}(t)E(z_{0})\right\rangle
 =
\sum_{n=m}^{\infty}e^{-\varepsilon^{-1}}
\frac{\varepsilon^{-n}}{n!}
a_{n}\left(\varepsilon^{-1}\right)
\\
\text{with}
&&
a_{n}\left(\varepsilon^{-1}\right)
=
\varepsilon^{m/2}\sqrt{n(n-1)\ldots (n-m+1)}
 \left\langle z_{0}^{\otimes n-m}\,,\,
   U_{\varepsilon}(t)^{*} b^{Wick} U_{\varepsilon}(t)
   z_{0}^{\otimes n}
\right\rangle
\end{eqnarray*}
By Lemma~\ref{wick-estimate} the quantity
$a_{n}\left(\varepsilon^{-1}\right)$
satisfies
$$
|a_{n}\left(\varepsilon^{-1}\right)|\leq
\left(n\varepsilon\right)^{\frac{p+q+m}{2}}
\left|\tilde{b}\right|_{\L(\bigvee^{p}\Z,\bigvee^{q}\Z)}
\leq
\left\langle
  n\varepsilon\right\rangle^{p}\left|\tilde{b}\right|_{\L(\bigvee^{p}\Z,\bigvee^{q}\Z)}\,.
$$
Hence Lemma~\ref{le.app} applied here with $\lambda=\varepsilon^{-1}$
and $\mu=p$ reduces the problem to the proof of
$$
\lim_{\lambda\to\infty}\int_{\mathbb{R}}
a_{[\sqrt{\lambda}s+\lambda]}(\lambda)\frac{e^{-\frac{s^{2}}{2}}}{\sqrt{2\pi}}~ds\,.
$$
The uniform estimate
$$
\left|a_{[\sqrt{\lambda}s+\lambda]}(\lambda)\right|\leq
C_{p}\left\langle 1+\frac{|s|}{\sqrt{\lambda}}\right\rangle^{p}
\leq C_{p}'\left\langle s\right\rangle^{p}
$$
and the pointwise convergence induced by Theorem~\ref{th.main2.2}
with $z=z_{0}$, $k=[\sqrt{\lambda}s+\lambda]$ and
$\varepsilon=\lambda^{-1}$
yields the result.

\noindent For a general $\left|z_{0}\right|>0$, write
$$
E(z_{0})=e^{-\frac{1}{2\varepsilon'}}\sum_{n=0}^{\infty}\frac{(\varepsilon')^{-n/2}}{\sqrt{n!}}(z_{0}')^{\otimes
n}
=E'(z_{0}')
$$
with $z_{0}'=\frac{z_{0}}{\left|z_{0}\right|}$ and $\varepsilon'=\frac{\varepsilon}{\left|z_{0}\right|^{2}}$\,.
By replacing the $\varepsilon$-quantization by the
$\varepsilon'$-quantization, with
\begin{eqnarray*}
&&  b^{Wick,\varepsilon'}= \left|z_{0}\right|^{-p-q}b^{Wick}\quad
\text{for}\quad
b\in \P_{p,q}(\Z)
\\
&&
H_{\varepsilon}=|z_{0}|^{2}
d\Gamma_{\varepsilon'}(-\Delta)+
|z_{0}|^{4}Q^{Wick, \varepsilon'}
\\
\text{and}&&
\left(i\varepsilon\partial_{t}u=H_{\varepsilon}u\right)
\Leftrightarrow
\left(i\varepsilon'\partial_{t}u=
d\Gamma_{\varepsilon'}(-\Delta)u+\left|z_{0}\right|^{2}Q^{Wick,\varepsilon'}u
\right)\,.
\end{eqnarray*}
Hence the previous result applied with $E'(z_{0}')$,
$\left|z_{0}'\right|=1$ and the $\varepsilon'$-quantization implies
$$
  \lim_{\varepsilon\to 0}\left\langle
  U_{\varepsilon}(t)E(z_{0})\,,\,b^{Wick}U_{\varepsilon}(t)E(z_{0})\right\rangle
=
\left|z_{0}\right|^{p+q}b(z_{t}')
$$
where $z_{t}'$ solves
$$
i\partial_{t}z_{t}'=-\Delta z_{t}' +
\left|z_{0}\right|^{2}(V*\left|z'_{t}\right|^{2})z_{t}'\quad,\quad
\quad z_{t=0}'=z_{0}'=\frac{z_{0}}{\left|z_{0}\right|}\,.
$$
Since this mean field equation preserves the norm
$\left|z'_{t}\right|$ like \eqref{hartree} does for $|z_{t}|$, this implies
\begin{eqnarray*}
  z_{t}'=\left|z_{0}\right|^{-1}z_{t}=\left|z_{t}\right|^{-1}z_{t}\;\;
\text{and } \;
\left|z_{0}\right|^{p+q}b(z_{t}')=b(z_{t})\,.
\end{eqnarray*}
\fin
\begin{remark}
 Another proof can be obtained directly from
  Proposition~\ref{pr.hepp} after checking  uniform number
estimates for $U_{2}(t,0)\Omega$. But working in this direction is
more efficient with the help of Wigner measures.
\end{remark}
\section{Wigner measures: Definition and first properties}
\label{se.WigMeas}
The notion of Wigner (or semiclassical) measures is well established
in the finite dimensional case. We refer the reader to
\cite{Bur}\cite{Ger1}\cite{GMMP}\cite{HMR}\cite{LiPa}\cite{Tar} for details.
The extension that we propose here to the infinite dimensional
case follows a projective approach.

\subsection{Wigner measure of a normal state}
\label{se.wigdef}
Consider the algebra of cylindrical sets
$\mathcal{B}_{cyl}(\Z)=\left\{X(p,E)=p^{-1}(E), \,p\in \p, \, E\in
  \mathcal{B}(p\Z)\right\}$ where $\mathcal{B}(p\Z)$
denotes for any $p\in\p$ the set of Borel subsets of $p\Z$.
 A cylindrical measure $\mu$ is a mapping defined on $\mathcal{B}_{cyl}(\Z)$ such that:
\begin{itemize}
    \item $\mu(\Z)=1$,
    \item For any $p\in \p$,
 $\mu_{p}(A)=\mu(p^{-1}(A))$ for $A\in \mathcal{B}(p\Z)$ defines a
      probability measure $\mu_{p}$ on $\mathcal{B}(p\Z)$.
\end{itemize}
The family of measures $\{\mu_p\}_{p\in\mathbb{P}}$ is often called a
weak distribution.

\smallskip

This notion is often introduced within the framework of real Hilbert
spaces (or more generally real topological vector spaces). This makes
no difference at this level.
The real structure on $\Z$, namely the real scalar product $S$, is
useful for the application of Bochner's theorem.
For any $\xi\in\Z$ the function  $z\mapsto e^{-2\pi i \,S(z,\xi)}$ is
a
cylindrical measurable function and the Fourier transform of
$\mu$ is well defined by
\begin{eqnarray*}
\F[\mu](\xi)=\int_\Z e^{-2\pi i \,S(z,\xi)}  \;d\mu.
\end{eqnarray*}
Bochner's theorem characterizes the Fourier transform of a
weak distribution. It says (see for example \cite{BSZ}) that a function
$G$ is the Fourier transform of a weak distribution if and only if
\begin{itemize}
\item $G$ is normalized: $G(0)=1$,
\item $G$ is of positive type:
  $\ds\sum_{i,j=1}^{N}\lambda_{i}\overline{\lambda_{j}}G(\xi_{i}-\xi_{j})\geq
  0$,
\item For any $p\in \p$, the restricted function $G|_{p\Z}$  is continuous.
\end{itemize}
An important point is that $\Z$ is a separable Hilbert space. Hence
the $\sigma$-algebra generated by the cylindrical sets, that is
containing $\mathcal{B}_{cyl}(\Z)$, is nothing but the Borel
$\sigma$-algebra, $\mathcal{B}(\Z)$, associated with the norm topology
on  $\Z$.
A probability measure well defined on $\mathcal{B}(\Z)$ will be shortly called
a probability measure on $\Z$.
The tightness Prokhorov's criterion (see \cite{Sch}) has
within this setting the next simple form.
\begin{lem}
\label{le.pro}
(See \cite{Sko})
  A cylindrical measure $\mu$ on $\Z$ extends to a probability measure
  on $\Z$ if and only if for any $\eta>0$ there exists
$R_{\eta}>0$ such that
$$
\forall p\in \p,\quad \mu\left(\left\{z\in \Z,\; \left|p z\right|\leq
    R_{\eta}\right\}\right)\geq 1-\eta\,.
$$
\end{lem}
By recalling that for any $R>0$ the ball $\left\{z\in\Z:\left|z\right|\leq
  R\right\}$ is weakly compact, this can be reinterpreted by saying
that a weak distribution $\mu$ extends as a Borel probability measure
if and only if its outer extension is a Radon measure on $\Z$ endowed
with the weak topology (see \cite{Sch}).

Consider  a family $(\rho^{\hbarr})_{\hbarr\in(0,\bar\hbarr)}$ of non
negative trace class operators on
$\H$ such that ${\rm Tr}[\rho^{\hbarr}]=1$, or equivalently  normal states
$\O\mapsto {\rm Tr}[\rho^\hbarr \O]$ on the space of all bounded
operators $\L(\H)$\,.
An additional number estimate assumption allows to associate with such
a family, Wigner probability measures on $\Z$.
\begin{thm}
\label{th.wig-measure}
Let $\left(\varrho^{\varepsilon}\right)_{\varepsilon\in (0,\bar\hbarr)}$ be a family
of normal states on $\L(\mathcal{H})$ parametrized by $\varepsilon$.
Assume   ${\rm Tr}[N^\delta \rho^{\hbarr}]\leq C_{\delta}$ uniformly
w.r.t. $\varepsilon\in (0,\overline{\varepsilon})$
 for some fixed $\delta>0$ and $C_{\delta}\in (0,+\infty)$.
Then for every sequence $(\hbarr_{n})_{n\in\nz}$ with $\lim_{n\to\infty}\hbarr_n= 0$
the exists a subsequence $(\hbarr_{n_k})_{k\in\nz}$ and a Borel probability measure $\mu$ on $\Z$ such that
\begin{eqnarray*}
\lim_{k\to\infty} \Tr[\rho^{\hbarr_{n_k}} b^{Weyl}]=
\lim_{k\to \infty} \Tr[\rho^{\hbarr_{n_k}} b^{A-Wick}]=
\int_{\Z} b(z) \; d\mu(z)\,,
\end{eqnarray*}
for all $b\in \ccup_{p\in\p}\mathcal{F}^{-1}\left(\mathcal{M}_{b}(p\Z)\right)$.\\
Moreover this probability measure $\mu$ satisfies $\ds \int_{\Z} |z|^{2\delta} \, d\mu(z) <\infty$.
\end{thm}
\begin{remark}
\textbf{a)} By introducing the reduced density matrix
$\varrho_{p}^{\varepsilon}\in \L^{1}(\Gamma_{s}(p\Z))$  defined for
$p\in \p$ as a partially traced operator
 $\Tr[\varrho_{p}^{\varepsilon}A]=
\Tr[\varrho^{\varepsilon}(A\otimes  I_{\Gamma_{s}(p^{\bot}\Z)})]$, one
could consider the Husimi function $\mu_{p}^{\varepsilon}$ of $\varrho_{p}^{\varepsilon}$
which is its finite dimensional Wick symbol. It is known that this
makes a weak probability distribution which admits weak limits
after extracting subsequences $\varepsilon_{n_{k}}\to\infty$. The
number estimate implies in finite dimension that such a limit is a
probability measure. Our results say essentially two  things: First after a
proper extraction  of subsequences, the family $(\mu_{p})_{p\in\mathbb{P}}$
makes a weak distribution, i.e. the convergence can hold
simultaneously for all the non countable family $p\in\p$. Secondly the
weak distribution is a Borel probability measure.\\
\noindent\textbf{b)} The estimate
$\int_{\Z}\left|z\right|^{2\delta}~d\mu(z)<+\infty$ will be proved in the more
precise form
$$
\int_{\Z}\left(1+\left|z\right|^{2}\right)^{\delta}~d\mu(z)\leq
\liminf_{\varepsilon_{n_{k}}\to\infty}~
\Tr\left[\varrho^{\varepsilon_{n_{k}}}(1+N)^{\delta}\right]\leq C_{\delta}'<+\infty\,.
$$
Contrary to the finite dimensional case, the first inequality is
not an equality even when the right-hand side converges.
Examples are given in Section~\ref{se.dimdefcomp}.
\end{remark}
\proof
\noindent\textbf{i)} The Proposition~\ref{pr.AWW2} implies
$$
\left|\Tr\left[\varrho^{\varepsilon}b^{Weyl}\right]-
\Tr\left[\varrho^{\varepsilon}b^{A-Wick}\right]\right|
\leq\left|b^{Weyl}-b^{A-Wick}\right|\stackrel{\varepsilon\to 0}{\to}0\,,
$$
for fixed $b\in \ccup_{p\in\p}\mathcal{F}^{-1}\left(\mathcal{M}_{b}(p\Z)\right)$.
Hence the result is true when it is proved after considering simply
the Anti-Wick observables.

\noindent\textbf{ii)}
Consider for $\varepsilon>0$ the function
$$
G_{\varepsilon}(\xi)=\Tr
\left[\varrho^{\varepsilon}W(\sqrt{2}\pi\xi)\right]e^{-\frac{\varepsilon\pi^{2}}{2}\left|\xi\right|^{2}}
=\Tr\left[\varrho^{\varepsilon}(e^{2i\pi S(\xi,.)})^{A-Wick}\right]\,.
$$
The positive type property and the normalization come from
\begin{eqnarray*}
  &&
G_{\varepsilon}(0)=\Tr\left[\varrho^{\varepsilon}\right]=1
\\
&&
\sum_{i,j=1}^{N}\lambda_{i}\overline{\lambda_{j}}G_{\varepsilon}(\xi_{i}-\xi_{j})=
\Tr
\left[
\varrho^{\varepsilon}
\left(
\left|\sum_{k=1}^{N}\lambda_{k}
e^{2i\pi S(\xi_{k},.)}\right|^{2}
\right)^{A-Wick}
\right]
\geq 0\,.
\end{eqnarray*}
The continuity when $\xi$ is restricted to any fixed  finite
dimensional $p\Z$ can be written with uniform estimates w.r.t
$\varepsilon\in (0,\bar\varepsilon)$. Consider the estimate
$\Tr\left[\varrho^{\varepsilon}(1+N)^{\delta_{1}}\right]\leq
C_{\delta_{1}}$ with $\delta_{1}\in (0,\min(1,2\delta))$.
Write for any $\xi, \eta\in
\Z$
\begin{eqnarray*}
|G_\hbarr(\eta)-G_\hbarr(\xi)|&=&
\left|{\rm Tr}\left[\rho^\hbarr
\frac{(N+1)^{\delta_{1}/2}}{(N+1)^{\delta_{1}/2}} [ W(\sqrt{2}\pi\eta)-
W(\sqrt{2}\pi\xi)]\frac{(N+1)^{\delta_{1}/2}}{(N+1)^{\delta_{1}/2}}\right] \right|
\\
&&\hspace{6cm}+ \left|e^{-\frac{\varepsilon
      \pi^{2}}{2}\left|\eta\right|^{2}}-e^{-\frac{\varepsilon
      \pi^{2}}{2}\left|\xi\right|^{2}}\right|
\\
&\leq& \left|[W(\sqrt{2}\pi\eta)- W(\sqrt{2}\pi\xi) ]
  (N+1)^{-\delta_{1}/2}\right|_{\L(\H)} \; {\rm
  Tr}[(N+1)^{\delta_{1}}\rho^\hbarr]
\\
&&\hspace{6cm}+ \left|e^{-\frac{\varepsilon
      \pi^{2}}{2}\left|\eta\right|^{2}}-e^{-\frac{\varepsilon
      \pi^{2}}{2}\left|\xi\right|^{2}}\right|\,.
\end{eqnarray*}
 We have found by Lemma \ref{est.weyl} two constants $\delta_{1}\in (0,1)$ and  $C'_{\delta_{1}}>0$ such that
\begin{equation}
  \label{eq.unifcont}
 \forall \xi,\eta\in \Z,\quad
\left|G_{\varepsilon}(\eta)-G_{\varepsilon}(\xi)\right|
\leq
C_{\delta_{1}}' \, |\eta-\xi|^{\delta_{1}} \; [(|\eta|^2+|\xi|^2)^{\delta_{1}/2}+1],
\end{equation}
holds uniformly w.r.t. $\varepsilon\in (0,\overline{\varepsilon})$ and
we recall the uniform estimate $\left|G_{\varepsilon}(\xi)\right|\leq
1$.
Hence for any $\varepsilon\in (0,\overline{\varepsilon})$, $G_{\varepsilon}$ is
the Fourier transform of a weak distribution $\mu^{\varepsilon}$ such
that
$$
\Tr\left[\varrho^{\varepsilon}b^{A-Wick}\right]=\int_{\Z}b(z)~d\mu^{\varepsilon}(z)
$$
holds for all $b\in \ccup_{p\in\p}\mathcal{F}^{-1}\left(\mathcal{M}_{b}(p\Z)\right)$.\\
\noindent\textbf{iii)}
Actually the uniform estimate \eqref{eq.unifcont} allows to apply an
Ascoli type argument
 after considering sequence $(\varepsilon_{n})_{n\in
  \nz}$ such that $\lim_{n\to\infty}\varepsilon_{n}=0$:
\begin{itemize}
\item Since $\Z$ is separable, it admits a countable dense set
  $\mathcal{N}=\left\{\xi_{\ell},\; \ell\in \nz \right\}$. For any
  $\ell\in\nz$ the sequence $G_{\varepsilon_{n}}(\xi_{\ell})$ remains
  in $\left\{\sigma\in \cz, \left|\sigma\right|\leq 1\right\}$. Hence
  by a diagonal extraction process there exists a subsequence
  $(\varepsilon_{n_{k}})_{k\in \nz}$ such that for all $\ell\in \nz$,
$G_{\varepsilon_{n_{k}}}(\xi_{\ell})$ converges in $\left\{\sigma\in
  \cz, \left|\sigma\right|\leq 1\right\}$ as $k\to \infty$. Set
$$
G(\xi_{\ell})=\lim_{k\to\infty}G_{\varepsilon_{n_{k}}}(\xi_{\ell})
$$
for all $\ell\in \nz$.
\item The uniform estimate \eqref{eq.unifcont} implies that the limit
  $G$ is uniformly continuous on any set $\mathcal{N}\cap
  \left\{z\in\Z:\left|z\right|\leq R\right\}$. Hence it admits a continuous
  extension still denoted $G$ in $(\Z,\left|~\right|_{\Z})$. An ``epsilon$/3$''-argument
  shows that for any $\xi\in\Z$
  $\lim_{k\to\infty}G_{\varepsilon_{n_{k}}}(\xi)$ exists and equals
  $G(\xi)$.
\item Finally $G$ is a normalized function of positive type as a limit
  of such functions.
\end{itemize}
Finally the uniform estimates $\left|G_{\varepsilon}(\xi)\right|\leq 1$
and $\left|G(\xi)\right|\leq 1$ allow to test the convergence again
any $\nu\in \mathcal{M}_{b}(p\Z)$ and to apply the Parseval identity with $b=\mathcal{F}^{-1}(\nu)$.
From any sequence $(\varepsilon_{n})_{n\in\nz}$ such that
$\lim_{n\to\infty}\varepsilon_{n}=0$, one can extract a subsequence
$\left(\varepsilon_{n_{k}}\right)_{k\to\infty}$ and find a weak
distribution such that the limit
$$
\lim_{n_{k}\to
  \infty}\Tr\left[\varrho^{\varepsilon_{n_{k}}}b^{Weyl}\right]=
\lim_{n_{k}\to
  \infty}\Tr\left[\varrho^{\varepsilon_{n_{k}}}b^{A-Wick}\right]=\int_{\Z}b(z)~d\mu(z)
$$
holds for any $b\in \mathcal{F}\left(L^{1}(p\Z, L_{p}(dz))\right)$ and
therefore for any $b\in \mathcal{S}_{cyl}(\Z)$.
\\
\noindent\textbf{iv)} The Prokhorov's criterion for $\mu$ in the
form stated in Lemma~\ref{le.pro} is again a consequence of
the uniform number estimate
$\Tr\left[N^{\delta}\varrho^{\varepsilon}\right]\leq
C_{\delta}$.
Fix any $p\in \p$ and set $d=\textrm{dim}p$.
The operators $N_{p}=N_{p\Z}\otimes I_{\Gamma_{s}(p^{\bot}\Z)}=\left(d\Gamma(I_{p\Z})\otimes
  I_{\Gamma_{s}(p^{\bot}\Z)}\right)=d\Gamma(p)$,
$N_{p^{\bot}}=\left(I_{p\Z}\otimes
  d\Gamma(I_{p^{\bot}\Z})\right)=d\Gamma(p^{\bot})$ and $N=d\Gamma(I)$
make a commuting family of non negative operators such that
$N=N_{p}+N_{p^{\bot}}$. Thus the inequality
$$
(1+\frac{d\varepsilon}{2}+N)^{s}\geq (1+\frac{d\varepsilon}{2}+N_{p})^{s}
$$
holds for any $s\geq 0$.  Hence the estimate
$\Tr\left[\varrho^{\varepsilon}N^{\delta}\right]\leq
C_{\delta}$ implies
\begin{equation*}
\Tr\left[\varrho^{\varepsilon}(1+\frac{d\varepsilon}{2}+N_{p})^{\delta}\right]
\leq
\Tr\left[\varrho^{\varepsilon}(1+\frac{d\varepsilon}{2}+N)^{\delta}\right]\leq
\Tr\left[\varrho^{\varepsilon}(2+N)^{\delta}\right]\leq
 C_{\delta}'\,,
\end{equation*}
with $C_{\delta}'>0$ independent of $\varepsilon$ and $p$ as soon as
$\varepsilon\leq \frac{1}{d}$.\\
Let $\chi\in \mathcal{C}^{\infty}(p\Z)$ be a non negative function on
$p\Z$, such that $\chi\equiv 0$ in a neighborhood of
$\left\{\left|z\right|\leq 1\right\}$. For any $R\geq 1$ the estimates
$$
\frac{(1+R^{2})^{\delta}}{(1+|z|^{2})^{\delta}}\chi(R^{-1}z) \leq 1
$$
holds with uniform estimates of the left-hand side in
$S_{p\Z}(1,\frac{\left|dz\right|^{2}}{\left\langle
    z\right\rangle^{2}})$. The pseudodifferential calculus in
$p\Z$
with the metric $\frac{|dz|^{2}}{\langle z\rangle^{2}}$, provides the
inequality of bounded operators on $\Gamma_{s}(p\Z)$
\begin{eqnarray*}
&&
(1+R^{2})^{\delta}A\circ B_{R}\circ A -C\varepsilon \leq
\left[\frac{(1+R^{2})^{\delta}}{(1+|z|^{2})^{\delta}}\chi(R^{-1}z)\right]^{Weyl}
\leq 1+C\varepsilon
\\
\text{with}
&&A=\left[(1+|z|^{2})^{-\delta/2}\right]^{Weyl}
\quad,\quad
B_{R}=\left[\chi(R^{-1}z)\right]^{Weyl}
\quad\text{and}
\quad
\left|B_{R}\right|_{\L(\Gamma_{s}(p\Z))}\leq C\,,
\end{eqnarray*}
with a constant $C>0$ independent of $\varepsilon\in (0,\frac{1}{d})$ and
$R\geq 1$. By Proposition~\ref{pr.calcfunc}, there exists a constant
$C'>0$ independent of $\varepsilon\in (0,\frac{1}{d})$ (and $R\geq 1$)
such that
$$
\left|A^{2}\circ
  (1+\frac{d\varepsilon}{2}+N_{p\Z})^{\delta}-I_{\Gamma_{s}(p\Z)}\right|_{\L(\Gamma_{s}(p\Z))}\leq C'\varepsilon\,.
$$
Hence the inequality
$$
(1+R^{2})^{\delta}\chi(R^{-1}pz)^{Weyl}\leq (1+2C\varepsilon) A^{-\delta}
$$
after tensorization with $I_{\Gamma_{s}(p^{\bot}\Z)}$ and testing on
the normal state $\varrho^{\varepsilon}$ yields
$$
(1+R^{2})^{\delta}\Tr\left[\varrho^{\varepsilon}
\chi(R^{-1}pz)^{Weyl}
\right]
\leq C_{\delta}''
$$
with a uniform constant $C_{\delta}''$ with respect to $\varepsilon\in
(0,\frac{1}{d})$ and $R\geq 1$. After taking the limit $n_{k}\to
\infty$, $\varepsilon_{n_{k}}\to 0$, we get
$$
\int_{\Z}1_{\{\left|pz\right|\geq R\}}(z)~d\mu(z)\leq
\int_{\Z}\chi(R^{-1}pz)~d\mu(z)
=\lim_{n_{k}\to\infty}\Tr\left[\varrho^{\varepsilon_{n_{k}}}
\chi(R^{-1}pz)^{Weyl}
\right]\leq C_{\delta}''(1+R^{2})^{-\delta}\,.
$$
This inequality is valid for any $p\in\p$ and the Prokhorov's
criterion of Lemma~\ref{le.pro} is satisfied. The weak distribution
$\mu$ is a probability measure on $\Z$.\\
\noindent\textbf{v)}  First the function $\left\langle
  z\right\rangle^{2\delta}$ is Borel measurable in $\Z$.
Take $p\in\p$ and $R\geq 1$ and take now $\chi_{0}\in
\mathcal{C}^{\infty}_{0}(p\Z)$, such that $0\leq \chi_{0}\leq 1$
and $\chi_{0}\equiv 1$ in a neighborhood of $0$.
Consider the estimates
\begin{eqnarray*}
(1+N)^{\delta}\geq
(1+N_{p})^{\delta}&\geq&
(1+N_{p})^{\delta/2}\chi_{0}(R^{-1}pz)^{Weyl}(1+N_{p})^{\delta/2}-C_{p}\varepsilon
(1+N_{p})^{\delta}\\
&\geq &
\left[\left((1+\left|pz\right|^{2})\right)^{\delta}\chi_{0}(R^{-1}pz)\right]^{Weyl}
-C_{p}'\varepsilon (1+N)^{\delta}
\end{eqnarray*}
where the two last inequalities are again derived from the finite dimensional
Weyl calculus (with a uniform control w.r.t. $R\geq 1$). After taking
the limit $n_{k}\to\infty$, $\varepsilon_{n_{k}}\to 0$, this implies
\begin{eqnarray*}
\int_{\Z}\left(1+\left|pz\right|^{2}\right)^{\delta}\chi_{0}(R^{-1}pz)~d\mu(z)
&=&\lim_{n_{k}\to \infty}\Tr\left[\varrho^{\varepsilon_{n_{k}}}
  \left[\left((1+\left|pz\right|^{2})\right)^{\delta}\chi_{0}(R^{-1}pz)\right]^{Weyl}\right]
\\
&\leq&
\liminf_{n_{k}\to\infty}\Tr\left[\varrho^{\varepsilon_{n_{k}}}(1+N)^{\delta}\right]
\leq C_{\delta}'\,.
\end{eqnarray*}
Taking the supremum w.r.t $R\geq 1$ and then w.r.t  a countable
increasing sequence $(p_{n})_{n\in\nz}$, $p_{n}\in \p$, such that $\sup_{n\in\nz}p_{n}=I_{\Z}$, yields
$$
\int_{\Z}(1+|z|^{2})^{\delta}d\mu(z)\leq C_{\delta}'<+\infty\,.
$$
\fin

\subsection{Complex Wigner measures, pure sequences}

More general families of trace class operators can be considered
by  linear decomposition
\begin{equation}
  \label{eq.decomptr}
\varrho^{\varepsilon}=\lambda_{R+}^{\varepsilon}\varrho_{R+}^{\varepsilon}-\lambda_{R-}^{\varepsilon}\varrho_{R-}^{\varepsilon}
+i\lambda_{I+}^{\varepsilon}\varrho_{I+}^{\varepsilon}-i\lambda_{I-}\varrho_{I-}^{\varepsilon}
\end{equation}
with $\lambda_{\bullet}^{\varepsilon}\geq 0$, $\varrho_{\bullet}^{\varepsilon}\geq
0$, $\Tr\left[\varrho_{\bullet}^{\varepsilon}\right]=1$ and
$$
\lambda_{R+}^{\varepsilon}
+
\lambda_{R-}^{\varepsilon}
+
\lambda_{I+}^{\varepsilon}
+
\lambda_{I-}^{\varepsilon}\leq 4\Tr\left[|\varrho^{\varepsilon}|\right]\,.
$$
\begin{prop}
  \label{pr.mucp}
Let $(\varrho^{\varepsilon})_{\varepsilon\in
  (0,\overline{\varepsilon})}$ be a family of trace class operators
such that
\begin{eqnarray}
\label{hyptraclassop}
\left|(1+N)^{\delta/2}\varrho^{\varepsilon}(1+N)^{\delta/2}\right|_{\mathcal{L}^{1}(\H)}\leq
C_{\delta}
\end{eqnarray}
uniformly for some $\delta>0$ and some
$C_{\delta}<+\infty$. Then for any sequence
$(\varepsilon_{n})_{n\in\nz}$ such that
$\lim_{n\to\infty}\varepsilon_{n}=0$, one can extract a subsequence
$\left(\varepsilon_{n_{k}}\right)_{k\in\nz}$ and find a (complex)
Borel measure $\mu$ on $\Z$ such that
\begin{equation}
\label{eq.limfaible}
\lim_{k\to\infty} \Tr[\rho^{\hbarr_{n_{k}}} b^{Weyl}]=
\lim_{k\to \infty} \Tr[\rho^{\hbarr_{n_k}} b^{A-Wick}]=
\int_{\Z} b(z) \; d\mu(z)\,,
\end{equation}
for all $b\in \ccup_{p\in\p}\mathcal{F}^{-1}\left(\mathcal{M}_{b}(p\Z)\right)$.\\
Moreover this measure satisfies $\int_{\Z}\left\langle
  z\right\rangle^{2\delta}~d\left|\mu\right|(z)<+\infty$.
\end{prop}
\proof
The decomposition \eqref{eq.decomptr} implies
\begin{eqnarray*}
&&
(1+N)^{\delta/2}\varrho^{\varepsilon}(1+N)^{\delta/2}=
\lambda_{R+}^{\varepsilon}r_{R+,\varepsilon}^{\varepsilon}-\lambda_{R-}^{\varepsilon}r_{R-,\delta}
+i\lambda_{I+}^{\varepsilon}r_{I+,\delta}^{\varepsilon}-i\lambda_{I-}r_{I-,\delta}^{\varepsilon}
\\
\text{with}
&&
r_{\bullet,\delta}^{\varepsilon}=(1+N)^{\delta/2}\varrho_{\bullet}^{\varepsilon}(1+N)^{\delta/2}\geq
0
\\
\text{and}
&&
\Tr\left[(1+N)^{\delta}\varrho_{\bullet}^{\varepsilon}\right]
=
\Tr\left[r_{\bullet,\delta}^{\varepsilon}\right]\leq
\left|(1+N)^{\delta/2}\varrho^{\varepsilon}(1+N)^{\delta/2}\right|_{\L^{1}(\H)}\,.
\end{eqnarray*}
Hence the symmetric writing with
$(1+N)^{\delta/2}\varrho^{\varepsilon}(1+N)^{\delta/2}$ of the
uniform weighted estimate ensures that
every term $\varrho_{\bullet}$ in fulfills the assumptions of
Theorem~\ref{th.wig-measure}.
It suffices to extract a subsequence which provides the convergence for
all the four terms.
\fin
\begin{definition}
 \label{de.Msc} For a family
 $\left(\varrho^{\varepsilon}\right)_{\varepsilon\in
   (0,\overline{\varepsilon})}$, satisfying (\ref{hyptraclassop}), the set of Borel measures $\mu$ which satisfy
\eqref{eq.limfaible}
is denoted $\mathcal{M}(\varrho^{\varepsilon},\varepsilon\in
(0,\overline{\varepsilon}))$ or simply $\mathcal{M}(\varrho^{\varepsilon})$.\\
Such a family $(\varrho^{\varepsilon})_{\varepsilon\in
  (0,\overline{\varepsilon})}$ (resp. a sequence
$(\varrho^{\varepsilon_{n}})_{n\in\nz}$) is said pure if $\mathcal{M}(\varrho^{\varepsilon},\varepsilon\in
(0,\overline{\varepsilon}))$ (resp.
$\mathcal{M}(\varrho^{\varepsilon_{n}},n\in\nz)$) has a single element
$\mu$.
\end{definition}
When the family $(\varrho^{\varepsilon})_{\varepsilon\in
  (0,\overline{\varepsilon})}$ is pure the limit in
\eqref{eq.limfaible} can be written with $\lim_{\varepsilon\to 0}$
instead of $\lim_{n_{k}\to \infty}$. This provides a characterization of
$\mathcal{M}(\varrho^{\varepsilon})=\left\{\mu\right\}$.
 For simplicity, we shall often assume that the family
$(\varrho^{\varepsilon})_{\varepsilon\in (0,\overline{\varepsilon})}$
is pure, when the reduction to such a case can be done after extracting a
suitable sequence.
\subsection{Countably separating sets of observables}
In order to identify a Wigner measure of
$\mu\in\mathcal{M}(\varrho^{\varepsilon})$ it is sufficient to test
on a ``dense set'' of observables. The good notion is given by the
Stone-Weierstrass theorem for $L^{1}$ spaces. It can be
  recovered from the standard Stone-Weierstrass theorem for continuous
functions in our case.
\begin{lem}
(cf \cite{Cou})
Let $\nu$ be a Borel probability measure on a separable Banach space $X$
and let $\left\{f_{n}, n\in \nz\right\}$ be a countable set of bounded
$\nu$-measurable functions which separates the points
$$
\forall x,y\in X,\exists n\in \nz, \quad f_{n}(x)\neq f_{n}(y)\,.
$$
Then for any $p\in [0,\infty)$, the algebra generated by
$\left\{f_{n}, n\in\nz\right\}$ is dense in $L^{p}(X,d\nu)$.
\end{lem}
Since ``the'' Wigner measure is not known a priori, the good notion of
``dense set'' that we shall use  is the following.
\begin{definition}
  A subset $\mathcal{D}\subset
  \ccup_{p\in\p}\mathcal{F}^{-1}(\mathcal{M}_{b}(p\Z))$ is said
  countably separating whenever it contains a countable subset,
  $\mathcal{D}\supset \mathcal{D}_{0}\sim\nz$, which
  separates the point of $\Z$:
$$
\forall x,y\in\Z,\exists f\in \mathcal{D}_{0},\quad f(x)\neq f(y)\,.
$$
\end{definition}
\begin{prop}
Let $\mu_{1}$ be a bounded Borel measure on $\Z$ and let
$(\varrho^{\varepsilon})_{\varepsilon\in(0,\overline{\varepsilon})}$
be a family of operators which fulfills the assumptions of
Definition~\ref{de.Msc}.
The two next statements are equivalent:
\begin{enumerate}
\item $\mathcal{M}(\varrho^{\varepsilon})=\left\{\mu_{1}\right\}$.
\item There exists a countably separating subset
  $\mathcal{D}\subset\ccup_{p\in\p}\mathcal{F}^{-1}(\mathcal{M}_{b}(p\Z))$
 such that
$$
\forall b\in \mathcal{D}\,,\quad
\lim_{\varepsilon\to 0}\Tr\left[\varrho^{\varepsilon}b^{Weyl}\right]
=\lim_{\varepsilon\to 0}\Tr\left[\varrho^{\varepsilon}b^{A-Wick}\right]
=\int_{\Z}b(z)~d\mu_{1}(z)\,.
$$
\end{enumerate}
\end{prop}
\begin{remark}
  A similar equivalence is obtained for $\mu_{1}\in
  \mathcal{M}(\varrho^{\varepsilon})$ after a subsequence extraction.
\end{remark}
\proof
Assume $\mu\in \mathcal{M}(\varrho^{\varepsilon})$.
There exists a sequence $(\varepsilon_{n_{k}})_{k\in \nz}$ and a
Borel measure $\mu$ such
that \eqref{eq.limfaible} holds for any $b\in
\ccup_{p\in\p}\mathcal{F}^{-1}\mathcal{M}_{b}(p\Z)$.
In particular this holds for any $b\in\mathcal{D}$:
$$
\int_{\Z}b(z)~d\mu(z)=\lim_{k\to\infty}\Tr\left[\varrho^{\varepsilon_{n_{k}}}b^{Weyl}\right]=
\int_{\Z}b(z)~d\mu_{1}(z)\,.
$$
The set $\mathcal{D}$ is dense in $L^{1}(\Z,d|\mu_{1}|)$ and in
$L^{1}(\Z,d|\mu|)$ so that the above equality of the extreme sides
extend to any bounded Borel function. This implies $\mu=\mu_{1}$.
\fin

The next examples will be useful in the application and allow to
reconsider an inductive point of view.
\begin{prop}
  \label{pr.exdense}
Let $(p_{\ell})_{\ell\in\nz}$ be an increasing sequence of projectors in
$\p$ such that $\sup_{\ell}p_{\ell}=I_{\Z}$ and let the family of operators
$(\varrho^{\varepsilon})_{\varepsilon\in(0,\overline{\varepsilon})}$
satisfy the assumptions of Definition~\ref{de.Msc}. Then
the identity $\mathcal{M}(\varrho^{\varepsilon})=\left\{\mu\right\}$
is equivalent to any of the next statement
\begin{enumerate}
\item For all $b\in \ccup_{\ell\in \nz}\mathcal{S}(p_{\ell}\Z)$,
the quantity $\Tr[\varrho^{\varepsilon}b^{Weyl}]$ converges to
$\int_{\Z}b(z)~d\mu(z)$
as $\varepsilon\to 0$.
\item For all $b\in \mathcal{S}_{cyl}(\Z)$,
the quantity $\Tr[\varrho^{\varepsilon}b^{Weyl}]$ converges to
$\int_{\Z}b(z)~d\mu(z)$
as $\varepsilon\to 0$.
\end{enumerate}
\end{prop}
\proof
It suffices to notice that  $\ccup_{\ell\in\nz}\mathcal{S}(p_{\ell}\Z)$,
and therefore $\mathcal{S}_{cyl}(\Z)$, is countably separating because
the weak topology separates the points.
\fin
\subsection{Orthogonality argument}
Complex  Wigner measures are especially interesting while considering
the joint measure associated with two families of vectors
$(u^{\varepsilon})_{\varepsilon\in(0,\overline{\varepsilon})}$ and
$(v^{\varepsilon})_{\varepsilon\in
  (0,\overline{\varepsilon})}$. Introduce the notation
$$
\varrho_{uv}^{\varepsilon}=|u^{\varepsilon}\rangle\langle
v^{\varepsilon}|\,.
$$
\begin{prop}
\label{pr.orth1}
  Assume that the family of vectors
$(u^{\varepsilon})_{\varepsilon\in(0,\overline{\varepsilon})}$ and
$(v^{\varepsilon})_{\varepsilon\in (0,\overline{\varepsilon})}$
satisfy the uniform estimates
$$
\left|(1+N)^{\delta/2}u^{\varepsilon}\right|_{\H}+\left|(1+N)^{\delta/2}v^{\varepsilon}\right|_{\H}
\leq C
\quad, \quad
\left|u^{\varepsilon}\right|_{\H}=\left|v^{\varepsilon}\right|_{\H}=1
$$
for some fixed $\delta>0$ and $C>0$. Assume further that  any
$\mu\in \mathcal{M}(\varrho_{uu}^{\varepsilon})$ and any $\nu\in
\mathcal{M}(\varrho_{vv}^{\varepsilon})$ are mutually orthogonal. Then
the family $(\varrho_{uv}^{\varepsilon})_{\varepsilon\in
  (0,\overline{\varepsilon})}$ is pure with
\begin{eqnarray*}
  && \mathcal{M}(\varrho^{\varepsilon}_{uv}, \varepsilon\in (0,\overline{\varepsilon}))=\left\{0\right\}\\
i.e.
&&
\lim_{\varepsilon\to 0}\left\langle
  u^{\varepsilon}\,,\,b^{Weyl}v^{\varepsilon}\right\rangle
=
\lim_{\varepsilon\to 0}\left\langle u^{\varepsilon}\,,\,b^{A-Wick}v^{\varepsilon}\right\rangle
=0
\end{eqnarray*}
for any $b\in \mathcal{F}^{-1}(\mathcal{M}_{b}(p\Z))$ and any $p\in \p$.
\end{prop}
\proof
Assume $\mathcal{M}(\varrho_{uu})=\left\{\mu\right\}$ and
$\mathcal{M}(\varrho_{vv}^{\varepsilon})=\left\{\nu\right\}$
with $\mu\perp\nu$. Take $\eta>0$. There exist two bounded closed subset $K_{1}$ and
$K_{2}$ such that
$$
\mu(K_{1})\geq 1-\eta
\quad,\quad
\nu(K_{2})\geq 1-\eta
\quad,\quad
K_{1}\cap K_{2}=\emptyset\,.
$$
Since $K_{1}$ and $K_{2}$ are compact in the weak topology,
$K_{1}\subset \complement K_{2}$, $\complement K_{2}$ open in the weak
topology, there exists a finite covering of $K_{1}$ of the form
$$
K_{1}\subset \ccup_{k=1}^{K}\left\{|p_{k}(z-z_{k})|\leq r_{k}\right\}
\quad,\quad
\ccup_{k=1}^{K}\left\{|p_{k}(z-z_{k})|\leq 2r_{k}\right\}\cap K_{2}=\emptyset
$$
with $p_{k}\in\p$, $z_{k}\in \Z$ and $r_{k}>0$ for all $k\in
\left\{1,\ldots, K\right\}$. By choosing for any $k$ a function
$\chi_{k}\in \mathcal{C}^{\infty}_{0}(p_{k}\Z)$ such that
 $\chi_{k}(p_{k}(z))\equiv 1$ when $\left|p_{k}(z-z_{k})\right|\leq r_{k}$ and $\chi_{k}(p_{k}z)= 0$ when
 $\left|p_{k}(z-z_{k})\right|\geq 2r_{k}$ the sum
$\chi(z)=\sum_{k=1}^{N}\frac{\chi_{k}(p_{k}z)}{\sum_{k'}\chi_{k'}(p_{k'}z)}$ defines a cylindrical
function $\chi\in \S_{cyl}(\Z)$ such that $\chi\equiv 1$ on $K_{1}$
and $\chi\equiv 0$ on $K_{2}$.\\
Take now any $b\in \S_{cyl}(\Z)$ and write
\begin{eqnarray*}
\left|\left\langle u^{\varepsilon}\,,b^{Weyl}v^{\varepsilon}\right\rangle\right|
&=&
\left|\left\langle
    u^{\varepsilon}\,,(b\chi)^{Weyl}v^{\varepsilon}\right\rangle\right|
+
\left|\left\langle
  u^{\varepsilon}\,,(b(1-\chi))^{Weyl}v^{\varepsilon}\right\rangle\right|\\
&\leq&
\left|(\overline{b}(1-\chi))^{Weyl}u^{\varepsilon}\right|_{\H}+ \left|(b\chi)^{Weyl}v^{\varepsilon}\right|_{\H}\,.
\end{eqnarray*}
From the Weyl pseudodifferential calcul we get
$$
\left|(\overline{b}(1-\chi))^{Weyl}u^{\varepsilon}\right|^{2}_{\H}\leq
\Tr\left[\varrho_{uu}^{\varepsilon}\left((1-\chi)^{2}|b|^{2}\right)^{Weyl}\right]
+C_{b\chi}
$$
where the right-hand side converges to
$\int_{\Z}|b|^{2}(1-\chi)^{2}(z)~d\mu(z)$ as $\varepsilon\to 0$.
The property $\chi\equiv 1$ on $K_{1}$ with $\mu(K_{1})\geq 1-\eta$
implies
$$
\limsup_{\varepsilon\to 0}
\left|(\overline{b}(1-\chi))^{Weyl}u^{\varepsilon}\right|^{2}_{\H}
\leq
\eta\left|b\right|_{L^{\infty}}^{2}
$$
and with the symmetric argument $\limsup_{\varepsilon\to 0}
\left|(b\chi)^{Weyl}v^{\varepsilon}\right|^{2}_{\H}
\leq
\eta\left|b\right|_{L^{\infty}}^{2}$. Hence we get
$$
\forall \eta>0, \quad \limsup_{\varepsilon\to 0}
\left|\left\langle
    u^{\varepsilon}\,,b^{Weyl}v^{\varepsilon}\right\rangle\right|
\leq 2\left|b\right|_{L^{\infty}}\sqrt{\eta}
$$
for any $b\in \mathcal{S}_{cyl}(\Z)$. This implies
$\mathcal{M}(\varrho_{uv}^{\varepsilon}, \varepsilon\in (0,\overline{\varepsilon}))=\left\{0\right\}$\,.
\fin

A straightforward consequence is the next proposition.
\begin{prop}
\label{pr.orth2}
  Make the same assumptions as in Proposition~\ref{pr.orth1} with the
  additional condition
  $\mathcal{M}(\varrho_{uu}^{\varepsilon})=\left\{\mu_{u}\right\}$
and
$\mathcal{M}(\varrho_{vv}^{\varepsilon})=\left\{\mu_{v}\right\}$. Then
the family of trace class operators
$(\varrho_{u+v,u+v}^{\varepsilon})_{\varepsilon\in
  (0,\overline{\varepsilon})}$
satisfies
$$
\mathcal{M}(\varrho_{u+v,u+v}^{\varepsilon})=\left\{\mu_{u}+\mu_{v}\right\}\,.
$$
\end{prop}
\proof
Write simply
\begin{eqnarray*}
\left\langle u^{\varepsilon}+v^{\varepsilon}\,,\,
  b^{Weyl}(u^{\varepsilon}+v^{\varepsilon})\right\rangle
&=&
\left\langle u^{\varepsilon}\,,\,
  b^{Weyl}u^{\varepsilon}\right\rangle
+
\left\langle v^{\varepsilon}\,,\,
  b^{Weyl}v^{\varepsilon}\right\rangle
\\
&&
+
\left\langle u^{\varepsilon}\,,\,
  b^{Weyl}v^{\varepsilon}\right\rangle
+
\left\langle v^{\varepsilon}\,,\,
  b^{Weyl}u^{\varepsilon}\right\rangle\,,
\end{eqnarray*}
and take the limit of every term as $\varepsilon\to 0$.
\fin

\subsection{Wigner measure and Wick observables}
\label{se.wigwick}

Up to some additional assumption on the state and by restricting the class of Wick
observables, we check in this subsection that testing with Weyl,
(or Anti-Wick) and Wick observables provides the same asymptotic information as
$\varepsilon\to 0$.\\
 Fix once and for all $p\in\p$, the choice of the metric
$g_{p}=|dz|^{2}$ or $g_{p}=\frac{|dz|^{2}}{\langle z\rangle^{2}}$.
From Proposition~\ref{pr.weylwick} we know that the class of symbols
$\ccup_{p\in\p, s\in \rz}S_{p\Z}(\langle z\rangle^{s},g_{p})$ and
$\oplus_{m,q\in\nz}^{\rm alg}\P_{m,q}(\Z)$ both contain
all the classes $\P_{m,q}(p\Z)$, with a good comparison of Weyl and
Wick quantizations on these smaller sets. In the limit $\varepsilon\to
0$, this comparison can be carried out to any $b\in \oplus_{m,q\in\nz}^{\rm alg}\P_{m,q}^{\infty}(\Z)$.
\begin{thm}
\label{th.wigwick}
Assume that the family of operators
$(\varrho^{\varepsilon})_{\varepsilon\in (0,\overline{\varepsilon})}$
satisfies
$$\left|(1+N)^{\delta/2}\varrho^{\varepsilon}(1+N)^{\delta/2}\right|_{\L^{1}(\H)}\leq
C_{\delta}$$ uniformly w.r.t $\varepsilon\in
(0,\overline{\varepsilon})$ for any $\delta>0$.
\begin{enumerate}
\item For any fixed $\beta\in \ccup_{p\in \p, s\in \rz}S_{p\Z}(\left\langle
    z\right\rangle^{s}, g_{p})$, the families
$(\beta^{Weyl}\varrho^{\varepsilon})_{\varepsilon\in
  (0,\overline{\varepsilon})}$
and
$(\beta^{A-Wick}\varrho^{\varepsilon})_{\varepsilon\in
  (0,\overline{\varepsilon})}$ satisfy the assumptions of
Definition ~\ref{de.Msc} and
\begin{equation}
  \label{eq.wigpol}
\mathcal{M}(\beta^{Weyl}\varrho^{\varepsilon})=\mathcal{M}(\beta^{A-Wick}\varrho^{\varepsilon})
=\left\{\beta\mu\,,\;\mu\in \mathcal{M}(\varrho^{\varepsilon})\right\}
\end{equation}
\item For any fixed $\beta\in \oplus_{m,q\in\nz}^{\rm alg}\P_{m,q}^{\infty}(\Z)$
  the family $(\beta^{Wick}\varrho^{\varepsilon})_{\varepsilon\in
  (0,\overline{\varepsilon})}$ satisfies the assumptions of
Definition ~\ref{de.Msc} and
\begin{equation}
  \label{eq.wigpol2}
\mathcal{M}(\beta^{Wick}\varrho^{\varepsilon})
=
\left\{\beta\mu\,,\;\mu\in \mathcal{M}(\varrho^{\varepsilon})\right\}
\,.
\end{equation}
\end{enumerate}
\end{thm}
A particular case holds when the measure is tested with $b=1$.
\begin{cor}
  \label{co.beg1}
Assume the uniform estimate
$\left|(1+N)^{\delta/2}\varrho^{\varepsilon}(1+N)^{\delta/2}\right|_{\L^{1}(\H)}\leq
C_{\delta}$ for all $\delta>0$ and further
$\mathcal{M}(\varrho^{\varepsilon})=\left\{\mu\right\}$.
\begin{enumerate}
\item
The
equality
$$
\lim_{\varepsilon\to
  0}\Tr\left[\beta^{Weyl}\varrho^{\varepsilon}\right]
=
\lim_{\varepsilon\to
  0}\Tr\left[\beta^{A-Wick}\varrho^{\varepsilon}\right]
=\int_{\Z}\beta(z)~d\mu(z)\,
$$
holds when $\beta\in \ccup_{p\in\p, s\in\rz} S_{p\Z}(\langle z\rangle^{s},g_{p})$
\item
The limit
$$
\lim_{\varepsilon\to
  0}\Tr\left[\beta^{Wick}\varrho^{\varepsilon}\right]
=\int_{\Z}\beta(z)~d\mu(z)\,
$$
\end{enumerate}
holds for any $\beta\in \oplus_{m,q\in\nz}^{\rm alg}\P_{m,q}^{\infty}(\Z)$.
\end{cor}
\noindent\textbf{Proof of Theorem~\ref{th.wigwick}:} 1)
The relation \eqref{eq.AWW1} extends to any $b\in
S_{p\Z}(\langle z\rangle^{s},g_{p})$ and implies
$\varepsilon^{-1}(b^{Weyl}-b^{A-Wick})=c(\varepsilon)^{Weyl}$ with
$c(\varepsilon)$ uniformly bounded in $S_{p\Z}(\langle
z\rangle^{s-2},g_{p})$. The result for $\beta^{A-Wick}$ can be deduced
from the one for $\beta^{Weyl}$.\\
Take $p\in \p$, $s\geq 0$ (this contains the case $s<0$) and $\beta\in S_{p\Z}(\langle
z\rangle^{s},g_{p})$.
Let $N_{p}=N_{p\Z}\otimes I_{\Gamma_{s}(p^{\bot}\Z)}$ and
$N_{p^{\bot}}=I_{\Gamma_{s}(p\Z)}\otimes N_{p^{\bot}\Z}$.
Our assumption on $(\varrho^{\varepsilon})_{\varepsilon\in
  (0,\overline{\varepsilon})}$
and the commutations
$[N_{p^{\bot}},N_{p}]=[N_{p^{\bot}},\beta^{Weyl}]=0$ imply for any $\delta>0$
\begin{eqnarray*}
&& (1+N)^{\delta/2}\beta^{Weyl}\varrho^{\varepsilon}(1+N)^{\delta/2}
=
ABA'RC
\\
\text{with}
&&
A=(1+N)^{\delta/2}(1+N_{p})^{-\delta/2}(1+N_{p^{\bot}})^{-\delta/2}\\
&&
B= (1+N_{p})^{\delta/2}\beta^{Weyl}(1+N_{p})^{-\delta/2-s}\\
&&
A'=(1+N_{p})^{\delta/2+s}(1+N_{p^{\bot}})^{\delta/2}(1+N)^{-\delta-s}\\
&& R=(1+N)^{\delta+s}\varrho^{\varepsilon}(1+N)^{\delta+s}
\quad\text{and}\quad
C=(1+N)^{-\delta/2-s}\,.
\end{eqnarray*}
The factors $A$, $A'$ and $C$ are uniformly
bounded operators when $\delta>0$ (and $s$) is fixed. The trace class norm of
the factor $R$ is uniformly bounded by $C_{\delta+s}$. Finally the
Weyl pseudodifferential calculus on $p\Z$ implies that $B=\gamma^{Weyl}$
with $\gamma(\varepsilon)$ uniformly bounded in $S_{p\Z}(1,g_{p})$ and
therefore $\left|B\right|_{\L(\H)}\leq C_{\delta,s}'$ uniformly w.r.t
$\varepsilon\in (0,\overline{\varepsilon})$.\\
Hence the family $(\beta^{Weyl}\varrho^{\varepsilon})_{\varepsilon\in
  (0,\overline{\varepsilon})}$ satisfies the assumptions of
Def.~\ref{de.Msc}. Let $\mu_{1}$ belong to
$\mathcal{M}(\beta^{Weyl}\varrho^{\varepsilon})$. After extracting the proper
sequence $(\varepsilon_{n})_{n\in\nz}$ such that
$\lim_{n\to\infty}\varepsilon_{n}=0$, one can assume
\begin{eqnarray*}
&&
\lim_{n\to\infty}\Tr\left[b^{Weyl}\beta^{Weyl}\varrho^{\varepsilon_{n}}\right]=\int_{\Z}b(z)~d\mu_{1}(z)\\
\text{and}
&&
\lim_{n\to\infty}\Tr\left[b^{Weyl}\varrho^{\varepsilon_{n}}\right]=\int_{\Z}b(z)~d\mu(z)
\end{eqnarray*}
for any $b\in \mathcal{S}_{cyl}(\Z)$. But the finite dimensional
pseudodifferential calculus implies
$b^{Weyl}\beta^{Weyl}=(b\beta)^{Weyl}+O_{\L(\H)}(\varepsilon_{n})$
with $b\beta\in \mathcal{S}_{cyl}(\Z)$. This implies
$$
\int_{\Z}b(z)~d\mu_{1}(z)=\int_{\Z}b(z)\beta(z)~d\mu(z)
$$
for all $b\in \mathcal{S}_{cyl}(\Z)$. According to
Proposition~\ref{pr.exdense} this implies $\mu_{1}=\beta\mu$.\\
\noindent 2) Since the $\ccup_{p\in\p,s\in\rz}S_{p\Z}(\langle
z\rangle^{s}, g_{p})$ contains
$\ccup_{p\in\P}\left(\oplus_{m,q\in\nz}^{\rm alg}\P_{m,q}(p\Z)\right)$, the
result is proved for any polynomial symbol $b\in
\P_{m,q}^{\infty}(\Z)$ such that $\tilde{b}=\Gamma(p)\tilde{b}\Gamma(p)$ for some finite dimensional
projector $p\in\p$. Consider now a general $b\in
\P_{m,q}^{\infty}(\Z)$ with $m,q\in\nz$. By Lemma~\ref{wick-estimate},
the operator
$$
(1+N)^{\delta/2}b^{Wick}(1+N)^{-\delta/2-m/2-q/2}
$$
is uniformly bounded for any $\delta>0$.  Since the trace class norm of
$(1+N)^{\frac{\delta+m+q}{2}}\varrho^{\varepsilon}(1+N)^{\frac{\delta+m+q}{2}}$
is uniformly bounded w.r.t $\varepsilon\in
(0,\overline{\varepsilon})$,
the family $(\beta^{Wick}\varrho^{\varepsilon})$ satisfies the
assumptions of Definition~\ref{de.Msc}. Introduce now an increasing
sequence  $(p_{\ell})_{\ell\in\nz}$ of $\p$ such that
$\sup_{\ell\in\nz}p_{\ell}=I$ and consider for $\ell\in \nz$
$$
\beta_{\ell}(z)=\beta(p_{\ell}z)\quad,\quad
\tilde{\beta}_{\ell}=p_{\ell}^{\otimes
  q}\circ\tilde{b}\circ p_{\ell}^{\otimes m}\,.
$$
Since $\tilde{\beta}$ is a compact operator, the finite rank operator
$\tilde{\beta}_{\ell}$ converges to $\tilde{\beta}$ in the norm
topology in $\L(\bigvee^{m}\Z,\bigvee^{q}\Z)$. The uniform
estimates
\begin{eqnarray*}
  &&
\left|(\beta-\beta_{\ell})^{Wick}(1+N)^{-m/2-q/2}\right|_{\L(\H)}\leq
C\left|\tilde{\beta}-\tilde{\ell}\right|_{\L(\bigvee^{m}\Z,\bigvee^{q}\Z)}\,,
\\
&&
\left(1+\left|z\right|^{2}\right)^{m/2+q/2}
\left(\left|\beta(z)\right|+\left|\beta_{\ell}(z)\right|\right) \leq
C \quad\text{with}\quad \lim_{\ell\to \infty}\beta_{\ell}(z)=\beta(z)\,,
\end{eqnarray*}
and the convergence
\begin{eqnarray*}
\forall b\in \mathcal{S}_{cyl}(\Z),\quad
\lim_{n\to\infty}\Tr\left[b^{Weyl}\beta_{\ell}^{Wick}\varrho^{\varepsilon_{n}}\right]
=\int_{\Z}b(z)\beta_{\ell}(z)~d\mu(z)
\end{eqnarray*}
 after extracting a sequence $(\varepsilon_{n})_{n\in \nz}$,
 $\lim_{n\to\infty}\varepsilon_{n}=0$, with
 $\int_{\Z}(1+|z|^{2})^{m/2+q/2}~d\mu(z)<+\infty$,  lead to
$$
\forall b\in \mathcal{S}_{cyl}(\Z),\quad
\lim_{n\to\infty}\Tr\left[b^{Weyl}\beta^{Wick}\varrho^{\varepsilon_{n}}\right]
=\int_{\Z}b(z)\beta(z)~d\mu(z)\,.
$$
\fin

The previous results provide the behaviour of $\lim_{\varepsilon\to
  0}\Tr\left[\beta^{Wick}\varrho^{\varepsilon}\right]$
for $\beta\in \oplus_{m,q\in\nz}^{\rm alg}\P_{m,q}^{\infty}(\Z)$ when
$\mathcal{M}(\varrho^{\varepsilon})=\left\{\mu\right\}$.
The next result checks the other way.
\begin{prop}
\label{pr.oppway}
Assume that the family $(\rho^\hbarr)_{\hbarr\in(0,\bar\hbarr)}$
 satisfies (\ref{hyptraclassop}) and that for any $C>0$  there exist $K_{C}>0$ such that
\begin{eqnarray*}
\sum_{k=0}^\infty \frac{C^k}{[k/2]!} \, \Tr[N^k \rho^\hbarr]\leq K_{C}<\infty
\end{eqnarray*}
holds uniformly w.r.t $\varepsilon\in (0,\overline{\varepsilon})$.
Assume that there exists a Borel measure $\mu$ such that
$$
\lim_{\varepsilon\to 0}\Tr\left[b^{Wick}\varrho^{\varepsilon}\right]=\int_{\Z} b(z)~d\mu(z)
$$
holds for any $b\in \oplus_{m,q}^{\rm alg}\P_{m,q}^{\infty}(\Z)$. This
implies
$$
\mathcal{M}(\varrho^{\varepsilon})=\left\{\mu\right\}\,.
$$
\end{prop}
\proof It is enough to prove the following statement:
\begin{eqnarray*}
\lim_{ \hbarr\to 0 }
{\rm Tr}[W(\xi)\rho^\hbarr]=\int_{\Z} e^{\sqrt{2} i S(\xi,z)}  d\mu.
\end{eqnarray*}
It is done when the right-hand side of
\begin{eqnarray}
\label{trace}
{\rm Tr}[W(\xi)\rho^\hbarr]&=&
\sum_{n=0}^\infty \frac{|\sqrt{\hbarr}\xi|^n}{2^n n!} \;\; {\rm Tr}\left[
\; h_n\left(\frac{i\sqrt{2}S(\xi,z)}{|\sqrt{\hbarr}\xi|}\right)^{Wick}
\rho^\hbarr\right]
\end{eqnarray}
is proved to be an absolutely convergent series,
 uniformly w.r.t. $\hbarr\in(0,\bar\hbarr)$.
With
\begin{eqnarray}
\nonumber
\Tr[W(\xi) \rho^\hbarr]&=& \lim_{M\to\infty} \Tr[W(\xi) 1_{[0,M]}(N)\, \rho^\hbarr]\\
\label{traceeq1}
&=& \lim_{M\to\infty} \sum_{n=0}^{\infty} \frac{|\sqrt{\hbarr}\xi|^n}{2^n n!} \;\; {\rm Tr}\left[
\; h_n\left(\frac{i\sqrt{2}S(\xi,z)}{|\sqrt{\hbarr}\xi|}\right)^{Wick} \; 1_{[0,M]}(N)\,
\rho^\hbarr\right]
\end{eqnarray}
and
\begin{eqnarray*}
\left|\Tr\left[h_n\left(\frac{i\sqrt{2}S(\xi,z)}{|\sqrt{\hbarr}\xi|}\right)^{Wick}
 1_{[0,M]}(N)\,\rho^\hbarr\right]\right|\leq  M_n \left|(N+1)^{-n/2}
\, h_n\left(\frac{i\sqrt{2}S(\xi,z)}{|\sqrt{\hbarr}\xi|}\right)^{Wick}
(N+1)^{-n/2}\right|_{\mathcal{L}(\mathcal{H})},
\end{eqnarray*}
with $M_{n}=\Tr\left[(1+N)^{n}\varrho^{\varepsilon}\right]$,
Lemma \ref{weyl-estimate} implies
\begin{eqnarray*}
\left|(N+1)^{-n/2} \; h_n\left(\frac{i\sqrt{2}S(\xi,z)}{|\sqrt{\hbarr}\xi|}\right)^{Wick}
(N+1)^{-n/2}\right|_{\mathcal{L}(\mathcal{H})}
&\leq& \sup_{k,j\in\mathbb{N}} \frac{(1+2\sqrt{2 (k+j)\varepsilon })^n}{(k\hbarr+1)^{n/2}
(j\hbarr+1)^{n/2}} \; \frac{n!}{[n/2]!}\\
&\leq& 8^n \frac{n!}{[n/2]!}.
\end{eqnarray*}
This leads to
 \begin{eqnarray}
 \label{traestimate} \sum_{n=0}^\infty
\frac{|\sqrt{\hbarr}\xi|^n}{2^n n!} \;\left|\Tr[ h_n\left(\frac{i\sqrt{2}
S(\xi,z)}{|\sqrt{\hbarr}\xi|}\right)^{Wick} \;1_{[0,M]}(N)\,\rho^\hbarr]\right|
\label{eq.6}
&\leq& \sum_{n=0}^\infty \frac{(4\sqrt{\overline{\varepsilon}}|\xi|)^n}{[n/2]!}
M_n  <\infty\,\end{eqnarray}
uniformly w.r.t. $\hbarr\in(0,\bar\hbarr)$ and $M>0$. Hence we can take the limit
$M\to \infty$ inside in all the terms of (\ref{traceeq1}). This leads
to (\ref{trace}) with a uniformly
absolutely  convergent series in the right-hand side according to
\eqref{traestimate} and our initial assumption.

Thus the sum and the limit as $\varepsilon\to 0$ can be interchanged
in (\ref{trace}):
\begin{eqnarray*} \lim_{\hbarr\to 0} {\rm Tr}[
W(\xi)\rho^\hbarr]&=& \sum_{n=0}^\infty
\frac{|\xi|^n}{2^n n!} \;\; \lim_{\hbarr\to 0}
{\rm Tr}[\sqrt{\hbarr^n} \,h_n\left(\frac{i\sqrt{2}
S(\xi,z)}{|\sqrt{\hbarr}\xi|}\right)^{Wick}\rho^\hbarr]\\
&=& \sum_{n=0}^\infty
\frac{1}{n!} \;\; \int_\Z (i\sqrt{2}
S(\xi,z))^n \, d\mu\\
&=& \int_\Z e^{\sqrt{2} i S(\xi,z)} d\mu.
\end{eqnarray*}
The last equality follows owing to the dominated convergence theorem and
\begin{eqnarray*}
\int_\Z e^{\delta |1_{p\Z}z|^2} d\mu =\lim_{\hbarr\to 0} \sum_{k=0}^\infty \frac{\delta^k}{k!}
\Tr[\rho^\hbarr \, d\Gamma(1_{p\Z})^k ]<\infty,
\end{eqnarray*}
for any  $\delta >0$ and any $p\in\mathbb{P}$. This completes the proof.
\hfill\fin

\section{Examples and applications of Wigner measures}
\label{se.exampleappli}
\subsection{Finite dimensional cases}

The first examples are given by Theorem~\ref{th.prodcoh}
\begin{enumerate}
  \item For any $z\in \Z$ the family of operators
    $\varrho^{\varepsilon}=|E(z)\rangle\langle E(z)|$  has a unique
    Wigner measure
$$
\mathcal{M}(|E(z)\rangle\langle E(z)|\,,\quad \varepsilon\in (0,\overline{\varepsilon}))=\left\{\delta_{z}\right\}\,.
$$
\item For any $z\in \Z$ and any $m\in \Z$ the family of operators
  $\varrho^{\varepsilon}=|z^{\otimes k_{\varepsilon}-m}\rangle\langle
  z^{\otimes k_{\varepsilon}}|$ with $\left|z\right|=1$ and
  $\lim_{\varepsilon\to 0}\varepsilon k_{\varepsilon}=1$ has a unique
  Wigner measure
$$
\mathcal{M}(|z^{\otimes k_{\varepsilon}-m}\rangle\langle
z^{\otimes k_{\varepsilon}}|\,,\quad \varepsilon\in
(0,\overline{\varepsilon}))=\frac{1}{2\pi}\int_{0}^{2\pi}e^{-im \theta}\delta_{e^{i\theta}z}~d\theta\,.
$$
\item In case 1) and 2) the convergence can be tested with Weyl,
  Anti-Wick of Wick observables according to Proposition~\ref{pr.mucp}
  and Theorem~\ref{th.wigwick}.
\end{enumerate}
Beside the explicit calculation of Theorem~\ref{th.prodcoh}
these results can be considered  through an inductive approach
since $E(z)$ or
$z^{\otimes n}$ lie in $\Gamma_{s}(\cz z)$. The natural extension
comes from Proposition~\ref{pr.exdense}-1) with a proper choice of
the first term in the increasing sequence $(p_{\ell})_{\ell\in \nz}$.
\begin{prop}
 \label{pr.findim}
Assume that the family $(\varrho^{\varepsilon})_{\varepsilon\in
  (0,\overline{\varepsilon})}$ satisfies the assumptions of
Definition~\ref{de.Msc}.
Assume further that there exists a finite dimensional space $p_{0}\in
\p$ such that
$$
\varrho^{\varepsilon}
=\Gamma(p_{0})\varrho\Gamma(p_{0})
=\varrho_{p_{0}}^{\varepsilon}
\otimes
|\Omega\rangle \langle \Omega|
$$
for all $\varepsilon\in (0,\overline{\varepsilon})$ with
$\varrho^{\varepsilon}_{p_{0}}\in \L^{1}(\Gamma_{s}(p_{0}\Z))$. Then
the Wigner measures of $(\varrho^{\varepsilon})_{\varepsilon\in
  (0,\overline{\varepsilon})}$ are given by
$$
\mathcal{M}\left(\varrho^{\varepsilon}\right)=\left\{\mu_{1}\otimes
  \delta_{0,p_{0}^{\bot}\Z}\,,\quad  \mu_{1}\in \mathcal{M}(\varrho_{p_{0}}^{\varepsilon})\right\}\,.
$$
\end{prop}

\subsection{Superpositions}

Two kinds of superpositions can be considered : 1) convex or linear
combination of trace class operators; 2) convex or linear
combination of wave functions.
The first one is the simplest.
\begin{prop}
\begin{enumerate}
  \item   Let $(M, \pi)$ be a probability space. Les
  $\left(\varrho^{\varepsilon}(m)\right)_{\varepsilon\in
    (0,\overline{\varepsilon}), m\in M}$ be a family of operators such
that
$$
\left|(1+N)^{\delta/2}\varrho^{\varepsilon}(m)(1+N)^{\delta/2}\right|_{\L^{1}(\H)}\leq C_{\delta}(m)
$$
for $\pi$-almost every $m\in M$ with  $C_{\delta}\in
L^{1}(M,d\pi)$ for some $\delta>0$. Assume further
$\mathcal{M}(\varrho^{\varepsilon}(m),\ \varepsilon\in
(0,\overline{\varepsilon}))=\left\{\mu(m)\right\}$
 for $\pi$-almost every $m\in M$, then the family
$(\int_{M}\varrho^{\varepsilon}(m)~d\pi(m))_{\varepsilon\in \in
  (0,\overline{\varepsilon})}$
satisfies the assumptions of Definition~\ref{de.Msc} and
$$
\mathcal{M}\left(\int_{M}\varrho^{\varepsilon}(m)~d\pi(m)\,,\quad
  \varepsilon\in (0,\overline{\varepsilon})\right)=
\left\{\int_{M}\mu(m)~d\pi(m)\right\}\,.
$$
\item Any bounded Borel measure on $\Z$ can be achieved as a Wigner
  measure.
\end{enumerate}
\end{prop}
\proof 1) Set $\varrho^{\varepsilon}=\int_{M}\varrho^{\varepsilon}(m)~
d\pi(m)$ and write
$$
\left|(1+N)^{\delta/2}\varrho^{\varepsilon}(1+N)^{\delta/2}\right|_{\L^{1}(\H)}
\leq \int_{M}C_{\delta}(m)~d\pi(m)\,.
$$
Then apply Lebesgue's convergence theorem to
$$
\Tr\left[b^{Weyl}\varrho^{\varepsilon}\right]=\int_{M}\Tr\left[b^{Weyl}\varrho^{\varepsilon}(m)\right]~d\pi(m)\,.
$$
\noindent 2) After reducing the problem to the case when $\mu$ is a
Borel probability measure on $\Z$, apply 1) with $M=\Z$, $\pi=\mu$, $m=z$ and
$\varrho^{\varepsilon}(z)=|E(z)\rangle\langle E(z)|$.
\fin

The second type of superposition requires an orthogonality
property. It is given by Proposition~\ref{pr.orth2}. Here are a few
examples
\begin{enumerate}
\item Take $u_{\ell}^{\varepsilon}=E(z_{\ell})$ for $\ell=1,\ldots, L$,
  with $L\in \nz$ fixed, and set
  $u^{\varepsilon}=L^{-1/2}\sum_{\ell=1}^{L}u_{\ell}^{\varepsilon}$.
When the $z_{\ell}$ are distinct, the family
$\left(|u^{\varepsilon}\rangle\langle
  u^{\varepsilon}|\right)_{\varepsilon\in (0,\overline{\varepsilon})}$
has a unique Wigner measure
$$
\mathcal{M}(|u^{\varepsilon}\rangle\langle
  u^{\varepsilon}|)=
\left\{L^{-1}\sum_{\ell=1}^{L}\delta_{z_{\ell}}\right\}\,.
$$
\item Take for any $\ell\in \left\{1,\ldots,L\right\}$,
$u_{\ell}^{\varepsilon}=z_{\ell}^{\otimes k_{\varepsilon}}$
with $|z_{\ell}|=1$ and $\ds\lim_{\varepsilon\to 0}\varepsilon k_{\varepsilon}=1$.
The family
$\left(|u^{\varepsilon}\rangle\langle
  u^{\varepsilon}|\right)_{\varepsilon\in (0,\overline{\varepsilon})}$
has a unique Wigner measure:
$$
\mathcal{M}(|u^{\varepsilon}\rangle\langle
  u^{\varepsilon}|)=
\left\{(2\pi
  L)^{-1}\sum_{\ell=1}^{L}\int_{0}^{2\pi}\delta_{e^{i\theta}z_{\ell}}~d\theta
\right\}\,.
$$
\item For $z\in \Z$ and  $u^{\varepsilon}=\frac{E(z)+|z^{\otimes k_{\varepsilon}}\rangle}{\sqrt{2}}$
 with $|z|=1$ and  $\lim_{\varepsilon\to
    0}\varepsilon k_{\varepsilon}=1$,
the family $\left(|u^{\varepsilon}\rangle\langle
  u^{\varepsilon}|\right)_{\varepsilon\in (0,\overline{\varepsilon})}$
has a unique Wigner measure:
$$
\mathcal{M}(|u^{\varepsilon}\rangle\langle
  u^{\varepsilon}|)=
\left\{\frac{1}{2}\delta_{z}+ \frac{1}{4\pi}
 \int_{0}^{2\pi}\delta_{e^{i\theta}z}~d\theta
\right\}\,.
$$
\item All this examples can be tested with Weyl, Anti-Wick or Wick
  observables
according to Proposition~\ref{pr.mucp} and Theorem~\ref{th.wigwick}.
\end{enumerate}

\subsection{Propagation of chaos and propagation of (squeezed) coherent states}
\label{se.propa}
Let us go back to the example of
Section~\ref{se.anex} where
$U_{\varepsilon}(t)=e^{-i\frac{t}{\varepsilon}H_{\varepsilon}}$
with $H_{\varepsilon}=d\Gamma(-\Delta) +Q^{Wick}$,
$\tilde{Q}=\frac{1}{2}V(x_{1}-x_{2})$
and $z_{t}$ solution to $i\partial_{t}z_{t}=
-\Delta z+(V*|z_{t}|^{2})z_{t}$
Theorem~\ref{th.main2.2}, Proposition~\ref{pr.cohwick} and
Proposition~\ref{pr.oppway}
imply:
\begin{enumerate}
\item For any $z_{0}\in \Z$ with $|z_{0}|=1$, the family
  $(|U_{\varepsilon}(t)z_{0}^{\otimes k_{\varepsilon}}\rangle\langle
  U_{\varepsilon}(t)z_{0}^{\otimes k_{\varepsilon}}|)_{\varepsilon\in
    (0,\overline{\varepsilon})}$ with $\lim_{\varepsilon\to
    0}\varepsilon k_{\varepsilon}=1$ is pure with
$$
\mathcal{M}\left(|U_{\varepsilon}(t)z_{0}^{\otimes k_{\varepsilon}}\rangle\langle
  U_{\varepsilon}(t)
z_{0}^{\otimes k_{\varepsilon}}|\right)=
\left\{\frac{1}{2\pi}\int_{0}^{2\pi}\delta_{e^{i\theta}z_{t}}~d\theta\right\}
=\mathcal{M}\left(|z_{t}^{\otimes k_{\varepsilon}}\rangle\langle
z_{t}^{\otimes k_{\varepsilon}}|\right)
$$
\item For any $z_{0}\in \Z$, the family
  $(|U_{\varepsilon}(t)E(z_{0})\rangle\langle
  U_{\varepsilon}(t)E(z_{0})|)_{\varepsilon\in
    (0,\overline{\varepsilon})}$ is pure
with
$$
\mathcal{M}\left(|U_{\varepsilon}(t)E(z_{0})\rangle\langle
  U_{\varepsilon}(t)E(z_{0})|\right)
=
\left\{\delta_{z_{t}}\right\}
=
\mathcal{M}\left(|E(z_{t})\rangle\langle
E(z_{t})|\right)\,.
$$
\end{enumerate}
These results are derived from the results for product states after
testing with Wick observable (any $b\in \oplus_{m,q}^{\rm alg}\P_{m,q}(\Z)$)\,.
Actually it is possible to recover the second one directly from the Hepp
method. For any $b\in \mathcal{S}_{cyl}(\Z)$,
Proposition~\ref{pr.hepp} implies
$$
\lim_{\varepsilon\to 0}\Tr\left[b^{Weyl}\left(|U_{\varepsilon}(t)E(z_{0})\rangle\langle
  U_{\varepsilon}(t)E(z_{0})|-
|W(\frac{\sqrt{2}}{i\varepsilon}z_{t})U_{2}(t,0)\Omega\rangle\langle
W(\frac{\sqrt{2}}{i\varepsilon}z_{t})U_{2}(t,0)\Omega|
\right)\right]=0\,.
$$
By the finite dimensional Weyl quantization, the second term equals
$$
\langle U_{2}(t,0)\Omega\,, b(.-z_{t})^{Weyl}U_{2}(t,0)\Omega\rangle\,.
$$
And it suffices to check that the family
$(|U_{2}(t,0)\Omega\rangle\langle U_{2}(t,0)\Omega|)_{\varepsilon\in
  (0,\overline{\varepsilon})}$
admits the unique Wigner measure $\delta_{0}$. This is a consequence
of Lemma~\ref{le.bogo} which first says
$|N^{k}U_{2}(t,0)\Omega|_{\H}\leq C_{k}$ for any $k\geq 0$
and then $\lim_{\varepsilon\to 0}\langle
U_{2}(t,0)\Omega\,,\,b^{Wick}U_{2}(t,0)\Omega\rangle=0$
when $b(0)=0$\,.

\subsection{Dimensional defect of compactness}
\label{se.dimdefcomp}
In the last example the mean field propagation of Wigner measure
attached with $U_{\varepsilon}(t)E(z_{0})$ can be proved directly
without using the result on Wick observables. As a corollary, this
provides the result for Wick observables $b^{Wick}$ when $b\in
\oplus_{m,q}^{\rm alg}\P_{m,q}^{\infty}(\Z)$
according to Theorem~\ref{th.wigwick}. The result for a general
$b\in \oplus_{m,q}^{\rm alg}\P_{m,q}(\Z)$ is still true but comes from a direct
proof or from Proposition~\ref{pr.cohwick}.\\
A natural question is whether the result of  Theorem~\ref{th.wigwick}
can be extended to any
observable $b^{Wick}$ with $b\in \oplus_{m,q}^{\rm alg}\P_{m,q}(\Z)$. The answer
is no, because in the infinite dimensional case there can be some
defect of compactness w.r.t to the dimension variable.
\\
Here is a typical example.
Consider a family $(z_{\varepsilon})_{\varepsilon\in
  (0,\overline{\varepsilon})}$ such that $z_{\varepsilon}$ converges
\underline{weakly} to $0$. There exists a constant $C>0$ such that
$\left|z_{\varepsilon}\right|\leq C$ for all $\varepsilon\in
(0,\overline{\varepsilon})$
and the family $(E(z_{\varepsilon}))_{\varepsilon\in
  (0,\overline{\varepsilon})}$ satisfies the assumptions of
Proposition~\ref{pr.oppway}. The Wigner measures $\mu\in
\mathcal{M}(|E(z_{\varepsilon}\rangle\langle E(z_{\varepsilon})|))$
are determined by testing on any $b\in \P_{m,q}^{\infty}(\Z)$. But
Theorem~\ref{th.prodcoh} says
$$
\left\langle
  E(z_{\varepsilon})\,,b^{Wick}E(z_{\varepsilon})\right\rangle=b(z_{\varepsilon})=
\langle z_{\varepsilon}^{\otimes q}\,,\tilde{b}
z_{\varepsilon}^{\otimes m}\rangle\,.
$$
When $m+q\geq 1$ the operator $\tilde{b}$ is compact, the right-hand side
converges to $0$  as $\varepsilon\to 0$. According to
Proposition~\ref{pr.oppway} this implies
$$
\mathcal{M}\left(|E(z_{\varepsilon})\rangle\langle E(z_{\varepsilon})|\right)=\left\{\delta_{0}\right\}\,.
$$
Meanwhile testing with $N=d\Gamma(I)=\left(|z|^{2}\right)^{Wick}$
implies
$$
\left\langle E(z_{\varepsilon})\,, N E(z_{\varepsilon})\right\rangle=\left|z_{\varepsilon}\right|^{2}
$$
where the right-hand side can reach any possible limit in $[0,C]$.

\subsection{Bose-Einstein condensates}
\label{se.Bocon}

The thermodynamic limit of the ideal Bose Gas presented within a
local algebra presentation in \cite{BrRo2} can be reconsidered by
introducing a small parameter $\varepsilon\to 0$.
Namely, the  large domain limit where bosonic particles are
moving freely in a domain $\Lambda$, with volume
$\left|\Lambda\right|\to \infty$, can be formulated with
$\left|\Lambda\right|=\frac{1}{\varepsilon}$ and $\varepsilon\to 0$.
For a fixed particle density the total number of particle is
$O(\frac{1}{\varepsilon})$ coherent with a mean field approach.
Before considering any dynamical problem, Wigner measures of
$\varepsilon$-dependent Gibbs states bring some interesting
presentation of the Bose-Einstein condensation.

Consider the Laplace operator $H_{0}=-\Delta_{x}$ on the $\varepsilon$-dependent torus
$\rz^{d}/(\varepsilon^{-1/d}\zz)^{d}$ with spectrum
$\sigma(H_{0})=\left\{\varepsilon^{2/d}|2\pi n|, n\in
  \zz^{d}\right\}$. The one particle space is
$\Z^{\varepsilon}=L^{2}(\rz^{d}/(\varepsilon^{-1/d}\zz)^{d})$ and the bosonic Fock
space is $\mathcal{H}^{\varepsilon}=\Gamma_{s}(\Z^{\varepsilon})$.
 For the inverse temperature $\beta=\frac{1}{k_{B}T}>0$ and a
chemical potential $\mu$, the Gibbs grand canonical equilibrium
state is associated with the operator $e^{-\beta d\Gamma(H_{0}-\mu
  I)}=\Gamma(e^{-\beta(H_{0}-\mu I)})$,
which is trace class if and only if $\mu<0$ (see
\cite[Proposition~5.2.27]{BrRo2}). This Gibbs state on
$\mathcal{L}(\mathcal{H}^{\varepsilon})$ is given  by
$$
\omega_{\varepsilon}(A)=\Tr\left[\varrho_{\varepsilon}A\right]
\quad\text{with}\quad
\varrho_{\varepsilon}=\frac{1}{\Tr\left[\Gamma(e^{-\beta(H_{0}-\mu)})\right]}\Gamma(e^{-\beta(H_{0}-\mu)})\,,
\quad \mu<0\,.
$$
It is convenient to introduce the parameter $z=e^{\beta\mu}$ and this
Gibbs state restricted to the CCR-algebra (the $C^{*}$-algebra generated by the Weyl
operators $W_{1}(f)$, $f\in \Z^{\varepsilon}$) is the gauge-invariant
quasi-free state given by the two-point function:
$\omega_{\varepsilon}(a_{1}^{*}(f)a_{1}(g))
=
\left\langle g\,,
  ze^{-\beta H_{0}}(1-ze^{-\beta H_{0}})^{-1} f
\right\rangle$\,. The index $_{1}$ means that the CCR are
written at this level in their initial form:
$[a_{1}(g),a_{1}^{*}(f)]=\langle g\,,\,f\rangle$.
This is proved in \cite[Proposition~5.2.28]{BrRo2} with the straightforward rewritting
$$
\omega_{\varepsilon}(W_{1}(f))
=
\exp\left[-\langle f\,,\,(1+ze^{-\beta
    H_{0}})(1-ze^{-\beta H_{0}})^{-1}f\rangle/4\right]
$$
The mean field analysis consists here in introducing
$a(f)=\varepsilon^{1/2}a_{1}(f)$ and
$W(f)=W_{1}(\varepsilon^{1/2}f)$:
\begin{eqnarray*}
&&  \omega_{\varepsilon}(a^{*}(f)a(g))=\varepsilon \langle g\,,\,
  ze^{-\beta H_{0}}(1-ze^{-\beta H_{0}})^{-1}f\rangle
\\
&&
\omega_{\varepsilon}(W(f))=\exp\left[-\varepsilon\langle f\,,\,(1+ze^{-\beta
    H_{0}})(1-ze^{-\beta H_{0}})^{-1}f\rangle/4\right]\,.
\end{eqnarray*}
 Further a rescaling
 motivated by the observation of the phenomena on a
large scale, is implemented with
$f(x)=\varepsilon^{1/2}\varphi(\varepsilon^{1/d}x)=D_{\varepsilon}\varphi$. After
conjugating with the
unitary transform
$\Gamma(D_{\varepsilon}):\mathcal{H}=\Gamma_{s}(\Z)\to
\mathcal{H}^{\varepsilon}=\Gamma_{s}(\Z^{\varepsilon})$, with
$\Z=L^{2}(\rz^{d}/\zz^{d})$ we are led to consider the asymptotic
behaviour  as $\varepsilon\to 0$ of the normal state
$$
\varrho^{\varepsilon}=\Gamma(D_{\varepsilon})^{*}\varrho_{\varepsilon}\Gamma(D_{\varepsilon})
=\frac{1}{\Tr\left[\Gamma(e^{-\beta(-\varepsilon^{2/d}\Delta-\mu)})\right]}
\Gamma(e^{-\beta(-\varepsilon^{2/d}\Delta-\mu)})
$$
which satisfies
\begin{eqnarray*}
\Tr\left[\varrho^{\varepsilon}W(f)\right]
&=&
\exp\left[-\frac{\varepsilon}{4}\langle f\,,\,(1+ze^{\beta
    \varepsilon^{2/d}\Delta})(1-ze^{\beta \varepsilon^{2/d}\Delta})^{-1}f\rangle_{\Z}\right]
\\
&=&
e^{-\frac{\varepsilon}{4}\left|f\right|_{\Z}^{2}}\exp\left[-\frac{\varepsilon}{2}\langle f\,,\,ze^{\beta
    \varepsilon^{2/d}\Delta}(1-ze^{\beta
    \varepsilon^{2/d}\Delta})^{-1}f\rangle_{\Z}\right]
\\
\Tr\left[\varrho^{\varepsilon}a^{*}(f)a(g)\right]&=&\varepsilon\langle
g\,,\,ze^{\beta\varepsilon^{2/d}\Delta}(1-ze^{\beta\varepsilon^{2/d}\Delta})^{-1}f\rangle_{\Z}\,.
\end{eqnarray*}
The above expressions are explicit after the decomposition in the Fourier basis
$f=\sum_{n\in\zz^{d}}f_{n}e^{2i\pi n.z}$ of any element $f\in\Z$\,.
For a given $z<1$ and $\beta>0$ the rescaled  particle density is
given by
\begin{equation}
  \label{eq.densbos}
\frac{\varepsilon z}{1-z}+\varepsilon
\sum_{n\in\zz^{d}\setminus\left\{0\right\}}
\frac{ze^{-\beta\varepsilon^{2/d}|2\pi
  n|^{2}}}{(1-ze^{-\beta\varepsilon^{2/d}|2\pi
  n|^{2}})}=
\frac{\varepsilon z}{1-z}+\nu_{\varepsilon}(\beta,z)\,.
\end{equation}
One checks easily for $\varepsilon'\geq \varepsilon$ and $z'\leq z< 1$
\begin{eqnarray*}
  &&
\nu_{\varepsilon'}(\beta,z)\leq
\nu_{\varepsilon}(\beta,z)\stackrel{\varepsilon\to
  0}{\to}\nu_{0}(\beta,z)=\int_{\rz^{d}}\frac{ze^{-\beta|2\pi
    u|^{2}}}{1-ze^{-\beta |2\pi u|^{2}}}~du\\
\text{and}&&
\forall \varepsilon\in [0,1),
\quad \nu_{\varepsilon}(\beta,z)\geq \nu_{\varepsilon}(\beta,z')\,.
\end{eqnarray*}
Here comes the discussion about the Bose-Einstein condensation. In
dimension $d\geq 3$ (this restriction may change with an alternative
Hamiltonian  $H_{0}=\lambda(D_{x})$), the quantity
$$
\nu_{0}(\beta,1)=\int_{\rz^{d}}\frac{e^{-\beta|2\pi
    u|^{2}}}{1-e^{-\beta |2\pi u|^{2}}}~du <+\infty\,.
$$
is well defined.\\
We focus on the case $d\geq 3$.\\
The previous discussion imply
$$
\forall \varepsilon>0, \forall z\in (0,1),\quad
\nu_{\varepsilon}(\beta,z)\leq \nu_{0}(\beta,1)
$$
while any total density can be achieved by \eqref{eq.densbos}.
The Bose-Einstein condensation occurs while considering the limit
$\varepsilon\to 0$ with the constraint
$\frac{z_{\varepsilon}\varepsilon}{1-z_{\varepsilon}}+\nu_{\varepsilon}(z_{\varepsilon},\beta)=\nu$
with $\beta>0$ and $\nu>0$ fixed. There are two possible cases:
\begin{description}
\item[$\bullet\nu\leq \nu_{0}(\beta,1)$:] Then $\lim_{\varepsilon\to
    0}z_{\varepsilon}=z<1$ and $\lim_{\varepsilon\to
    0}\frac{\varepsilon z_{\varepsilon}}{1-z_{\varepsilon}}=0$\,.
\item[$\bullet\nu > \nu_{0}(\beta,1)$:] The inequality
  $\nu-\nu_{0}(\beta,1)\leq \frac{\varepsilon
  z_{\varepsilon}}{1-\varepsilon}\leq \nu$  leads to
$z_{\varepsilon}=1-\frac{\varepsilon}{\nu-\nu_{0}(\beta,1)}+o(\varepsilon)$\,.
The proportion $1-\nu_{0}(\beta,1)/\nu$ of the gas lies in the ground
state  $n=0$ of the one-body Hamiltonian. This is the Bose-Einstein
condensation phenomenon.
\end{description}
It is interesting to reconsider this limit $\varepsilon\to 0$ with
 $\beta>0$ and $\nu>0$ fixed ($d\geq 3$)  within the Wigner measure point of
view. This is possible owing to the explicit formula
\begin{equation}
  \label{eq.explicit}
\Tr\left[\varrho^{\varepsilon}W(\sqrt{2}\pi f)\right]
=
e^{-\varepsilon \pi^{2}\left|f\right|_{\Z}^{2}}
\exp\left[-\varepsilon \pi^{2}
\sum_{n\in\zz^{d}}\left|f_{n}\right|^{2}\frac{z_{\varepsilon}
e^{-\beta\varepsilon^{2/d}\left|2\pi n
  \right|^{2}}}{(1-z_{\varepsilon}e^{-\beta\varepsilon^{2/d}|2\pi n|^{2}})}\right]\,,
\end{equation}
where $f=\sum_{n\in\zz^{d}}f_{n}e^{2i\pi n.x}$.
Remember that the charactistic function of Wigner measures are
determined after considering the limit $\varepsilon\to 0$ of the above
expression for any \underline{fixed} $f\in \Z$. Hence the problem
is reduced to the application of Lebesgue's theorem in the argument of
the exponential.\\
For any $n\neq 0$ the quantity
$\frac{z_{\varepsilon}
e^{-\beta\varepsilon^{2/d}\left|2\pi n
  \right|^{2}}}{(1-z_{\varepsilon}e^{-\beta\varepsilon^{2/d}|2\pi
  n|^{2}})}$ converges to $0$ as $\varepsilon\to 0$ because $d/2<1$
and $z_{\varepsilon}\leq 1$. Hence we get
$$
\lim_{\varepsilon\to 0}\Tr\left[\varrho^{\varepsilon}W(\sqrt{2}\pi f)\right]=
\lim_{\varepsilon\to 0}\exp\left[-\frac{\varepsilon \pi^{2}
    z_{\varepsilon}}{1-z_{\varepsilon}}\left|f_{0}\right|^{2}\right]\,.
$$
With the constraint $\frac{\varepsilon
  z_{\varepsilon}}{1-z_{\varepsilon}}\leq \nu <+\infty$, there are two possibilities
\begin{itemize}
  \item First $\lim_{\varepsilon \to 0}\frac{\varepsilon
  z_{\varepsilon}}{1-z_{\varepsilon}}= 0$ implies $\nu\leq \nu_{0}(\beta,1)$ and
$\mathcal{M}(\varrho^{\varepsilon})=\left\{\delta_{0}\right\}$.
   \item The second case $\lim_{\varepsilon\to
       0}\frac{\varepsilon z_{\varepsilon}}{1-z_{\varepsilon}}=\nu-\nu_{0}(\beta,1)>0$
     implies
$$
\lim_{\varepsilon\to 0}\Tr\left[\varrho^{\varepsilon}W(\sqrt{2}\pi f)\right]
=
e^{-\pi^{2}(\nu-\nu_{0}(\beta,
    1))\left|f_{0}\right|^{2}}=e^{-\pi^{2}(\nu-\nu_{0}(\beta,1))|\langle f,1\rangle|^{2}}\,.
$$
Hence the Wigner measure  of the family
$(\varrho^{\varepsilon})_{\varepsilon>0}$ equals $\gamma_{\nu}\times
\delta_{0}$ on $\Z=\cz 1\times \left\{1\right\}^{\perp}$
where $\gamma_{\nu}$ is the gaussian measure
\begin{eqnarray*}
  \gamma_{\nu}(z_{1})=\frac{e^{-\frac{\left|z_{1}\right|^{2}}{\nu-\nu_{0}(\beta,1)}}}{
(\pi(\nu-\nu_{0}(\beta,1))^{d/2}}\quad,\quad z_{1}\in \cz\,.
\end{eqnarray*}
\end{itemize}
Our  scaled observables  can
measure asymptotically only  the Bose-Einstein phase
in a non trivial way. The rest
of the state  provides the factor $\delta_{0}$.
 While testing with the observable
$(|z|^{2})^{Wick}=N$, the dimensional defect of compactness phenomenon
already illustrated in Subsection~\ref{se.dimdefcomp} occurs again:
only the density of the condensate remains.
\begin{remark}
  \begin{description}
  \item[i)] It is possible to consider various dispersion relations
    $H_{0}=\lambda(D_{x})$ and the discussion about the dimension may
    change. Other boundary conditions (here periodic boundary
    conditions are considered) and the discussion about the
    convergence of $\lim_{\varepsilon\to 0}z_{\varepsilon}=1$ may
    change a little bit. We refer the reader to
 \cite{BrRo2} for the case of Dirichlet boundary conditions.
\item[ii)] From \eqref{eq.explicit} it is possible to consider the
  limit for any fixed $f\in\Z$ as $\varepsilon\to 0$ with various
  behaviours of $z_{\varepsilon}$. This provides asymptotically
a weak distribution. But the uniform tightness assumption
$\Tr\left[\varrho^{\varepsilon}(1+N)^{\delta}\right]\leq C$ is not
satisfied. The scaling has to be  adapted differently to the
dimension $d=2$ or $d=1$ by taking care of the singularity
at the momentum $0$,  in order to allow a non trivial Wigner
measure in the
thermodynamic and mean field limit.
  \end{description}
\end{remark}

\subsection{Application 1: From the propagation of coherent states to
  the propagation of chaos via Wigner measures}
\label{se.appli1}
In the previous sections we showed how the propagation of (squeezed)
coherent states can be derived from the propagation of Hermite states
or directly via the Hepp method. The Hepp method is very flexible (see
\cite{GiVe} for example) and therefore it is interesting to know
whether a result for coherent states provides an information for
product states or more general states.
Here is a simple and abstract result which relies on some gauge
invariance argument.
\begin{thm}
\label{th.cohchaos}
Let $U_{\varepsilon}$ be a unitary operator on $\H$ possibly depending
on $\varepsilon\in (0,\overline{\varepsilon})$ which commutes
with the number operator $[N,U_{\varepsilon}]=0$.
Assume that for a given $z\in \Z$ such that $\left|z\right|=1$, there exists
$z_{U}\in \Z$ such that
$$
\mathcal{M}\left(|U_{\varepsilon}E(z)\rangle\langle
  U_{\varepsilon}E(z)|\right)=\left\{\delta_{z_{U}}\right\}\,.
$$
Then for any non negative function
$\varphi \in L^{1}(\rz, ds)$ such that
$\int_{\rz}\varphi(s)(1+|s|)^{\delta}~ds<\infty$ for some $\delta>0$ and
$\int_{\rz}\varphi(s)~ds=1$, the state
$$
\varrho^{\varepsilon}_{\varphi}=\sum_{n=0}^{\infty}\varepsilon^{1/2}
\varphi(\varepsilon^{1/2}(n-\varepsilon^{-1}))
|U_{\varepsilon}z^{\otimes n}\rangle\langle U_{\varepsilon}z^{\otimes n}|
$$
satisfies the conditions of Definition~\ref{de.Msc} and
$$
\mathcal{M}\left(\varrho^{\varepsilon}_{\varphi}\right)=
\frac{1}{2\pi}\int_{0}^{2\pi}\delta_{e^{i\theta}z_{U}}~d\theta\,.
$$
\end{thm}
\proof
Owing to the relation
$$
\Gamma(e^{-i\theta})b^{Weyl}\Gamma(e^{i\theta})=
e^{-i\theta N}b^{Weyl}e^{i\theta N}=b(e^{-i\theta}.)^{Weyl}\;.
$$
Our assumptions imply
$$
\mathcal{M}
\left(
\Gamma(e^{i\theta})|U_{\varepsilon}E(z)\rangle\langle U_{\varepsilon}
E(z)|\Gamma(e^{-i\theta})\right)=\delta_{e^{i\theta}z_{U}}
$$
for any $\theta\in \rz$. The assumptions of Definition~\ref{de.Msc}
are satisfied because $U_{\varepsilon}$ preserves the number. After
  taking the average w.r.t $\theta\in
[0,2\pi]$:
$$
\sigma^{\varepsilon}=\frac{1}{2\pi}\int_{0}^{2\pi}
\Gamma(e^{i\theta})|U_{\varepsilon}E(z)\rangle\langle U_{\varepsilon}E(z)|\Gamma(e^{-i\theta})~d\theta
$$
this implies
$$
\mathcal{M}
\left(
\sigma^{\varepsilon}
\right)=\frac{1}{2\pi}\int_{0}^{2\pi}\delta_{e^{i\theta}z_{U}}~d\theta
$$
where the right-side is an extremal point of the convex set of Borel probability
measure which are invariant after the natural action of $S^{1}$ on
$\Z$: $S^{1}\times \Z\ni(\gamma,z)\to \gamma z\in \Z$.

Again the commutation $[U_{\varepsilon},N]=0$ and the expression
\eqref{coherent-vect} for $E(z)$ imply
\begin{eqnarray*}
  \sigma^{\varepsilon}
&=&(2\pi)^{-1}\int_{0}^{2\pi}U_{\varepsilon}|\Gamma(e^{i\theta})E(z)\rangle\langle
  \Gamma(e^{i\theta})E(z)|U_{\varepsilon}^{*}~d\theta
\\
&=&(2\pi)^{-1}\int_{0}^{2\pi}
U_{\varepsilon}|E(e^{i\theta}z)\rangle\langle
  E(e^{i\theta}z)|
U_{\varepsilon}^{*}~d\theta
\\
&=&
\sum_{n=0}^{\infty} \frac{e^{-\frac{1}{\hbarr}}}{\hbarr^n n!}
|U_{\varepsilon} z^{\otimes n}\rangle\langle
 U_{\varepsilon}z^{\otimes n}|.
\end{eqnarray*}
For any $b\in \mathcal{S}_{cyl}(\Z)$, the quantity
$$
\sum_{n=0}^{\infty}\frac{e^{-\frac{1}{\varepsilon}}}{\varepsilon^{n}n!}\langle
U_{\varepsilon}z^{\otimes n}\,,\,
b^{Weyl}U_{\varepsilon}z^{\otimes}\rangle
=\Tr\left[b^{Weyl}\sigma^{\varepsilon}\right]
$$
converges as $\varepsilon\to 0$ to
$(2\pi)^{-1}\int_{0}^{2\pi}b(e^{i\theta}z_{U})~d\theta$\,.
By Lemma~\ref{le.app} this implies
$$
\forall b\in \mathcal{S}_{cyl}(\Z)\,,
\quad
\lim_{\varepsilon\to 0}
\int_{\rz} a_{[\varepsilon^{-1/2}
  s+\varepsilon^{-1}]}(\varepsilon^{-1}) \;
\frac{e^{-\frac{s^2}{2}}}{\sqrt{2\pi}}
=(2\pi)^{-1}\int_{0}^{2\pi}b(e^{i\theta}z_{U})~d\theta\,,
$$
where $[t]$ is the integer part of $t\in\rz$ and
$$
a_{n}(\varepsilon^{-1})=\langle U_{\varepsilon}z^{\otimes
  n}\,,b^{Weyl} U_{\varepsilon}z^{\otimes n}\rangle
\,.
$$
Call $\gamma$ the Gaussian measure
$e^{-\frac{s^{2}}{2}}\frac{ds}{\sqrt{2\pi}}$ on $\rz$.
For any finite subdivision $\mathcal{I}=\{I_{1}\,\ldots,I_{L}\}$
of $\rz=I_{1}\sqcup\ldots\sqcup I_{L}$ with intervals,  the states
$$
\sigma_{I_{\ell}}^{\varepsilon}=
(\gamma(I_{\ell}))^{-1}\int_{I_{\ell}}
|U_{\varepsilon}z^{\otimes[\varepsilon^{-1/2}s+\varepsilon^{-1}]}\rangle
\langle
U_{\varepsilon}z^{\otimes[\varepsilon^{-1/2}s+\varepsilon^{-1}]}|~d\gamma(s)
$$
satisfy the assumptions of Definition~\ref{de.Msc} with the gauge
invariance
$$
\Gamma(e^{i\theta})\sigma_{I_{\ell}}^{\varepsilon}\Gamma(e^{-i\theta})= \sigma_{I_{\ell}}^{\varepsilon}\,.
$$
Moreover the state
$$
\underline{\sigma}^{\varepsilon}=\int_{\rz}
|U_{\varepsilon}z^{\otimes[\varepsilon^{-1/2}s+\varepsilon^{-1}]}\rangle
\langle
U_{\varepsilon}z^{\otimes[\varepsilon^{-1/2}s+\varepsilon^{-1}]}|~d\gamma(s)
=\sum_{\ell=1}^{L}\gamma(I_{\ell})\sigma_{I_{\ell}}^{\varepsilon}
$$
is a finite barycenter of the $\sigma_{I_{\ell}}^{\varepsilon}$ with a
unique
Wigner measure
$(2\pi)\int_{0}^{2\pi}\delta_{e^{i\theta}z_{U}}~d\theta$.
Since $\mathcal{I}$ is finite (or countable),
from any sequence $(\sigma^{\hbarr_n}_{I_\ell})$ with
$\lim_{n\to\infty}\varepsilon_{n}=0$, one
can extract a
subsequence $(\varepsilon_{n_{k}})_{k\in\nz}$ such that
$$
\mathcal{M}(\sigma_{I_{\ell}}^{\varepsilon_{n_{k}}}, k\in\nz)=\left\{\nu_{\ell}\right\}\,.
$$
Since the measure $\mu_{U}$ is an
extremal point in the convex set of gauge invariant probability
measures, all the $\nu_{\ell}$ have to be identical to $\mu_{U}$.
Since this holds for any sequence $(\varepsilon_{n})_{n\in\nz}$, we
have proved for any interval $I=(\alpha,\beta)$ with $\alpha<\beta$,
$\mathcal{M}\left(\sigma_{I}^{\varepsilon},\varepsilon\in
  (0,\overline{\varepsilon})\right)=\left\{\mu_{U}\right\}$.\\
Now take $\psi\in L^{1}(\rz, \gamma)$ and consider the state
$$
\underline{\sigma}_{\psi}^{\varepsilon}=\int_{\rz}
|U_{\varepsilon}z^{\otimes[\varepsilon^{-1/2}s+\varepsilon^{-1}]}\rangle
\langle
U_{\varepsilon}z^{\otimes[\varepsilon^{-1/2}s+\varepsilon^{-1}]}|~d\gamma(s)
=\sum_{\ell=1}^{L}\gamma(I_{\ell})\sigma_{I_{\ell}}^{\varepsilon}\,.
$$
If there exists $\delta>0$ such that
$\int_{\rz}(1+|s|)^{\delta}\psi(s)~d\gamma(s)<+\infty$, the family
$(\underline{\sigma}_{\psi}^{\varepsilon})_{\varepsilon\in (0,\overline{\varepsilon})}$
satisfy the assumption of Definition~\ref{de.Msc}. Let
$(\varepsilon_{n})_{n\in\nz}$ be a sequence such that
$\mathcal{M}(\underline{\sigma}_{\psi}^{\varepsilon_{n}},
n\in\nz)=\left\{\nu\right\}$.
Fix $b\in \mathcal{S}_{cyl}(\Z)$. The function $\psi$ can be
approximated in $L^{1}(\rz,d\gamma)$ by $\psi_{c}\in
\mathcal{C}^{0}_{c}(\rz)$. After choosing a finite subdivision
$\mathcal{I}$ such that the diameter of any $I_{\ell}$ intersecting
the support of $\psi_{c}$ is bounded by $\Delta$ one gets
$$
\left|\Tr\left[b^{Weyl}\underline{\sigma}_{\psi}^{\varepsilon_{n}}\right]
-\Tr\left[b^{Weyl}\sum_{\ell=0}^{L}\frac{\int_{I_{\ell}}
      \psi_{c}(t)~dt}{\gamma(I_{\ell})}
\sigma_{I_{\ell}}^{\varepsilon}\right]
\right|
\leq C_{b}\left[\omega(\psi_{c})\Delta
+
\left\|\psi-\psi_{c}\right\|_{L^{1}(\rz,\gamma)}
\right]
$$
where $\omega(\psi_{c})$ is the continuity modulus of
$\psi_{c}$. Hence
the right-hand side can be made arbitrarily small, uniformly with
respect to $\varepsilon_{n}$, while we know that the second term
of the left-hand side converges when $\psi_{c}$ and $\mathcal{I}$ are
fixed.
We have proved
$$
\int_{\Z}b(z)~d\nu(z)=\lim_{n\to\infty}\Tr\left[b^{Weyl}\varrho^{\varepsilon_{n}}\right]
=\int_{\Z}b(z)~d\mu_{U}(z)
$$
for any $b\in \mathcal{S}_{cyl}(\Z)$ and this proves $\nu=\mu_{U}$.
Since this holds for any $\nu\in
\mathcal{M}(\underline{\sigma}_{\psi}^{\varepsilon})$, we obtain
$$
\mathcal{M}(\underline{\sigma}_{\psi}^{\varepsilon})=\left\{\mu_{U}\right\}\,.
$$
The result for $\varrho_{\varphi}^{\varepsilon}$ comes from
$$
\left|
\varrho_{\varphi}^{\varepsilon}-\underline{\sigma}_{\psi}^{\varepsilon}
\right|_{\L^{1}(\H)}\leq \left|\varphi-\sum_{k\in\zz}
\varepsilon^{-1/2}\left(\int_{I_{k}^{\varepsilon}}\varphi(t)~dt\right)
1_{I_{k}^{\varepsilon}}\right|_{L^{1}(\rz,ds)}\stackrel{\varepsilon\to
0}\to 0
$$
with $I_{k}^{\varepsilon}=[\varepsilon^{1/2}k-\varepsilon^{-1/2},
\varepsilon^{1/2}(k+1)-\varepsilon^{-1/2}]$ and
$\psi(s)=\varphi(s)\sqrt{2\pi}
e^{\frac{s^{2}}{2}}$. The condition
$\int_{\rz}(1+|s|)^{\delta}$ $\varphi(s) ds<+\infty$
ensures that $\mathcal{M}(\varrho_{\varphi}^{\varepsilon})$ is well defined.
\fin

\subsection{Application 2: Propagation of correlated states}
\label{se.appli2}

This a simple application of the orthogonality of Wigner measures
combined with the results of Subsection~\ref{se.propa}.

  Let $H_{\varepsilon}=d\Gamma(-\Delta)+Q^{Wick}$ be the Hamiltonian
  studied in Section~\ref{se.anex} and let $z_{t}$ denote the solution
  to $i\partial_{t}z_{t}=-\Delta
  z_{t}+(V*\left|z_{t}\right|^{2})z_{t}$. The family of integers
$(k_{\varepsilon})_{\varepsilon\in (0,\overline{\varepsilon})}$ is
assumed to satisfy $\lim_{\varepsilon\to 0}\varepsilon k_{\varepsilon}=1$.
\begin{enumerate}
\item Let $z_{0,\ell}\in \Z$, $\ell=1,\ldots, L$, satisfy
  $\left|z_{0,\ell}\right|=1$ and set
$u^{\varepsilon}=L^{-1/2}\sum_{\ell=1}^{L}z_{0,\ell}^{\otimes
  k_{\varepsilon}}$,
$u^{\varepsilon}(t)=e^{-i\frac{t}{\varepsilon}H_{\varepsilon}}u_{\varepsilon}$.
At any time $t\in \rz$ the identity
$$
\mathcal{M}(|u_{\varepsilon}(t)\rangle\langle
u_{\varepsilon}(t)|)=\left\{(2\pi
  L)^{-1}\sum_{\ell=1}^{L}\int_{0}^{2\pi}\delta_{e^{i\theta}z_{t,\ell}}~d\theta\right\}
$$
as soon as $ z_{1,t},\ldots, z_{\ell,t}$ are linearly independent.
In particular this holds for any $t\in \rz$ when $L=2$ and $z_{0,1}$ and
$z_{0,2}$ are linearly independent.
\item Let $z_{0}\in \Z$ satisfy $\left|z_{0}\right|=1$ and set
$u^{\varepsilon}=2^{-1/2}z_{0}^{\otimes
  k_{\varepsilon}}+2^{-1/2}E(z_{0})$ and
$u^{\varepsilon}(t)=e^{-i\frac{t}{\varepsilon}H^{\varepsilon}}u_\hbarr$.
Then
$$
\mathcal{M}(|u^{\varepsilon}(t)\rangle\langle
u^{\varepsilon}(t)|)=\left\{\frac{1}{2}\delta_{z_{t}}+ \frac{1}{4\pi}
 \int_{0}^{2\pi}\delta_{e^{i\theta}z_{t}}~d\theta\
\right\}\,.
$$
\item Moreover the convergence can be tested with Weyl, Anti-Wick and
Wick operators according to Theorem~\ref{th.wig-measure} and Theorem~\ref{th.wigwick}\,.
\end{enumerate}

\appendix
\section{Normal approximation}
\label{se.app1}
We prove a technical lemma which is a slight adaptation of the normal approximation to the Poisson
distribution. Recall that for all $-\infty\leq \alpha<\beta\leq\infty$ we have the well known fact:
\begin{eqnarray}\ds
\label{poisson}
\lim_{\lambda\to\infty} \sum_{1+\frac{\alpha}
{\sqrt{\lambda}}\leq\frac{n}{\lambda}\leq 1+\frac{\beta}{\sqrt{\lambda}}}
\frac{\lambda^n}{n!} \; e^{-\lambda}=\int_\alpha^\beta \frac{e^{-\frac{s^2}{2}}}{\sqrt{2\pi}} \, ds.
\end{eqnarray}

\begin{lem}
\label{le.app}
Let $\{a_n(\lambda)\}_{n\in\mathbb{Z},\lambda>0}$ be a family of complex
numbers with $a_n(\lambda)=0$ if $n<0$. Assume that there exist $\mu\in \nz$ and $C_{\mu}>0$ such
that:
$$
\sup_{n\in\nz,\lambda>0}\left|a_{n}(\lambda)\right|\left\langle
\frac{n}{\lambda}\right\rangle^{-\mu} \leq C_{\mu}\,.
$$
Then the equality
\begin{eqnarray}
\label{unif-poisson}
\lim_{\lambda\to\infty} \sum_{n=0}^\infty \frac{\lambda^n}{n!} e^{-\lambda} \; a_n(\lambda)
=\lim_{\lambda\to\infty}\int_{\rz} a_{[\sqrt{\lambda}
  s+\lambda]}(\lambda) \; \frac{e^{-\frac{s^2}{2}}}{\sqrt{2\pi}}
\,ds\,.
\end{eqnarray}
holds whenever one of the two limits exists.
\end{lem}
\proof
Notice that both the series and the integral in (\ref{unif-poisson}) are absolutely convergent
for finite values of $\lambda$. By hypothesis
 $\tilde a_n(\lambda)= a_n(\lambda) \la \frac{n}{\lambda}\ra^{-\mu}$ are bounded and moreover they satisfy
\begin{eqnarray}
\label{limitequi}
&&\lim_{\lambda\to\infty} \sum_{n=0}^\infty \frac{\lambda^n}{n!} e^{-\lambda}
\tilde a_n(\lambda)\; \left(1-\left\langle
\frac{n}{\lambda}\right\rangle^{\mu}\right) =0,\\\label{limitequi1}
&&\lim_{\lambda\to\infty}\int_{\rz} \tilde a_{[\sqrt{\lambda}
  s+\lambda]}(\lambda) \; \left(1-\left\langle
\frac{[\sqrt{\lambda} s+\lambda]}{\lambda}\right\rangle^{\mu}\right)
  \; \frac{e^{-\frac{s^2}{2}}}{\sqrt{2\pi}}
\,ds=0\,
\end{eqnarray}
since we may bound uniformly for $\lambda$ large each of the terms inside the sum and the integral respectively by
\begin{eqnarray*}
&&C^1_\mu \sum_{n=0}^\infty \frac{\lambda^n}{n!} e^{-\lambda}
\; n^{\mu}<C_\mu^0, \hspace{.2in} \mbox{ and } \hspace{.1in}
C^2_\mu \int_{\rz} \; |s|^\mu \; \frac{e^{-\frac{s^2}{2}}}{\sqrt{2\pi}}
\,ds<C_\mu^0, \;\; \forall \lambda >1.
\end{eqnarray*}
Therefore there is no restriction if we assume all $a_n(\lambda)$ bounded by $1$ since if we prove
 (\ref{unif-poisson}) for $\tilde a_n(\lambda)$ then it holds for $a_n(\lambda)$ by the limits
 (\ref{limitequi})-(\ref{limitequi1}).

For all $h>0$ there exists $\alpha<\beta$ such that
\begin{eqnarray*}
\int_\beta^\infty\frac{e^{-\frac{s^2}{2}}}{\sqrt{2\pi}} \, ds <h/7, \hspace{.4in}
\int_{-\infty}^\alpha\frac{e^{-\frac{s^2}{2}}}{\sqrt{2\pi}} \, ds <h/7.
\end{eqnarray*}
Now by (\ref{poisson}) we have
\begin{eqnarray*}
\lim_{\lambda\to\infty}\sum_{1+\frac{\beta}{\sqrt{\lambda}} \leq \frac{n}{\lambda}} \frac{\lambda^n}{n!} e^{-\lambda}
=\int_\beta^\infty\frac{e^{-\frac{s^2}{2}}}{\sqrt{2\pi}} \, ds, \hspace{.4in}
\lim_{\lambda\to\infty}\sum_{ \frac{n}{\lambda} \leq 1+\frac{\alpha}{\sqrt{\lambda}}} \frac{\lambda^n}{n!} e^{-\lambda}
=\int_{-\infty}^\alpha\frac{e^{-\frac{s^2}{2}}}{\sqrt{2\pi}} \, ds
\end{eqnarray*}
Therefore there exists $\lambda_1$ such that for all $\lambda >\lambda_1$ we have
\begin{eqnarray*}
\sum_{1+\frac{\beta}{\sqrt{\lambda}} \leq \frac{n}{\lambda}} \frac{\lambda^n}{n!} e^{-\lambda} \leq h/6,
\hspace{.4in}
\sum_{\frac{n}{\lambda} \leq 1+\frac{\alpha}{\sqrt{\lambda}}} \frac{\lambda^n}{n!} e^{-\lambda}\leq h/6.
\end{eqnarray*}
Let denote $\ds I_{\alpha,\beta}(\lambda)=\int_\alpha^\beta a_{[\sqrt{\lambda} s+\lambda]}(\lambda)
\frac{e^{-\frac{s^2}{2}}}{\sqrt{2\pi}} \, ds$. We obtain for all $\lambda>\lambda_1$:
\begin{eqnarray}
\label{est.poisson}
\ds\left|\sum_{n=0}^\infty \frac{\lambda^n}{n!} e^{-\lambda} a_n(\lambda)-
\int_{-\infty}^\infty a_{[\sqrt{\lambda} s+\lambda]}(\lambda)
\frac{e^{-\frac{s^2}{2}}}{\sqrt{2\pi}} \, ds\right| \leq
\underbrace{\left|\sum_{
\alpha<\frac{n-\lambda}{\sqrt{\lambda}}<\beta}
\frac{\lambda^n}{n!} e^{-\lambda} a_n(\lambda)- I_{\alpha,\beta}(\lambda)\right|}_{J_{\alpha,\beta}(\lambda)} +2h/3
\end{eqnarray}
Using the Stirling formula there exists $\lambda_2$ such that for all $\lambda>\lambda_2$ we have
\begin{eqnarray*}
\left| \sum_{
\alpha<\frac{n-\lambda}{\sqrt{\lambda}}<\beta}
\frac{\lambda^n}{n!} e^{-\lambda} [1-\frac{n!}{\sqrt{2\pi n} (n/e)^n})]\; a_n(\lambda) \right| \leq
h/9.
\end{eqnarray*}
This yields the following estimate
\begin{eqnarray}
\label{eq.14}
J_{\alpha,\beta}(\lambda) \leq
\left| \sum_{
\alpha<\frac{n-\lambda}{\sqrt{\lambda}}<\beta}
\frac{1}{\sqrt{2\pi n}} [e^{\lambda \varphi(\frac{n}{\lambda})}-e^{-(\frac{n-\lambda}{\sqrt{\lambda}})^2/2}] \right|
+\underbrace{\left| \sum_{
\alpha<\frac{n-\lambda}{\sqrt{\lambda}}<\beta} \frac{e^{-(\frac{n-\lambda}{\sqrt{\lambda}})^2/2}}{
\sqrt{2\pi n}} a_n(\lambda)-I_{\alpha,\beta}(\lambda)\right|}_{L_{\alpha,\beta}(\lambda)}+h/12,
\end{eqnarray}
where $\varphi(x)=x-1-x\ln(x)$. To complete the proof one needs to  estimate infinitesimally
the two terms in the r.h.s. of the above inequality. Notice that by means of Riemann sums we have
\begin{eqnarray}
\label{riemann}
\lim_{\lambda\to\infty}
\sum_{\alpha<\frac{n-\lambda}{\sqrt{\lambda}}<\beta}
\frac{e^{-(\frac{n-\lambda}{\sqrt{\lambda}})^2/2}}{\sqrt{2\pi n}}=
\lim_{\lambda\to\infty}
\sum_{\alpha<\frac{n-\lambda}{\sqrt{\lambda}}<\beta}
\frac{e^{- (\frac{n-\lambda}{\sqrt{\lambda}})^2/2}}{\sqrt{2\pi \lambda}}=
\int_{\alpha}^\beta \frac{e^{-s^2/2}}{\sqrt{2\pi}} \,ds.
\end{eqnarray}
 We have
\begin{eqnarray*}
\sum_{
\alpha<\frac{n-\lambda}{\sqrt{\lambda}}<\beta}
\frac{1}{\sqrt{2\pi n}} [e^{\lambda \varphi(\frac{n}{\lambda})}-e^{-(\frac{n-\lambda}{\sqrt{\lambda}})^2/2}]=
\sum_{
\alpha<\frac{n-\lambda}{\sqrt{\lambda}}<\beta}
\frac{e^{- (\frac{n-\lambda}{\sqrt{\lambda}})^2/2}}{\sqrt{2\pi n}} [e^{\lambda \tilde\varphi(\frac{n}{\lambda})}-1],
\end{eqnarray*}
where $\tilde\varphi(x)=x-1-x\ln(x)+(x-1)^2/2$ which is an increasing function null at $1$. Therefore one obtains
\begin{eqnarray}
\label{eq.13}
\left|\sum_{
\alpha<\frac{n-\lambda}{\sqrt{\lambda}}<\beta}
\frac{1}{\sqrt{2\pi n}} [e^{\lambda \varphi(\frac{n}{\lambda})}-e^{-(\frac{n-\lambda}{\sqrt{\lambda}})^2/2}]\right|
\leq \int_{\alpha}^\beta \frac{e^{-s^2/2}}{\sqrt{2\pi}} \,ds \; \; \;
[e^{\lambda \tilde\varphi(\frac{\beta}{\sqrt{\lambda}}+1)}-1],
\end{eqnarray}
with a r.h.s. converging to $0$ when $\lambda\to \infty$ since $
\lim_{\lambda\to \infty} e^{\lambda \tilde\varphi(\frac{\beta}{\sqrt{\lambda}}+1)}=1,$ which we bound
by $h/12$ for $\lambda $ larger than a given $\lambda_3$.
One can obtain the estimate
\begin{eqnarray*}
L_{\alpha,\beta}(\lambda) \leq  \left|\sum_{
\alpha<\frac{n-\lambda}{\sqrt{\lambda}}<\beta} \frac{e^{-(\frac{n-\lambda}{\sqrt{\lambda}})^2/2}}{
\sqrt{2\pi \lambda}} \,a_n(\lambda)-I_{\alpha,\beta}(\lambda)\right|+h/18,
\end{eqnarray*}
using the fact that
\begin{eqnarray*}
\sum_{
\alpha<\frac{n-\lambda}{\sqrt{\lambda}}<\beta} \frac{e^{-(\frac{n-\lambda}{\sqrt{\lambda}})^2/2}}{
\sqrt{2\pi \lambda}} \underbrace{\left|\frac{1}{\sqrt{(\frac{n-\lambda}{\sqrt{\lambda}})\frac{1}{\sqrt{\lambda}} +1}}-1\right|}_{(1)}
\leq h/18,
\end{eqnarray*}
since $\lim_{\lambda\to\infty}(1)=0$ and the sum is uniformly bounded by (Equ.~\ref{riemann}).
By  splitting the integral in $I_{\alpha,\beta}(\lambda)$ over the intervals
$[\frac{n-\lambda}{\sqrt{\lambda}},\frac{n+1-\lambda}{\sqrt{\lambda}})$ one can show that
\begin{eqnarray*}
\left|I_{\alpha,\beta}(\lambda)-\sum_{\alpha<\frac{n-\lambda}{\sqrt{\lambda}}<\beta} a_n(\lambda)
\int_{\frac{n-\lambda}{\sqrt{\lambda}}}^\frac{n+1-\lambda}{\sqrt{\lambda}} \frac{e^{-s^2/2}}{\sqrt{2\pi}}
 ds\right| \leq h/18.
\end{eqnarray*}
This yields
\begin{eqnarray}
\label{eq.15}
L_{\alpha,\beta}(\lambda) \leq h/9+
\sum_{\alpha<\frac{n-\lambda}{\sqrt{\lambda}}<\beta}
 [\frac{e^{-(\frac{n-\lambda}{\sqrt{\lambda}})^2/2}}{\sqrt{2\pi \lambda}} \; -
 \int_{\frac{n-\lambda}{\sqrt{\lambda}}}^\frac{n+1-\lambda}{\sqrt{\lambda}} \frac{e^{-s^2/2}}{\sqrt{2\pi}} ds]
\end{eqnarray}
with a r.h.s. converging to $0$ when $\lambda\to \infty$ which we bound by $h/18$ for $\lambda$
larger than $\lambda_4$. Combining  the estimates (\ref{eq.14}), (\ref{eq.13}) and (\ref{eq.15})   with (\ref{est.poisson})
we obtain that for all $ h>0$, there exists $\lambda_0$ such that for all $\lambda >\lambda_0$ we have
\begin{eqnarray*}
\ds\left|\sum_{n=0}^\infty \frac{\lambda^n}{n!} e^{-\lambda} a_n(\lambda)-
\int_{-\infty}^\infty a_{[\sqrt{\lambda} s+\lambda]}(\lambda)
\frac{e^{-\frac{s^2}{2}}}{\sqrt{2\pi}} \, ds\right| \leq
h.
\end{eqnarray*}
This gives the claimed result.
\hfill\fin

\bibliographystyle{empty}

\end{document}